\documentclass[aps,onecolumn,prd,superscriptaddress,floatfix,showpacs,showkeys]{revtex4-2}

\usepackage[latin1]{inputenc}
\usepackage[T1]{fontenc}
\usepackage{lmodern}
\usepackage{newtxtext}

\usepackage{amsmath}
\usepackage{amssymb}
\usepackage{amsthm}
\usepackage{amscd}
\usepackage{mathtools}
\usepackage{empheq}
\usepackage{mathrsfs}
\usepackage{dsfont}
\usepackage{scalerel}

\usepackage{graphicx}
\usepackage{grffile}
\usepackage{xcolor}
\usepackage{subcaption}
\usepackage{sidecap}

\usepackage{tikz}
\usetikzlibrary{arrows.meta, positioning, shapes.geometric, decorations.pathmorphing}

\usepackage{etoolbox}
\usepackage[mathlines]{lineno}
\usepackage[ruled]{algorithm2e}

\usepackage[colorlinks=true, pdfstartview=FitV, linkcolor=blue, citecolor=blue, urlcolor=blue]{hyperref}

\makeatletter
\renewcommand{\p@subsubsection}{\thesection\,}

\renewcommand{\subsubsection}{\@startsection{subsubsection}{3}{\z@}%
  {2ex \@plus 1ex \@minus .1ex}%
  {1ex \@plus .1ex}%
  {\normalfont\small\bfseries}}
\makeatother

\makeatletter
\def\appendix@toc@setup{%
  \let\oldl@section\l@section
  \def\l@section##1##2{%
    \begingroup
    \def\numberline####1{\makebox[6.5em][l]{Appendix ####1.\quad}}%
    \oldl@section{##1}{##2}%
    \endgroup
  }%
}
\apptocmd{\appendix}{\addtocontents{toc}{\protect\appendix@toc@setup}}{}{}
\makeatother

\newtheorem{thm}{Theorem}[section]
\newtheorem{prop}[thm]{Proposition}

\newtheorem{rem}{Remark}[section]

\def\bprop{\begin{prop}}
\def\eprop{\end{prop}}
\def\bt{\begin{thm}}
\def\et{\end{thm}}
\def\br{\begin{rem}}
\def\er{\end{rem}}

\usepackage{mathrsfs}
\usepackage{bm}

\def\be{\begin{equation}}
\def\ee{\end{equation}}
\def\bes{\begin{equation*}}
\def\ees{\end{equation*}}
\def\bea{\begin{equation} \begin{aligned}}
\def\eea{\end{aligned} \end{equation}}
\def\beas{\begin{equation*} \begin{aligned}}
\def\eeas{\end{aligned} \end{equation*}}

\def\bi{\begin{itemize}}
\def\ei{\end{itemize}}

\def\d{\, \mathrm{d}}
\newcommand{\x}{\mathbf{x}}

\definecolor{rred}{rgb}{0.7,0,0.1}
\definecolor{ccyan}{rgb}{0,.5,1}
\definecolor{greenrb}{rgb}{0.2,0.6,0.2}

\definecolor{ppink}{rgb}{1,0.1,0.6}

\renewcommand{\x}{\bm{x}}          

\def\W{{\boldsymbol{W}_t}}

\newcommand{\cG}{\mathcal{G}}

\newcommand{\G}{\bm{G}}
\def\adots{
  \mathinner{\mkern1mu\raise1pt\hbox{.}\mkern2mu\raise4pt\hbox{.}
  \mkern2mu\raise7pt\vbox{\kern7pt\hbox{.}}\mkern1mu}}
\def\Id{{\rm Id}}
\def\Re{{\rm Re \,}}

\usepackage{wrapfig,overpic,setspace}

\begin{document}

\title{Beyond Critical Slowing Down: Slow Modes, Extreme Tails, and Field Decoherence in Tipping Transitions}

\author{Micka{\"e}l D. Chekroun}
\email{mchekroun@atmos.ucla.edu}
\affiliation{Department of Atmospheric and Oceanic Sciences, University of California, Los Angeles, CA 90095-1565, USA}
\affiliation{Department of Earth and Planetary Sciences, Weizmann Institute of Science, Rehovot 76100, Israel}
\author{Valerio Lucarini}
\affiliation{School of Computing and Mathematical Sciences, University of Leicester, Leicester, LE17RH, UK}
\affiliation{School of Sciences, Great Bay University, Dongguan, P.R. China}

\date{\today}

\begin{abstract}
Tipping transitions are abrupt reorganizations in which a system leaves one statistical regime for another. Their anticipation is usually framed through critical slowing down: recovery becomes slower, autocorrelation rises, variance may increase, and spectral power shifts toward low frequencies. For noisy, spatially extended systems, however, these scalar signatures do not by themselves reveal which part of the dynamics is losing resilience, whether rare excursions toward a competing state are becoming more likely, or whether the spatial field is reorganizing before the transition. We examine these complementary questions in the stochastic Ghil--Sellers energy balance model, a latitudinally resolved climate model whose ice--albedo feedback supports warm and snowball metastable regimes. We follow three complementary viewpoints rather than aiming at a single universal indicator. Reduced Ruelle--Pollicott (RP) resonances and Kolmogorov modes diagnose relaxation and response in physically interpretable observables; Extreme Value Theory probes the accessibility and persistence of excursions into the tails; and Data-Adaptive Harmonic Modes (DAHMs) diagnose the frequency-resolved organization of the full temperature field. Near tipping, several reduced decay rates slow together and their modes become geometrically harmonized along a common transition direction. The associated Green functions develop delayed-recovery tails and enhanced low-frequency susceptibility only when the relevant response residues are nonzero. When approaching the warm-to-snowball bifurcation, the cold extremes of the Ghil--Sellers model become less sharply bounded and more clustered, while the full field becomes less compressible at fixed DAHM rank and its phase distribution broadens, even as a dominant low-frequency transition component emerges. The resulting picture goes beyond critical slowing down.
Early warning is more comprehensively viewed as a question of \emph{which component of the dynamics is reorganizing, through which observable, toward which transition channel, and at which level of spatial description}. Slow relaxation, directional tail excursions, and loss of full-field phase coherence are distinct physical manifestations of the
approach to a critical transition, and their joint interpretation provides a more informative basis for early-warning analysis in high-dimensional stochastic systems. We further show that a small localized change in the albedo parameterization can create an additional pair of folds, illustrating that a distance to tipping may carry structural model uncertainty in addition to statistical uncertainty.

\end{abstract}


\maketitle


\setcounter{tocdepth}{3}
\tableofcontents

\section{Introduction}

Critical transitions in climate dynamics - often referred to as \textit{tipping points \citep{Boers2025,Lenton2025}} - are commonly associated with the loss of stability of a metastable state and the subsequent displacement of the system toward a competing regime. In the climate context, such transitions are especially difficult to diagnose because the observed system is noisy, high-dimensional, spatially extended, and only partially observed. Different mechanisms can lead to tipping, depending on how stability is lost \citep{ashwin2012tipping,alkhayuon2018rate,Feudel2018,Halekotte2020,AlkhayuonTysonWieczorek2021_phase_tipping,Hastings2026}. They are of five types: (i)~B-tipping (bifurcation tipping), in which a slowly varying parameter crosses a critical threshold, so that the existing attractor disappears or becomes unstable and the system is forced into a new state; (ii)~N-tipping (noise-induced tipping), in which random fluctuations drive the system across a basin boundary before any deterministic threshold is reached, triggering an unexpected transition; (iii)~P-tipping (phase-space tipping), in which a perturbation or excursion in state space displaces the system into a different basin of attraction, despite no significant parameter change; (iv)~R-tipping (rate-induced tipping), in which the external forcing changes too rapidly for the system to track its moving equilibrium, leading to a transition without bifurcation or strong noise; and (v)~S-tipping (shock tipping), in which a sufficiently large abrupt disturbance pushes the system beyond a basin boundary, initiating a shift to an alternative stable regime. Together, these mechanisms highlight that tipping can occur because a threshold is crossed (B), random variability is strong (N), the state is displaced (P), forcing changes too quickly (R), or an external shock occurs (S). Many real climate and ecological tipping events may involve combinations of these mechanisms.%

In the case of B-tipping, which is the case specifically under investigation in this paper, classical early-warning signals rely on the idea of critical slowing down: as a stability threshold is approached, recovery from perturbations becomes slower, autocorrelations increase, variance may grow, and spectral power shifts toward low frequencies \citep{scheffer2012anticipating,boers2021critical}. This picture has been influential and physically transparent, and it is closely related to degenerate fingerprinting ideas in climate dynamics \citep{Held2004}. Yet, it also leaves open a central question: what exactly is slowing down, through which observable is it seen, and how does this scalar signature relate to rare excursions and to the organization of the full spatial field?

This question is particularly sharp for stochastic non-equilibrium systems. In such systems, the relevant statistical state is not a deterministic trajectory but an invariant probability measure, and perturbations are naturally studied through response theory. Following the stochastic climate paradigm initiated by \citet{hasselmann1976} and developed in modern response-theoretic form \citep{majda2005information,LucariniChekroun2023,Lucarini_Chekroun_PRL24,chekroun2025kolmogorov}, we consider It\^o diffusions (Eq.~\eqref{Eq_Gauss}), with invariant measure $\mu$, and use Green functions to describe the linear response of observables. The Green function (Eq.~\eqref{Eq_Green_intro}) gives the causal impulse response associated with an observable $\Psi$ and a perturbation pattern ${\bm G}$, and the response formula given in Eq.~\eqref{Eq_LRF} turns this kernel into a prediction for the perturbed expectation. This connects early-warning diagnostics to the same objects that control statistical sensitivity and climate response \citep{kubo1966,majda2005information,Santos2022,hasselmann1976}.

The spectral structure behind this response is provided by Ruelle--Pollicott (RP) resonances and Kolmogorov modes \citep{Chekroun_al_RP2,Santos2022,chekroun2025kolmogorov}. For stochastic systems with a smooth invariant density, these resonances are isolated eigenvalues of the Kolmogorov generator $\mathcal L_K$ in Eq.~\eqref{Eq_Kop}, and their eigenfunctions describe the dominant modes of relaxation of probability densities and observables \citep{eckmann2003spectral,Chekroun_al_RP2,dyatlov2015stochastic}. The correlation expansion \eqref{Eq_decomp_corr1} and the power-spectrum representation (Eq.~\eqref{Eq_PSD}) show how these modes organize temporal variability. The same modes also enter the Green-function expansion (Eq.~\eqref{GreenH}), with weights $\alpha_{j\ell}$ given by Eq.~\eqref{Eq_alpha}. Thus a resonance close to the imaginary axis is not, by itself, a universal warning signal: it matters for a given diagnostic only if its coefficient is non-negligible for the chosen observable and perturbation. This residue-conditioned viewpoint is the basis of the RP early-warning theory developed in \citep{Chekroun_Lucarini26_theoretic}; here we use it to interpret numerical evidence in a concrete climate model.

Our testbed is the stochastic Ghil--Sellers energy balance model (EBM) \citep{Ghil1976}, recalled in Section \ref{Sec_EBM_model}. This model is cast as a partial differential equation that describes the  time evolution of latitudinally averaged surface temperature via the competing effects of solar energy input and scattering, infrared energy emission to space, and energy transport across latitudinal belts and has historically played a major role in theoretical climate dynamics. It is simple enough to permit long integrations and repeated spectral diagnostics, but rich enough to provide a meaningful testbed for testing optimal fingerprinting for climate change detection and attribution \citep{Lucarini_Chekroun_PRL24} and for studying climate multistability through the ice-albedo feedback \citep{Ghil1976,Bodai2015,Lucarini2022NPG}. 
We consider a warm metastable climate state subject to stochastic forcing and to radiative perturbations designed to move the system in the proximity of a warm-to-snowball transition. The goal is not to reduce tipping to a single indicator, but to examine how the approach to transition is expressed through three complementary diagnostics: reduced RP resonances and Green functions, extreme-value statistics, and data-adaptive harmonic analysis of the full field.

The first diagnostic is a reduced transfer-operator analysis. Since within our discretization, the discretized EBM has $37$ temperature components, we estimate reduced RP resonances \citep{Chekroun_al_RP2} in the physically interpretable observable plane
\be
Z_k=
\left(
\Delta T_k-\overline{\Delta T},
T_{{\rm ave},k}-\overline{T_{\rm ave}}
\right).
\ee
Here $T_{\rm ave}$ measures the bulk thermal state, while $\Delta T$ measures the area-weighted contrast between low and mid--high latitudes. Within this reduced phase space, the reduced Markov matrix (Eq.~\eqref{eq:transition_matrix}) then provides a finite-dimensional approximation of the reduced propagator. This reduced construction does not claim to resolve the full Kolmogorov spectrum of the EBM; rather, it identifies the relaxation geometry visible through two climate-relevant observables. The numerical results show that, near tipping, several dominant reduced decay rates compress toward the origin and the corresponding reduced Kolmogorov modes become geometrically harmonized along a common slow direction. The consequence for response is described in Section~\ref{Sec_EBM_Green_consequences}: the slow cluster contributes to Green functions through Eq.~\eqref{eq:EBM_green_slow_cluster}, and the static susceptibility is amplified according to Eq.~\eqref{eq:EBM_static_susceptibility_residues} only when the relevant response residues are nonzero.

The second diagnostic probes a different part of the dynamics: the tails of the invariant distribution. RP resonances describe relaxation and response in the bulk of the statistical dynamics, but a metastable system approaching a transition may also visit the edge of its basin more often and linger there for longer times. Here, we take inspiration from \citep{Faranda2014} and use Extreme Value Theory (EVT) \citep{Gnedenko1943,dehaan2006extreme} to examine annual extremes of the global mean temperature anomaly. A key parameter of the Generalized Extreme Value (GEV) distribution (Eq.~\eqref{eq:GEV_distribution}) that is used to study the properties of extreme events is the tail index $\xi$, which indicates whether the tails are upper limited (or lower limited, in the case one studies negative fluctuations) or extend to infinity. A second parameter of great relevance in EVT  is the extremal index $\theta$, which measures the clustering of extremes \citep{lucarini2016extremes}. The results show a marked asymmetry: as the warm state approaches the transition, cold extremes become less sharply bounded and more persistent than warm extremes. This is consistent with the physical picture that cold excursions probe the escape direction toward the competing climate state, whereas warm excursions point deeper into the warm metastable regime.

The third diagnostic uses the full temperature field rather than a projection on a reduced observable space. For this purpose we rely on Data-Adaptive Harmonic Modes (DAHMs) \citep{chekroun2017data}, which combine the frequency organization of Fourier analysis with the multivariate pattern extraction of empirical decompositions; see also \citep{schmidt2019spectral}. The DAHM construction starts from the block-Hankel matrix of time-lagged cross-correlations (Eq.~\eqref{Hij}). Its eigenvectors are organized by Fourier frequency and can be written in a multivariate cosine form (Eq.~\eqref{Eq_DAHM}), with amplitudes and phases for each spatial channel. This phase information is intrinsic to the DAHM spectral representation, not a post-processing convention. It allows us to define the phase order parameter (Eq.~\eqref{Eq_DAH_phase_order_parameter}) and to diagnose whether the latitudinal temperature field remains phase-organized across channels and frequency bands.

The DAHM results reveal a complementary full-field signature of tipping. In the current-climate regime, a small number of dominant DAHM modes reconstructs most of the multivariate variability, indicating strong compression by coherent spatio-temporal harmonic patterns. Along the near-tipping trajectory, the leading DAHM modes still capture the large-scale trend, but the same fixed number of modes gives a poorer reconstruction. More modes are required to reach comparable accuracy, and the DAHM phase distribution broadens. We interpret this broadening as multivariate phase decoherence: the field retains a coherent transition component, but loses sharp phase organization across latitude. This is distinct from, and complementary to, the harmonization seen in the reduced RP modes. The reduced observables reveal an emerging slow transition geometry; the full-field DAHM analysis shows that the spatial phase organization of the actual temperature field becomes less coherent along the transition path.

\subsection{Key Message and Structure of the Paper}
The main message of the paper is therefore that metastable climate tipping in a spatially extended stochastic system is not faithfully represented by a single early-warning diagnostic. In the stochastic Ghil--Sellers model, the approach to transition is expressed across several spatio-temporal layers: a reduced slow spectral subspace in physically interpretable observables, an asymmetric and persistent cold tail in the global mean temperature statistics, and a loss of phase coherence in the full latitude-dependent temperature field. These three diagnostics answer different questions about the same transition. RP resonances and Green's functions identify the dominant relaxation and response channels visible through selected observables and perturbations. EVT identifies whether rare excursions toward the edge state become more accessible and persistent in time. DAHMs identify whether the full spatio-temporal field remains compressible into coherent harmonic patterns and whether its latitudinal channels retain organized phase relations. Taken together, they show that the onset of tipping is not only a temporal slowing-down phenomenon, but a reorganization of the system's spatio-temporal variability, tail dynamics, and field coherence.

Before any of this, it is useful to have a visualization of the phase-parameter-space skeleton organizing the dynamics of tipping. Figures~\ref{Fig_GS_surface} and~\ref{Fig_GS_bifurcation} display the tipping geometry of the model used throughout: the two coexisting climates, the unstable edge state that separates them, the folds at which each stable state is lost, and the two solar-forcing values at which all the stochastic experiments below are performed. They also make plain why a single scalar is an incomplete summary, since the mean temperature and the equator-to-pole contrast order the warm and edge states in opposite senses.

The paper is organized as follows. Section~\ref{Sec_EBM_mother} introduces the Ghil--Sellers EBM, its deterministic tipping geometry, and the stochastic numerical experiments that provide the physical laboratory for the rest of the paper. Section~\ref{Sec_Kolmo_tipping} then develops the first diagnostic lens by connecting Green functions, RP resonances, Kolmogorov modes, and the reduced transfer-operator construction directly to the relaxation and response geometry observed in the EBM. Section~\ref{Sec_EVT} turns to the tails and uses EVT to diagnose the accessibility and persistence of cold excursions toward the competing regime. Section~\ref{Sec_Multivariate} moves from reduced observables to the full latitude-dependent temperature field and develops the DAHM power and phase diagnostics leading to the decoherence interpretation. Section~\ref{Sec_Synthesis} distills the physical early-warning lessons of these three viewpoints without reference to the numerical details,  and  Section~\ref{Sec_Structural} then asks how robust the tipping geometry on which those lessons rest is to the model's own parameterization uncertainties, showing that a small, narrow-band change to the albedo closure suffices to add a pair of folds to the diagram. Finally, Section~\ref{Sec_Discussion_Outlook} discusses the broader scope, limitations, and extensions of the lessons gained by this work.


\section{The physical laboratory: tipping in the stochastic Ghil--Sellers model}
\label{Sec_EBM_mother}

Before introducing the spectral and statistical diagnostics, it is useful to make the physical transition itself explicit. The Ghil--Sellers model provides a deliberately simple setting in which the ingredients behind early warning can be separated without removing the spatial structure of the temperature field. Its deterministic dynamics contains competing climate states organized by the ice--albedo feedback, while stochastic forcing turns these states into metastable regimes and allows the system to sample fluctuations both within a regime and toward its boundary. The same model can therefore be viewed at three levels that will recur throughout the paper: recovery inside the warm regime, rare excursions toward the competing cold regime, and reorganization of the full latitude-dependent field.

\begin{figure}
  \centerline{\includegraphics[width=.95\textwidth]{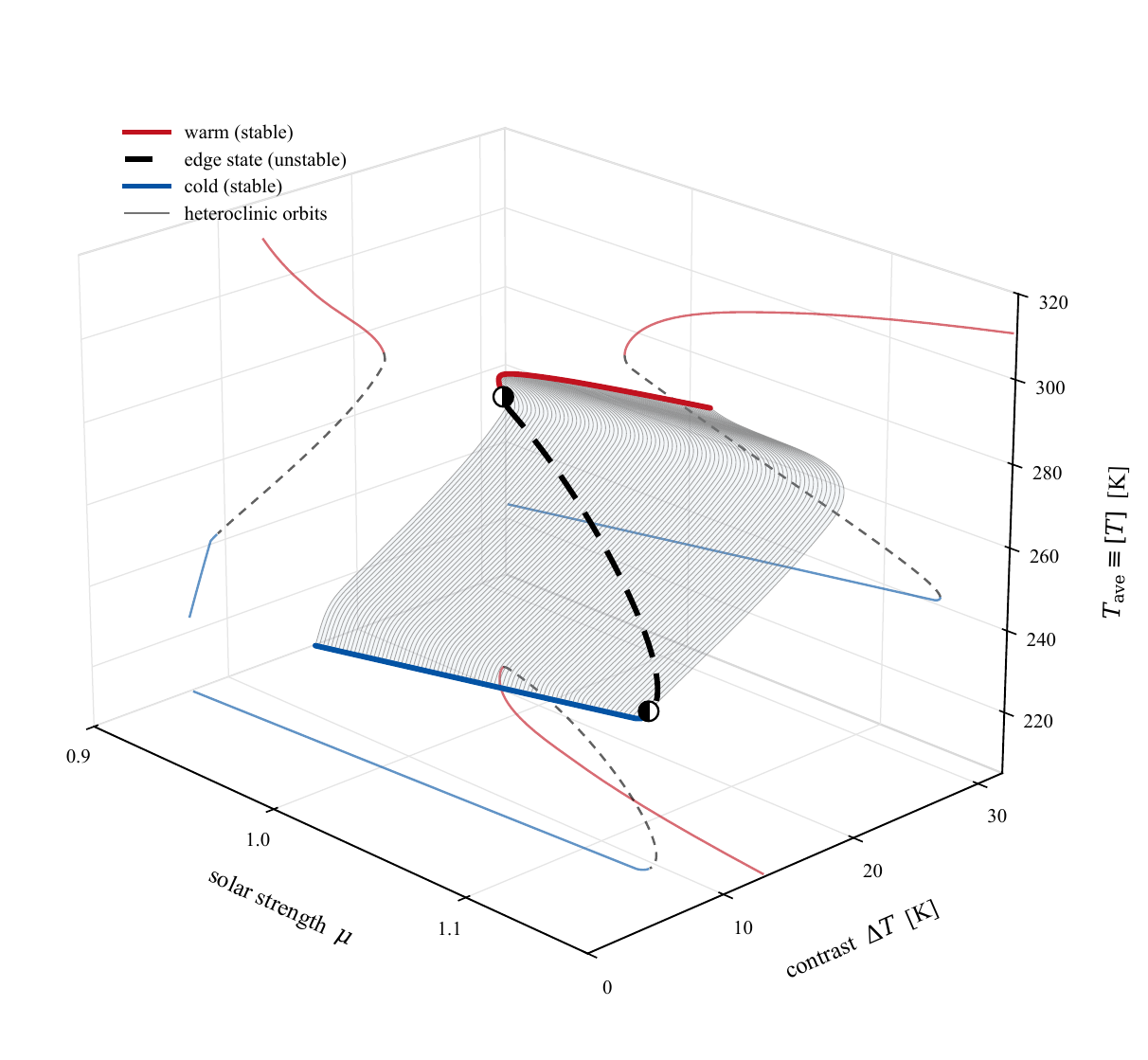}}
  \caption{ {\bf The deterministic tipping geometry of the Ghil--Sellers model.} The surface swept out by the transient trajectories of the model, shown in the space spanned by the solar-forcing factor $\mu$, the low-to-high-latitude temperature contrast $\Delta T$, and the global mean temperature $T_{\rm ave}$, for $\alpha_{\max}=0.6$. Thin grey curves are individual trajectories released from the unstable edge state; together they map out the sheet to which the dynamics rapidly aligns. Thick curves are the branches of steady states embedded in that sheet: warm (red, solid), unstable edge state (black, dashed), and cold (blue, solid). The two saddle-node folds, where the warm and edge branches merge at $\mu_{w\to c}$ and the edge and cold branches merge at $\mu_{c\to w}$, are marked with the half-filled circles of Arnold's convention. The same three branches are projected onto the two rear vertical planes and the floor of the box; the projection onto the $(\mu,T_{\rm ave})$ plane is the classical bifurcation diagram reproduced in Fig.~\ref{Fig_GS_bifurcation}(a). This figure is a higher-precision version of Fig.~6 in \citep{Bodai2015}; see Appendix \ref{App_numerics} for details.}
\label{Fig_GS_surface}
\end{figure}

Figure~\ref{Fig_GS_surface} sets the stage by displaying the whole deterministic geometry at once. Two scalar diagnostics organize it: the area-weighted global mean temperature $T_{\rm ave}$, which plays the role of an order parameter, and the low-to-high-latitude contrast $\Delta T$ between two equal-area boxes separated at $30^\circ$N, which is a simple measure of thermodynamic disequilibrium. The thin grey curves are the heteroclinic trajectories that connect the unstable edge state to either attractor; because the model is strongly diffusive, an arbitrary initial condition collapses onto this one-dimensional family very quickly, so the curves sweep out a thin sheet rather than filling the space \citep{Bodai2015}. Embedded in that sheet are the three branches of steady states, which meet pairwise at the two folds. The figure makes visible something a scalar diagram cannot: along any transient the contrast $\Delta T$ rises well above its value at either endpoint, so the climate \emph{in transition} is further from thermodynamic equilibrium than either of the states it connects. Appendix~\ref{App_numerics} summarizes how the figure was produced.

\subsection{The Ghil-Sellers energy balance model}\label{Sec_EBM_model}
The Ghil-Sellers model \citep{Ghil1976} describes succinctly the key processes of absorption and reflection of solar radiation, the emission to space of infrared radiation, and the large-scale transport of energy from low to high latitudes \citep{Peixoto1992}. The model has been key to  explaining the metastability of the Earth's climate  \citep{Bodai2015,Ghil2020,Lucarini2022NPG,Lucarini_Chekroun_PRL24} and describes the evolution of the zonally-averaged surface temperature $T(x,t)$, in which $x=2\phi/\pi\in[-1,1]$ denotes the normalised latitude $\phi$ and $t$ is time:
\bea\label{Eq_EBM}
 c(x) \partial_t T_S  =&\left(\frac{2}{\pi}\right)^2\frac{1}{\cos\left(\frac{\pi x}{2}\right)}\partial_x\left(\cos\left(\frac{\pi x}{2}\right)k(x,T_S)\partial_xT_S\right)  \\
  &\hspace{2.5cm}+\mu Q(x)\left(1 - \alpha(x,T_S)\right) - \sigma T_S^4\left(1 - m\tanh(c_3T_S^6)\right), 
\eea
with standard differentiation notation and boundary and initial conditions given by $\partial_xT_S(-1,t) = \partial_xT_S(1,t) = 0$ and $T_S(x,0) = T_0(x)$.
Here $c(x)$ is the effective heat capacity of the atmosphere, land, and ocean per unit surface area at $x$. The first term on the right hand side (RHS) is an effective diffusive representation of meridional energy transport. The coefficient $k(x,T_S)$ parameterizes, at the zonally averaged level, the net poleward redistribution associated with sensible and latent heat transport rather than a molecular diffusion process. We use throughout the tabulated functional form and parameter values of \citep{Bodai2015}; in particular, $k$ is not retuned as the control parameter is varied. This is useful physically because the approach to tipping studied below is then driven by the radiative control and the ice--albedo feedback, while the large-scale meridional transport closure is held fixed.

The solar forcing (second term on the RHS) is modulated by the solar constant $\mu$, the  irradiance $Q$ and the albedo $\alpha$. The longwave emission (third term on the RHS) is represented by Boltzmann's law, modified by the greenhouse effect, whose intensity is modulated by the constant $m$. The functional dependence of the albedo---which decreases with temperature as presence of ice requires low temperatures---allows for the presence  of the  ice-albedo feedback, which is key for climate multistability. Further details on the model are reported in Appendices \ref{App_params} and \ref{App_numerics}. 


The physical usefulness of this model for an early-warning study comes from the fact that its transition geometry is transparent. In the deterministic limit, the ice--albedo feedback allows a warm climate and a much colder, strongly glaciated climate, which is usually referred to as a snowball state \citep{Ghil2020}, to coexist over a finite range of the solar-forcing factor $\mu$, with an unstable edge state separating their basins of attraction \citep{Bodai2015,Ghil2020}. The parameter $\mu$ therefore plays the role of a control parameter, while an integrated quantity such as the area-weighted global mean temperature $T_{\rm ave}$ provides a natural order parameter. Moving $\mu$ along the warm branch toward its fold weakens the restoring dynamics before the warm state is lost. Adding stochastic forcing changes the operational problem: the deterministic attracting states become metastable statistical regimes, and sufficiently large fluctuations may explore the basin boundary, or even trigger an escape, before the deterministic branch disappears. This distinction between weakening recovery and growing escape risk is one reason why a single scalar critical-slowing-down indicator is not expected to tell the whole story \citep{Chekroun_Lucarini26_theoretic}.

\begin{figure}
  \centerline{\includegraphics[width=\textwidth]{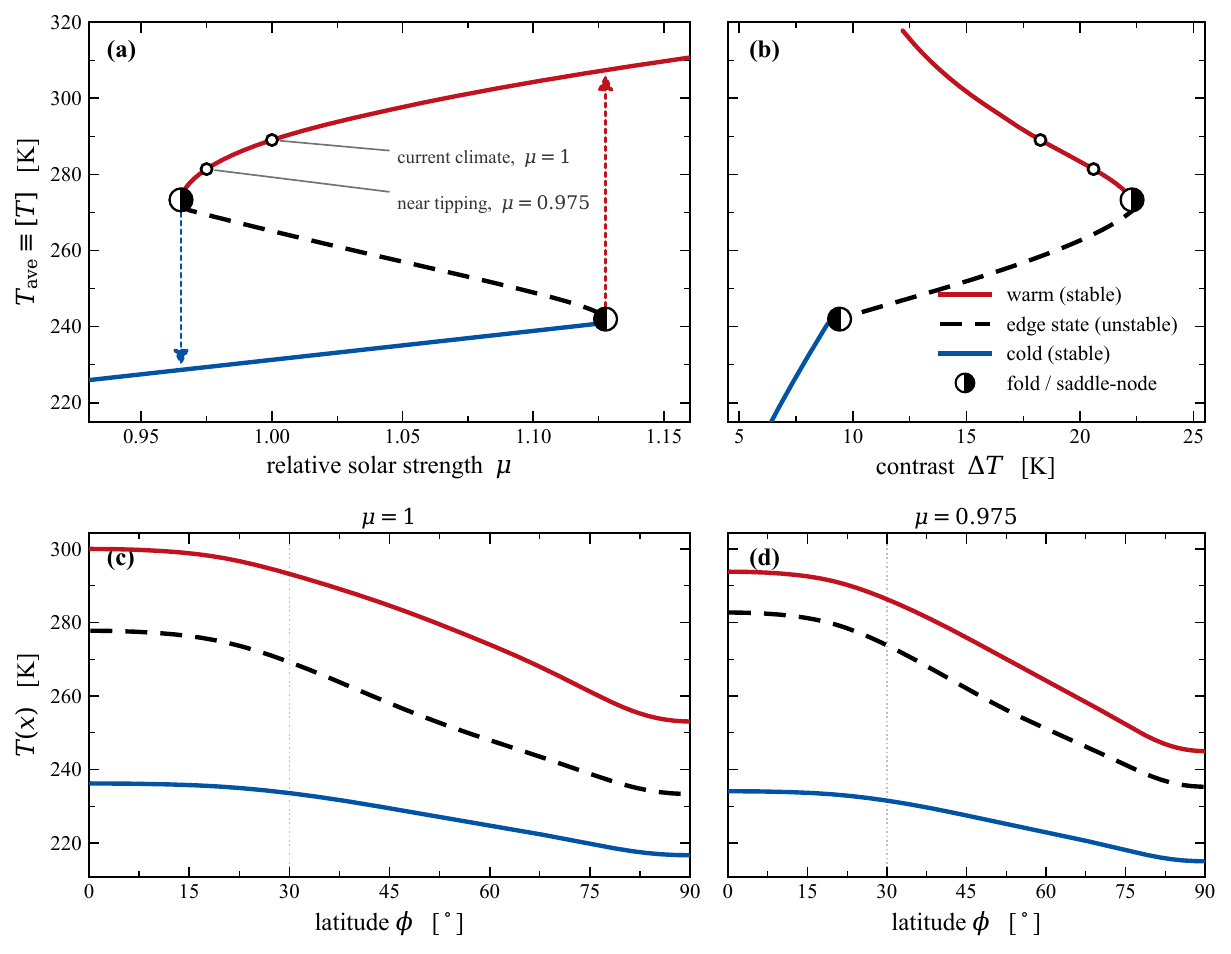}}
  \caption{ {\bf Deterministic tipping geometry of the Ghil--Sellers energy balance model.} (a) Bifurcation diagram in the $(\mu,T_{\rm ave})$ plane, with the solar-forcing factor $\mu$ as control parameter and the area-weighted global mean temperature $T_{\rm ave}$ as order parameter. Red: stable warm branch; black dashed: unstable edge branch; blue: stable cold branch; half-filled circles: the two saddle-node folds. Dotted arrows indicate the two branches of the hysteresis loop. Open circles mark the solar forcings of the stochastic experiments, $\mu=1$ (current climate) and $\mu=0.975$ (precritical). (b) The same equilibria in the $(\Delta T,T_{\rm ave})$ plane, sharing the ordinate with (a), where $\Delta T$ is the temperature difference between two equal-area boxes separated at $30^\circ$N. (c,d) Steady latitudinal temperature profiles of the three states at $\mu=1$ and $\mu=0.975$ (northern hemisphere; the model is symmetric about the equator); the dotted vertical line marks the box border $x=1/3$. Between (c) and (d) the warm and edge profiles visibly approach one another. In the deterministic model the warm state is lost at the fold; stochastic forcing broadens these states into metastable probability clouds and permits excursions toward the basin boundary before the fold is reached.}
\label{Fig_GS_bifurcation}
\end{figure}

 Figure~\ref{Fig_GS_bifurcation} makes this transition geometry quantitative and identifies the two settings used in the stochastic experiments below. Panel (a) is the classical bifurcation diagram: the warm branch is lost at a fold located at $\mu_{w\to c}=0.9652$, where $T_{\rm ave}=273.3$~K, and the cold branch is lost at $\mu_{c\to w}=1.1277$, so that the two regimes coexist over a window of width $\Delta\mu=0.163$ in solar strength. The dotted arrows close the hysteresis loop. Panel (b) shows the same family of steady states in the $(\Delta T,T_{\rm ave})$ plane; there the three branches form a single continuous curve, because the mean temperature parametrizes the family monotonically, and the folds become interior points at which stability changes rather than turning points. Panels (c) and (d) show the corresponding latitudinal temperature profiles at the two experimental settings. Two features matter for what follows. First, the vertical ordering of the warm and edge branches is reversed between $T_{\rm ave}$ and $\Delta T$: the edge state is colder than the warm state but is \emph{further} from thermodynamic equilibrium, a polar-amplification signature already emphasized by \citet{Bodai2015}. Second, moving from $\mu=1$ to $\mu=0.975$ leaves the warm state only $0.0098$ in $\mu$ above its fold, and the separation between the warm state and the edge state in $T_{\rm ave}$ falls from $23.9$~K to $12.0$~K, as the profiles in panels (c) and (d) show directly. The precritical experiment is therefore genuinely close to the deterministic threshold, in the specific sense that  the distance to the edge state in $T_{\rm ave}$ has roughly halved. 

\subsection{Numerics and tipping point experiments}\label{Sec_EBM_num_setting}
We numerically solve the Ghil--Sellers model on $M=37$ latitude grid points separated by $5^\circ$, with a time step of one day. The spatial derivatives are approximated with standard centered finite differences. To represent unresolved variability and continuously probe the stability of the warm regime, we add an additive white-noise forcing to the discretized temperature field with covariance matrix $\eta_0\mathbf{I}_M$, where $\eta_0\geq0$, and integrate the resulting stochastic system with the Euler--Maruyama scheme. The deterministic reference state is the warm climate obtained at the present-day solar forcing $\mu=1$ and $\eta_0=0$; see Fig.~1(a) of \citep{Bodai2015}. The strong-noise experiments use $\eta_0=0.2$, while a weaker-noise setting is introduced below for the spectral estimation.

It is worth recording what the deterministic model already says about these two settings, since it fixes the baseline against which the stochastic diagnostics of the following sections should be read. Linearizing about the warm state, the leading eigenvalue weakens from $\lambda_1=-4.73\times10^{-9}\,\mathrm{s}^{-1}$ at $\mu=1$ to $-2.25\times10^{-9}\,\mathrm{s}^{-1}$ at $\mu=0.975$; the corresponding linear relaxation time $\tau=1/|\lambda_1|$ therefore roughly doubles, from about $6.7$ to about $14.1$~years. Critical slowing down is thus present, and measurably so, in the precritical experiment. The point of the analysis that follows is not to dispute this, but to show that this single scalar rate is not the whole story: over the same interval  the observable-space separation from the edge state also halves (Fig.~\ref{Fig_GS_bifurcation}), and neither effect is visible in every observable. Near a fold one expects $\lambda_1\propto-(\mu-\mu_{w\to c})^{1/2}$, which is consistent with the two values quoted above.
 

\begin{figure}
  \centerline{\includegraphics[width=.85\textwidth, height=.45\textwidth]{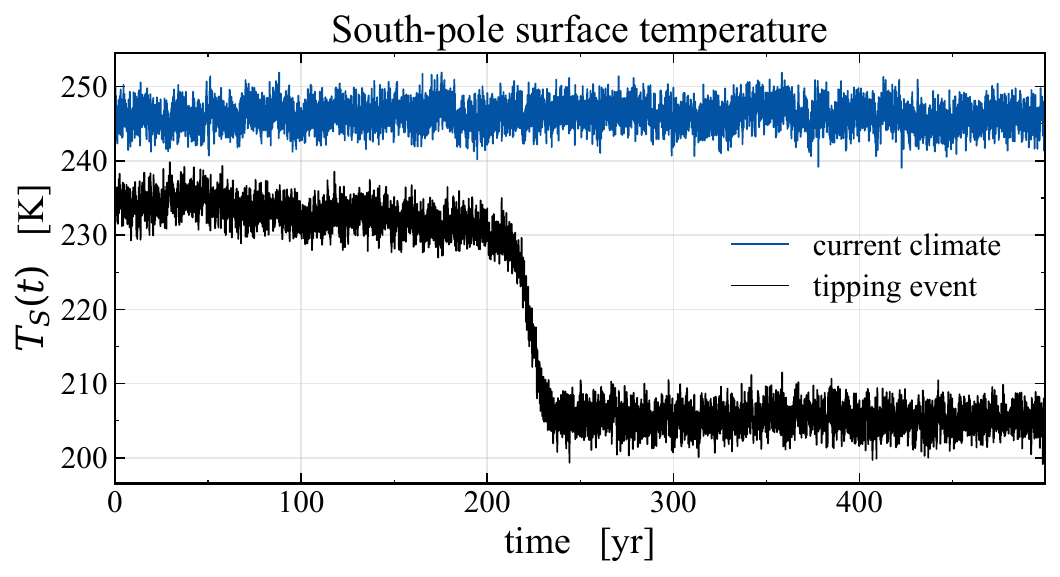}}
  \caption{{\bf Reference fluctuations and a warm-to-cold tipping trajectory in the stochastic energy balance model.}
The South-Pole surface temperature $T_S(t)\equiv T(x=-1,t)$, in kelvin,  is shown as a function of time  in years for the strong-noise experiments. The current-climate trajectory uses $\mu=1$, whereas the precritical experiment uses $\mu=0.975$ and undergoes an abrupt transition to the colder regime. The two records illustrate the distinction between stationary fluctuations within the warm metastable regime and a trajectory that ultimately leaves it; the corresponding latitude-dependent fields are analyzed in Section~\ref{Sec_Multivariate}.}
\label{Fig_EBM_dynamics}
\end{figure}


 In addition to the strong-noise transient experiment shown in Fig.~\ref{Fig_EBM_dynamics}, we perform low-noise experiments with $\eta_0=0.1$ in order to estimate the reduced RP spectrum in regimes where the metastable climate state is well resolved. We consider two autonomous reference regimes: the current-climate regime, obtained by keeping the parameters at their reference warm-state values, and a precritical near-tipping regime, obtained by freezing the parameters close to the transition threshold while the system still fluctuates around the warm branch ($\mu=0.975$);  both settings are marked on the deterministic diagram of Fig.~\ref{Fig_GS_bifurcation}(a), and the corresponding warm, edge and cold profiles are those of Fig.~\ref{Fig_GS_bifurcation}(c,d). In both cases, after discarding spin-up, long stationary time series are generated and projected onto the same reduced set of observables. These reduced trajectories are then used to estimate the reduced Markov matrix $M_\tau$ defined in Eq.~\eqref{eq:transition_matrix} below, from which the reduced RP resonances and Kolmogorov modes are extracted. The purpose of the low-noise setting is not to reproduce the full forced transition path, but to compare the intrinsic relaxation geometry of the current and near-critical stochastic climates.


\section{Slow modes, response, and reduced relaxation geometry}
\label{Sec_Kolmo_tipping}

Critical slowing down is fundamentally a statement about recovery. In a high-dimensional stochastic system, however, recovery need not be governed by one scalar rate, nor need every observable see the same slow structure. The first lens of our analysis therefore combines response theory, RP resonances, and a reduced transfer-operator construction in the EBM. The aim is not to separate ``method'' from ``application'' into two disconnected parts, but to move from the causal response problem to the spectral objects that organize it and then directly to the relaxation geometry visible in physically interpretable temperature observables.

Three pieces of information should be kept distinct throughout. A resonance gives a decay rate and hence a persistence time; its Kolmogorov mode---the eigenfunction of the Kolmogorov generator (Eq.~\eqref{Eq_Kop}) associated with that resonance, defined in Section~\ref{Sec_RPresonances}---gives the spatial or observable pattern that carries that decay; and the response residue determines whether a chosen forcing and observable actually activate that mode in a measured response. This is the intuitive content of the residue-conditioned viewpoint developed more formally in the companion study \citep{Chekroun_Lucarini26_theoretic}: a small denominator creates the possibility of amplification, but the geometry of the observable and perturbation decides whether that possibility is realized.

\subsection{Green's functions and response}\label{Sec_GreenFct}
We consider the following class of It\^o stochastic differential equation (SDEs):
\be\label{Eq_Gauss}
\d X_t = {\bm F} (X_t) \d t +{\bm \Sigma}(X_t) \d \W,
\ee  
in which $X_t$ is a $d$-dimensional state vector, ${\bm F}$ is a vector field on $R^d$, ${\bm \Sigma}$ is a $d\times p$ matrix-valued function on $R^d$, and $\W$ is a $p$-dimensional Brownian motion. We assume that ${\bm F}$  and ${\bm \Sigma}$ are sufficiently smooth to ensure the existence of a unique ergodic invariant measure $\mu$, representing the system's statistical equilibrium. Conditions for this to hold can be found in e.g.~\citep{Hairer_Majda,Chekroun_al_RP2} and references therein. The noise term in Eq.~\eqref{Eq_Gauss} often represents unresolved small-scale variables or processes. Such stochastic systems can be rigorously derived from chaotic systems under appropriate timescale separation assumptions  \citep{majda2001mathematical}. Hasselmann proposed in \citep{hasselmann1976} to use such stochastic systems to study the dynamics of slow chaotic climatic variables influenced by fast weather variables, modeled as the stochastic component in Eq.~\eqref{Eq_Gauss}. A recent review on this topic is provided in \citep{LucariniChekroun2023}.

Assume that ${\bm F}$ is perturbed to ${\bm F}+ \epsilon g(t) {\bm G}$, where $\epsilon$ is sufficiently small, $g$ is a bounded time-dependent function, and ${\bm G}$  is a smooth vector field on $\mathbb{R}^d$. Using standard perturbative arguments at the leading order in $\epsilon$ for the solution to the Fokker-Planck equation associated with Eq.~\eqref{Eq_Gauss}, we can derive a useful formulation of the FDT, often referred to as linear response theory (LRT) \citep{majda2005information,Santos2022}. 

The goal of response theory is to provide formulas that rely solely on the structural characteristics and statistics of the unperturbed system, enabling the prediction of the time evolution of the system's statistical quantities when a perturbation is applied. These statistical quantities are typically ensemble averages $\langle \Psi \rangle_{\rho_\epsilon^t}$ (of arbitrary observable $\Psi$) with respect to the system's probability distribution $\rho_\epsilon^t$ at time $t$, that satisfies the following (perturbed) Fokker-Planck equation:
\be\label{Eq_FKE}
\partial_t \rho_\epsilon= -\mbox{div} ({\bm F} \rho_\epsilon )-\epsilon g(t)\mbox{div}({\bm G} \rho_\epsilon ) + \frac{1}{2}\sum_{i,j=1}^d \partial_{ij} \big(a_{ij}(\x) \rho_\epsilon \big),
\ee
in which the $a_{ij}(\x)$ are the coefficients of the covariance matrix  $ {\bm \Sigma}(\x){\bm \Sigma}(\x)^T$. 
 In other words, one wishes to quantify the impact of the term $\epsilon g(t)\mbox{div}({\bm G} \cdot)$ on the ensemble average $\langle \Psi \rangle_{\mu}$, i.e.~when statistics are evaluated with respect to the unperturbed system's statistical equilibrium $\mu$, the stationary solution of Eq.~\eqref{Eq_FKE} when $\epsilon=0$.

The LRT provides this answer. It predicts that  
\be\label{Eq_LRT}
\langle \Psi \rangle_{\rho_\varepsilon^t} - \langle \Psi \rangle_{\mu}=\delta^{(1)}[\Psi] (t) +O(\epsilon^2),
\ee
in which the first-order response operator $\delta^{(1)}[\Psi] (t)$  is given, explicitly, by:
\be\label{Eq_LRF}
\delta^{(1)}[\Psi] (t) = \epsilon \int_{-\infty}^t  \cG_{\Psi,G} (t-s) g(s)\d s.
\ee
Here, $\G_{\Psi,G}$ is the system's Green function associated with the observable $\Psi$.  It is given as
\be\label{Eq_Green_intro}
\cG_{\Psi,G}(t)=\Theta(t) \hspace{-1ex}\int  \hspace{-1ex} \bigg(e^{t  \mathcal{L}_K}
 \Psi(\x)\big[L_{\bm G}\log(\mu)\big] (\x)\bigg)  \mu(\d \x),
\ee
in which $\Theta(t)$ is the Heaviside function ensuring causality  \citep{ruelle2009,Lucarini2017,Lucarini2018JSP},   the operator $L_{\G}$ is the following Liouville operator 
\be\label{Eq_Liouville}
L_{\G}= -\mbox{div} ({\bm G} \cdot ),
\ee
and $\mathcal{L}_K$ denotes the Kolmogorov operator associated with Eq.~\eqref{Eq_Gauss}: 
\begin{equation}\label{Eq_Kop}
\mathcal{L}_K \psi (\x) ={\bm F} (\x) \cdot \nabla \psi+ \sum_{i,j=1}^d a_{ij}(\x) \partial_{ij} \psi.
\end{equation} 
 Note that the linear character lies here in the linear dependence of the response operator $\delta^{(1)}[\Psi] (t)$ on the term $\epsilon g(t)$, controlling the magnitude of the perturbation of the drift term in Eq.~\eqref{Eq_Gauss}. We refer to  \citep{Santos2022} for formulas of response operators  when the (Gaussian) diffusion term is perturbed and to \citep{chekroun2025kolmogorov} for the case of non-Gaussian jump-diffusion perturbations.   
 At higher-order in $\epsilon$, these response operators involve nonlinear dependences on the perturbation terms; see \citep{ruelle_nonequilibrium_1998,lucarini2008,LucariniColangeli2012,Lucarini_Chekroun_PRL24}.

Equations \eqref{Eq_LRT}-\eqref{Eq_Kop} offer significant practical and theoretical insights. Firstly, they provide a generalized version of the FDT \citep{kubo1966,majda2005information}, as Eq.~\eqref{Eq_Green_intro} enables us to interpret  the Green's function in terms of time-lagged correlations \citep{majda2005information,Baiesi2013,Santos2022,chekroun2025kolmogorov}. 
Another important aspect  as demonstrated in \citep{Santos2022}  by relying on  the framework of  \citep{Chekroun_al_RP2}, is that they allow for decomposing the Green's function into contributions from the system's modes of (unforced) variability. The latter are the Kolmogorov modes that we describe in the next section. 


\subsection{Ruelle-Pollicott resonances, correlations and power spectra expansions}\label{Sec_RPresonances}

The theory of Ruelle--Pollicott (RP) resonances was first developed for deterministic chaotic systems; see \citep{ruelle1986locating,pollicott1986meromorphic,keller1999stability,baladi2000positive,gaspard2005chaos} and references therein. In that setting, RP resonances describe the discrete spectrum of transfer operators \citep{baladi2000positive,schutte1999direct,schutte2001transfer,Chek_al14_RP}, or dually of Koopman operators \citep{mezic2005spectral,budivsic2012applied,brunton2022modern}, once these operators are represented on appropriate functional spaces. They have since become a central object in the spectral theory of dynamical systems \citep[e.g.][]{faure2011upper,giulietti2013anosov,dyatlov2019mathematical}, precisely because they encode the relaxation and frequency content of key statistical diagnostics: correlation decay, mixing rates, peaks in power spectra \citep{baladi2000positive,gaspard2002dynamical,butterley2007smooth,melbourne2007power,lasota2013chaos,eisner2015operator}, and, in suitable settings, coherent or almost-invariant structures  \citep{froyland2007detection,froyland2013analytic,froyland2014almost,Froyland2021}. The deterministic theory is, however, technically delicate. The physically relevant invariant measures are often singular with respect to Lebesgue measure, and the corresponding spectral theory requires anisotropic or otherwise problem-adapted spaces \citep{butterley2007smooth,faure2011upper,gouezel2006banach,baladi2017quest,giulietti2019parabolic}.

For stochastic systems, the same organizing idea takes a concrete operator-theoretic form. As framed in \citep{Chekroun_al_RP2}, RP resonances can be identified with isolated eigenvalues of the Kolmogorov generator $\mathcal L_K$ in Eq.~\eqref{Eq_Kop}, and the associated eigenfunctions are the Kolmogorov modes, the stochastic analogue of Koopman-generator eigenfunctions in the deterministic setting \citep{froyland2013estimating}. The often-used statement that ``noise regularizes the problem'' has a precise meaning here. For a broad class of hypoelliptic It\^o diffusions, diffusion together with the bracket structure smooths transition probabilities and typically yields a sufficiently regular invariant density and Markov semigroup to formulate the spectral problem in standard spaces such as $L^2_\mu$. One can then rely on semigroup theory \citep{Engel_Nagel} rather than on the specialized anisotropic spaces needed for many deterministic chaotic systems \citep[e.g.][]{Hairer_Majda,Chekroun_al_RP2}. It is this smoothing of the probabilistic evolution---not a generic claim that randomness makes the dynamics simpler---that is meant by regularization. It leads directly to expansion formulas for temporal correlations and power spectra, such as Eq.~\eqref{Eq_decomp_corr1} below. The same viewpoint has recently been extended to jump-diffusion models driven by non-Gaussian noise disturbances \citep{chekroun2025kolmogorov}.

The relevance of these resonances for early-warning theory is immediate. RP resonances do not merely provide abstract spectral data; they identify the decay channels through which observables relax toward the system's statistical equilibrium. A resonance close to the imaginary axis produces a long-lived contribution to correlations and power spectra, but only for observables that project onto the corresponding Kolmogorov mode. Thus the RP framework naturally refines the usual critical-slowing-down picture: it specifies which mode is slowing down, at which frequency, and through which observable it is visible. This point becomes concrete in Section~\ref{Sec_EBM_reduced_Modes}, in which the reduced RP resonances and Kolmogorov modes of the EBM are interpreted as signatures of an emerging slow subspace rather than as a single scalar gap.

The scope of RP-based methods has broadened substantially in recent years. They have been used to study sensitivity to perturbations and response theory \citep{Chek_al14_RP,lucarini2016response,Santos2022,Lucarini_Chekroun_PRL24,chekroun2025kolmogorov}, critical slowing down and crises \citep{tantet2015early,tantet2018crisis}, stochastic bifurcations \citep{bagheri2014effects,RP_Hopf}, and the inference of homogenized coarse-grained processes from multiscale data \citep{crommelin2011diffusion}. They also provide diagnostics for stochastic parameterizations and reduced-order models of nonlinear systems with complex slow--fast interactions \citep{chekroun2021stochastic}, as well as for data-driven emulation of geophysical turbulence \citep{Kondrashov_al2018_QG}. These developments complement earlier work on metastability leveraging spectral properties of Fokker--Planck operators and Markov state models \citep[e.g.][]{matkowsky1981eigenvalues,schutte2013metastability,pavliotis2014stochastic}.

 In the present work, we use this RP/Kolmogorov-mode viewpoint as our first lens on the tipping phenomenon: it reveals how the bulk relaxation and response geometry reorganizes as the stochastic EBM approaches transition. We now recall the main formulas from the RP theory to articulate our analysis. 

Assume that the It\^o diffusion in Eq.~\eqref{Eq_Gauss} possesses a unique ergodic invariant measure $\mu$ and generates a $C_0$ semigroup $e^{t\mathcal{L}_K}$ on $L^2_\mu(\mathbb{R}^d)$. Then \citet[Corollary~1]{Chekroun_al_RP2} gives, for any observables $f$ and $g$ that are square-integrable with respect to $\mu$, 
\be\label{Eq_decomp_corr1}
C_{f,g}(t)= \sum_{j=1}^N \left(\sum_{k=0}^{m_j-1}a_{jk}^{f,g}  \frac{ t^k}{k!} (\mathcal{L}_K-\lambda_j \textrm{Id})^{k}\right) e^{\lambda_j t} + \mathcal{Q}(t),
\ee
in which the remainder satisfies $\mathcal{Q}(t)\underset{t\rightarrow \infty}\longrightarrow 0$, and the coefficients $ a_{jk}^{f,g}$ are given by \citep[Eq.~(2.11)]{Chekroun_al_RP2}
\bea\label{Eq_coeff_aij}
a_{jk}^{f,g}&=\langle\Phi_j^\ast, g \rangle_\mu \int f(\x) (\mathcal{L}_K-\lambda_j \textrm{Id})^{k}   \Phi_j(\x) \mu (\d \x),
\eea
in which $\Phi_j$ (resp.~$\Phi_j^\ast$) denotes the eigenfunction  (dual eigenfunction) associated with the RP resonances $\lambda_j$, $\langle \cdot,\cdot \rangle_\mu$ denotes the $L^2_\mu$-Hermitian inner product,  while $m_1,\cdots,m_N$ correspond to the (algebraic) multiplicity of these resonances. 
Note that in the above decomposition, we assumed that the observables have been centered, i.e.~that the mean with respect to the ergodic measure $\mu$ has been subtracted.

For these systems, the spectrum of the Kolmogorov operator $\mathcal{L}_K$  is typically contained in the 
left-half complex plane \citep{eckmann2003spectral,Chekroun_al_RP2}, $\{z\in \mathbb{C}\;:\: \Re(z)\leq 0\}$ and its resolvent  $R(z)=(z \mbox{Id}-\mathcal{L}_K)^{-1}$, is a well-defined operator that possesses the following integral representation
\be\label{Eq_resolvent}
R(z) f =\int_{0}^{\infty} e^{-z t} e^{t \mathcal{L}_K} f \d t.
\ee

The RP resonances correspond then to the $N$ poles of the resolvent $R(z)$, and the $m_1,\cdots,m_N$ in Eq.~\eqref{Eq_decomp_corr1} to the orders of these poles. 
The location of these poles are independent of the observables but their relative importance (through the coefficients  $a_j^{f,g}$ depend on them.

A useful  consequence of Eq.~\eqref{Eq_decomp_corr1} is the decomposition of the \emph{power spectrum}, $S_{f, g}(\omega)$, in terms of RP resonances and Kolmogorov modes . The latter is obtained by taking the Fourier transform of the correlation function $C_{f,g}(t)$ (Wiener-Khinchin theorem), which gives e.g.~in the case of poles of order $1$ ($m_1=\cdots=m_N=1$),
\be\label{Eq_PSD}
S_{f, g}(\omega) =\sum_{j=1}^N \frac{a_{j0}^{f,g}   \mathrm{Re}(\lambda_j) }{(\omega - \mathrm{Im}(\lambda_j))^2 + \mathrm{Re}(\lambda_j)^2}+\mathcal{D}(\omega),	\ee
in which $\omega$ is a complex frequency and  $\mathcal{D}(\omega)$ denotes some ``background'' function decaying to $0$ as $|\omega|\rightarrow \infty$; see also  \citep{melbourne2007power}. 
The decomposition given by Eq.~\eqref{Eq_PSD} shows that the meromorphic extension into the complex plane of the power spectrum $S_{f, g}(\omega) $ when $\omega$ is real,  present poles that are exactly given by the RP resonances. These poles manifest themselves in the (real) power spectrum as peaks that stand out over a continuous background at frequencies given by the imaginary
part of the eigenvalues, when $a_{j0}^{f,g}\neq 0$ and if  the $\lambda_j$ are close enough to the imaginary axis.
 
Thus, the RP resonances and Kolmogorov modes characterize the various ways a system's stochastic dynamics expresses its temporal variability through observables. 

\subsection{Kolmogorov modes and response}\label{Sec_Kolmo_modes}
As explained below, the Green's function $\cG_{\Psi,G}(t)$ can also be interpreted as  time-lagged correlations between the observables $\Phi=L_{\bm G}\log(\mu)$  and $\Psi$, involving only the statistics of the unperturbed system. In that sense, it provides a general version of the FDT \citep{kubo1966,majda2005information} valid for diffusion models covered by Eq.~\eqref{Eq_Gauss}. 

 Using then the decomposition of (temporal) correlations in terms of Kolmogorov modes recalled in Section \ref{Sec_RPresonances},  the Green function can be approximated as 
\begin{equation}\label{GreenH}
\cG_{\Psi,G}(t) \approx \sum_{j=1}^{N}\sum_{\ell=0}^{m_j-1} \frac{\alpha_{j\ell}(\Psi)}{\ell!} e^{\lambda_jt}t^{\ell}, \quad t\geq 0, 
\end{equation}
with $\cG_{\Psi,G}(t)=0$ when $t\leq 0$, and in which the coefficients $\alpha_{j\ell}$ are given by:
\begin{equation}\label{Eq_alpha} 
\alpha_{j\ell}(\Psi)=\langle \Phi_j^\ast, \Psi \rangle_{\mu} \int \big[L_{\bm G} \log(\mu)\big](\x)(\mathcal{L}_{K} - \lambda_j \mbox{Id})^{\ell} \Phi_j (\x) \mu(\d\x),
\end{equation} 
by application of Eq.~\eqref{Eq_coeff_aij} with $f=L_{\bm G} \log(\mu)$ and $g=\Psi$. 
Thus, the  $\alpha$'s  weight, according to Eq.~\eqref{Eq_alpha}, provide the contribution of the Kolmogorov modes to the response for a given observable $\Psi$ and forcing pattern ${\bm G}$ defining the operator $L_{\bm G}$ given by Eq.~\eqref{Eq_Liouville}.

These modes also inform on the proximity to tipping points through the spectral gap $\gamma=\Re(\lambda_1)$  associated with the slowest decaying mode $\varphi_1$.  Rough dependence of statistics on parameters is found as $\gamma \rightarrow 0$ \citep{Chek_al14_RP}.  When approaching a tipping point, $\gamma \rightarrow 0$,  and Eq.~\eqref{GreenH} indicates that 
any Green's function decays sub-exponentially, irrespectively of the observable and forcing, unless the corresponding $\alpha$-coefficient vanishes.
At criticality the response may  even diverge \citep{Chek_al14_RP,Tantet2018,AshwinJSP,Santos2022}.

If each  isolated eigenvalue has unitary multiplicity, the time-lagged correlation for $\Psi$ is $\mbox{Corr}_{\Psi}(t)=\sum_{j=1}^N a_j(\Psi) e^{\lambda_j |t|} $ with $a_j(\Psi)$ defined in Corollary 1 of \citep{Chekroun_al_RP2}.
If $\gamma \rightarrow 0$, $\mbox{Corr}_{\Psi}(t)$ decays subexponentially when $\Psi$ has a non-vanishing projection onto $\varphi_1$. This is the so-called critical slowing down associated with tipping behaviour  \citep{scheffer2012anticipating,boers2021critical}, which was originally discovered in the context of continuous phase transition \citep{stanley1971}. If $\Psi\propto\varphi_1$, $\mbox{Corr}_{\Psi}(t)\propto e^{\lambda_1|t|}$  and $\varphi_1$ is the critical mode invoked by degenerate fingerprinting \citep{Held2004},  encoding the ``natural'' tipping observable.


\subsection{EBM's reduced RP resonances and Kolmogorov modes}\label{Sec_EBM_reduced_Modes}

Leveraging the Ulam method \citep{ulam1960collection}, Markov matrices are crucial for both learning transfer operator properties  and estimating the system's propagator's long-term behavior;  e.g.~\citep{schutte1999direct,junge2009discretization,klus2018data,Froyland2021}. This propagator governs the evolution of probability densities and reveals also how correlations and other system aspects change over time \citep{froyland2003detecting,Chek_al14_RP, froyland2013estimating}. This approach applies to both deterministic and stochastic systems \citep{froylandapproximating1998,fishman2002,Chekroun_al_RP2}. 

Ulam's method provides thus a way to estimate the propagator $e^{t\mathcal{L}_K}$  using Markov transition matrices. Eigenvalues of the Kolmogorov-L\'evy  operator are then obtained through logarithm formulas \citep{crommelin2011diffusion,Chekroun_al_RP2}. 
This method consists of subdividing
the state space $\mathcal{X}$, typically shadowing the dynamics' forward attractor, 
 into $N_g$ non-intersecting \emph{cells} or \emph{boxes} $\{B_i\}_{i=1}^{N_g}$ and estimating the dynamics' probability transitions across these boxes. Mathematically, the propagator, $e^{t\mathcal{L}_K}$, for a given transition time $t=\tau$, is approximated by a $N_g\times N_g$ Markov transition matrix $M_{\tau}$,  whose entries are given by \citep{Chekroun_al_RP2}:
\begin{equation}\label{projected transfer operator}
\left[M_{\tau}\right]_{i,j}=\frac{1}{\rho_0(B_i)}\int_{B_i}e^{\tau \mathcal{L}_K}\mathds{1}_{B_j}(\x)\d \mu( \x),
\end{equation}
for $i,j=1,\ldots,N_g$, in which $\mathds{1}_{B_i}$ denotes the indicator function on the box $B_i$, $\mu$ denotes the system's ergodic  invariant probability measure and $\rho_0$, a given initial distribution. The transition matrix is then estimated  by computing a classical maximum likelihood estimator given by \citep{crommelin2011diffusion,schutte1999direct}:
\be\label{eq:transition_matrix}
 \left[M_{\tau}\right]_{i,j} = \frac{\#\bigg\{ \Big(\x_{k}\in B_j \Big) \land \Big(\x_{k+\ell} \in B_i \Big) \bigg\}}{\# \Big\{\x_{k} \in B_j\Big\}},
\ee
 for $i,j = 1,\ldots, N_g$. 
Given a sampling time $\delta t$ at  which the time series $\{\x_k\}_{k=1}^{N_d}$ solving Eq.~\eqref{Eq_Gauss}, is collected, the formula \eqref{eq:transition_matrix} counts the number of observations ($\#\{\cdot\}$)  visiting the box $B_i$ after $\ell= \lfloor \tau/\delta t \rfloor$ iterations, knowing that it already landed in $B_j$.
The resulting operation results into a coarse-graining of the dynamics encoded by $M_\tau$ \citep{crommelin2006b,Chekroun_al_RP2} that incorporates artificial diffusion \citep{froyland2013estimating}, of minor impact when $N_d$ and $N_g$ are sufficiently large. Alternatively, the probability transitions in Eq.~\eqref{eq:transition_matrix} can be estimated from many short-term trajectories instead of a single long-term trajectory \citep{dellnitz1999approximation,schutte1999direct,schutte2001transfer,froyland2010coherent,klus2018data}.

For high-dimensional systems, time series constructed from reduced coordinates (observables) are often used instead of the full-state vector $\x_k$ for obvious computational reasons. In this case, transition matrices can still be estimated in a reduced state space and their eigenvalues  can still be informative about the genuine system's transitions (in the full state space) through the notion of reduced RP resonances; see \citep[Theorem A]{Chek_al14_RP} and  \citep{Chekroun_al_RP2,RP_ENSO}.

\begin{figure}
\centerline{\includegraphics[width=.95\textwidth, height=.55\textwidth]{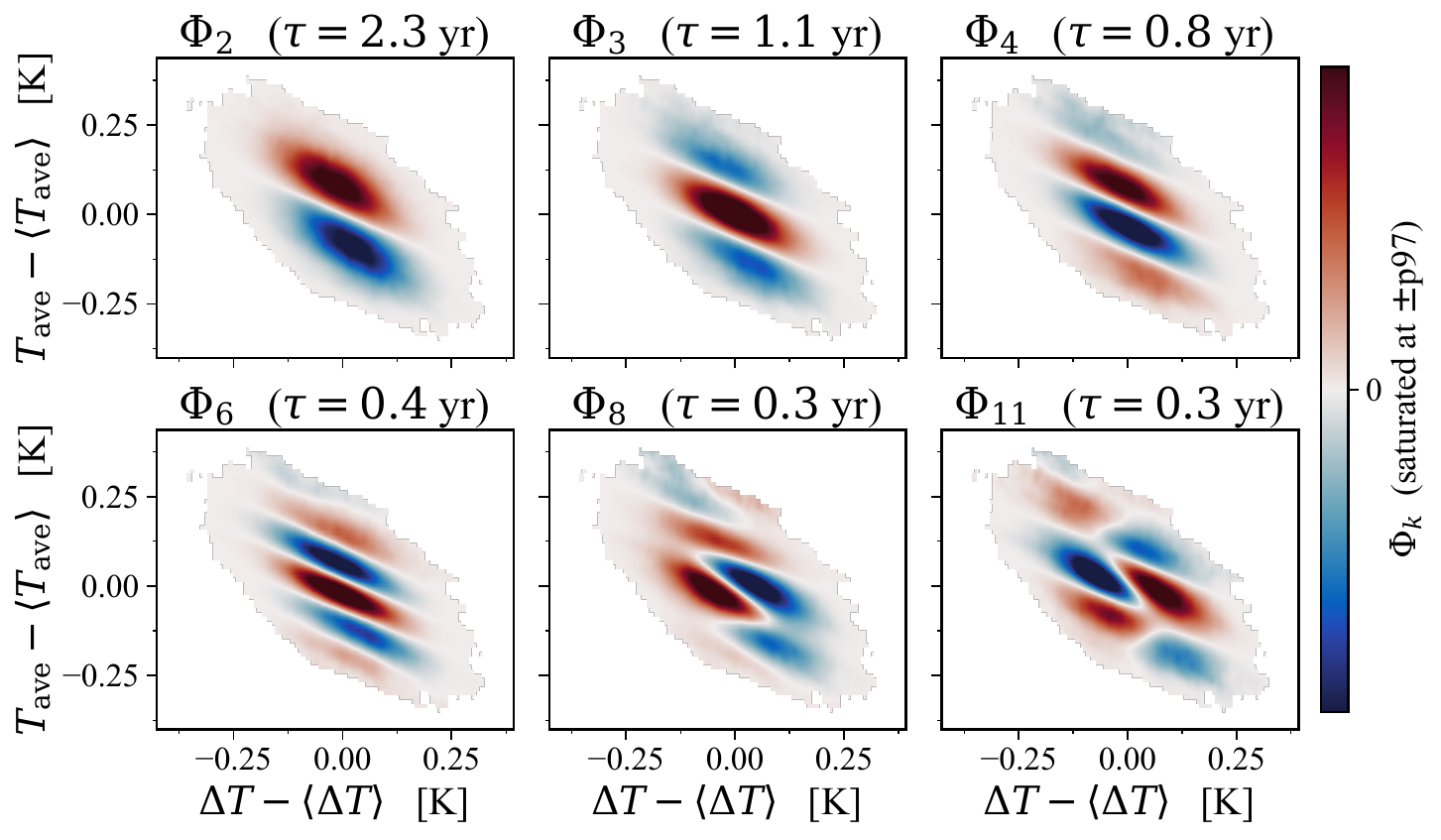}}
\caption{{\bf Reduced Kolmogorov modes in the current-climate regime, low-noise case.}
Representative right eigenvectors of the Markov approximation $M_\tau$ are shown in the two-dimensional observable plane $(\Delta T-\overline{\Delta T},T_{\rm ave}-\overline{T_{\rm ave}})$. Here $T_{\rm ave}$ is the area-weighted global mean temperature and $\Delta T$ is the area-weighted subtropical minus mid--high-latitude temperature contrast, using both hemispheres. The sign of each eigenvector is arbitrary; the physically relevant information is the geometry of its nodal set and positive--negative lobes. In the current-climate regime these structures remain comparatively localized and mode-dependent, indicating several distinct relaxation patterns within the warm-state probability cloud. Both axes are temperature anomalies and should be read in kelvin.}
\label{Fig_Kolmo_current}
\end{figure}

\begin{figure}
\centerline{\includegraphics[width=.95\textwidth, height=.55\textwidth]{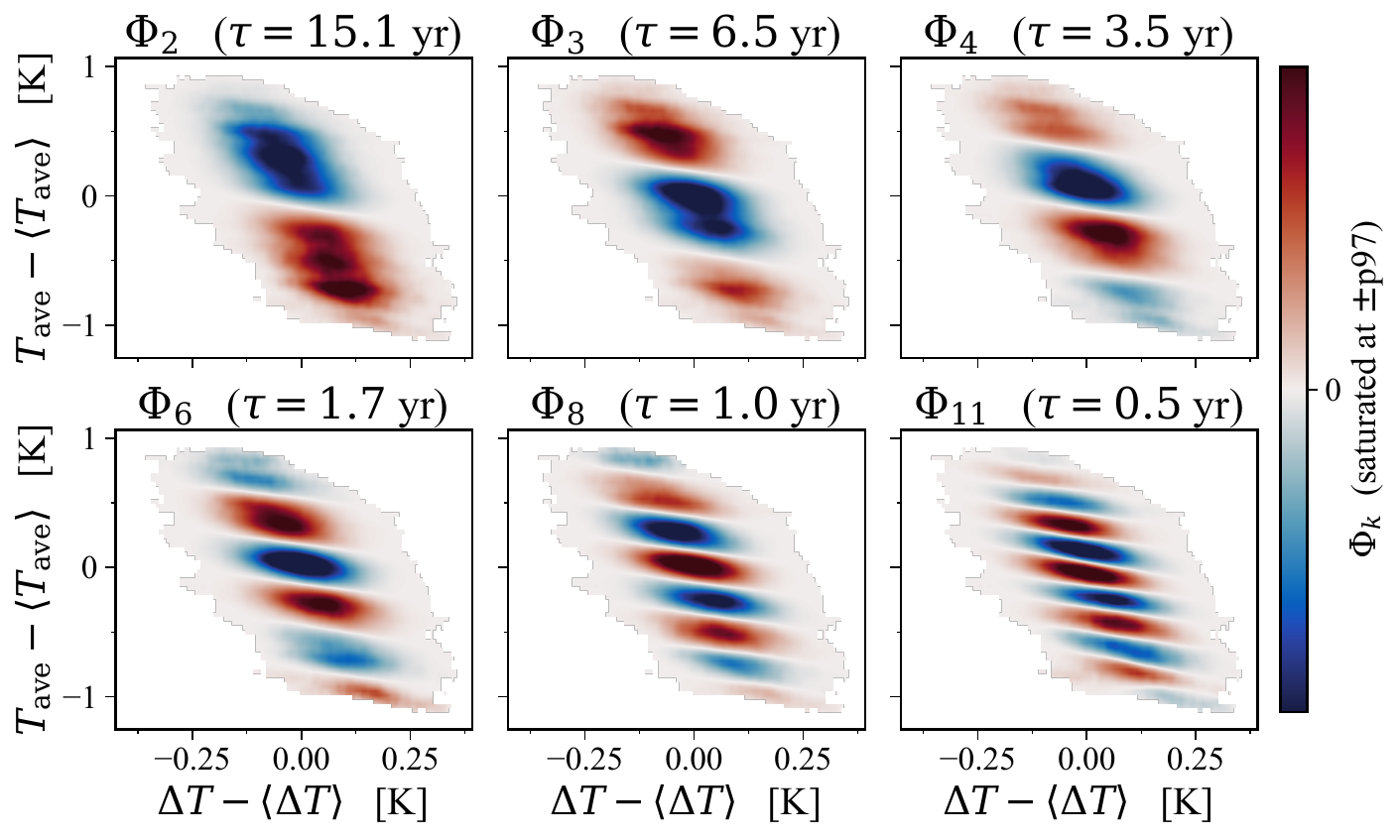}}
\caption{{\bf Reduced Kolmogorov modes in the precritical near-tipping regime, low-noise case.}
The construction is identical to Fig.~\ref{Fig_Kolmo_current}, with the reduced Markov matrix estimated at $\mu=0.975$ while the trajectory remains in the warm metastable regime. Several leading modes develop similarly oriented banded structures across the $(\Delta T,T_{\rm ave})$ plane. We refer to this geometric alignment as \emph{harmonization}: the relaxation structures seen by the two physical observables become increasingly organized around a common slow transition direction. The axes use the same temperature observables, units, and scale as in Fig.~\ref{Fig_Kolmo_current}.}
\label{Fig_Kolmo_tipping}
\end{figure}

\begin{figure}
  \includegraphics[width=.49\textwidth, height=.35\textwidth]{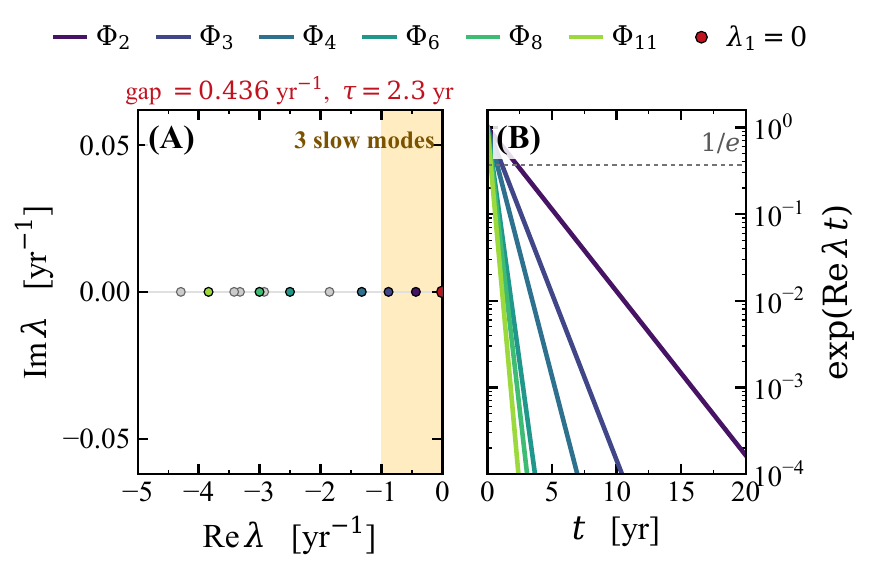}
  \includegraphics[width=.49\textwidth, height=.35\textwidth]{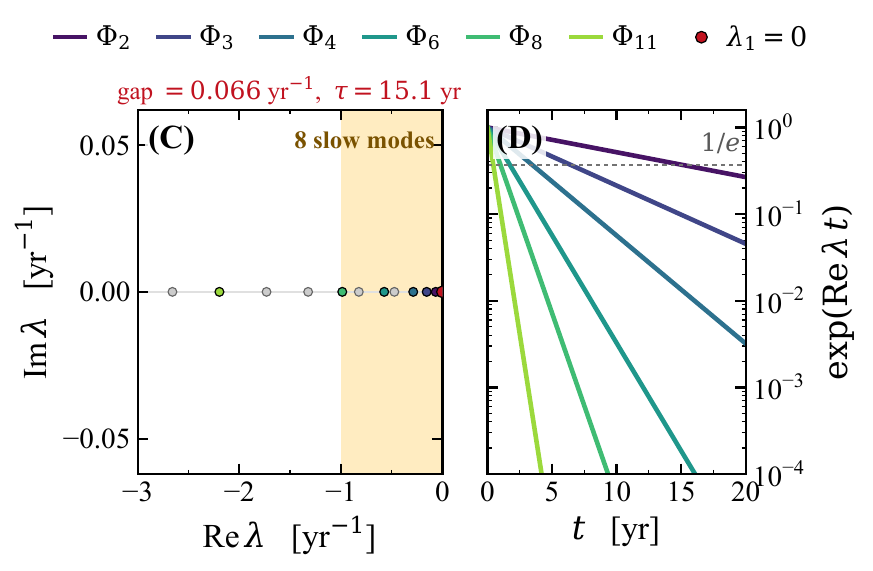}
    \caption{{\bf Reduced RP resonances and decay rates before and near tipping.}
Panels (A) and (C) show the reduced RP eigenvalues obtained from the Markov approximation in the current-climate and precritical regimes, respectively, reported as rates $\lambda=\log(E)/\tau_{\rm lag}$ in yr$^{-1}$; panels (B) and (D) show the corresponding decay $\exp(\Re\lambda\,t)$ of the dominant modes, on a semilogarithmic scale, so that each mode appears as a straight line whose slope is its resonance. In (A) and (C) the shaded band collects the resonances with $|\Re\lambda|<1$~yr$^{-1}$, whose number rises from three to eight; the red dot at the origin is $\lambda_1=0$, associated with the invariant measure, and the six colored resonances are the modes whose patterns are shown in Figs.~\ref{Fig_Kolmo_current} and~\ref{Fig_Kolmo_tipping}, color-matched throughout. The comparison is intended to emphasize the leading part of the spectrum: near tipping, a group of dominant decay rates is displaced toward zero rather than a single isolated rate doing so. This spectral compression is the rate counterpart of the geometric harmonization seen in Fig.~\ref{Fig_Kolmo_tipping} and motivates the interpretation of a slow reduced subspace rather than a single soft scalar mode.}      
  \label{Fig_RP_decay_rates}
\end{figure}

We now apply this reduced transfer-operator construction to the low-noise EBM experiments described above. The reduced state space is the two-dimensional observable plane
\be
Z_k=
\left(
\Delta T_k-\overline{\Delta T},
T_{{\rm ave},k}-\overline{T_{\rm ave}}
\right),
\label{eq:EBM_reduced_observable_plane}
\ee
in which $T_{\rm ave}$ is the area-weighted global mean temperature, while $\Delta T$ is the area-weighted difference between subtropical temperatures, equatorward of $30^\circ$, and mid--high-latitude temperatures, poleward of $30^\circ$, taking both hemispheres into account. Thus $\Delta T$ measures a large-scale meridional thermal contrast, whereas $T_{\rm ave}$ measures the bulk thermal state of the climate.

The reduced Markov matrix is therefore not built in the full $37$-dimensional temperature field, but in the plane spanned by two physically interpretable climate observables: one measuring the mean thermal state and one measuring the redistribution of heat between low and higher latitudes, keeping in mind that the meridional heat transport is approximately proportional to the temperature gradient. The Ulam boxes are placed on the observed support of this reduced trajectory, and the transition matrix $M_\tau$ is estimated from transitions in this reduced plane. In the computations shown below, the time series is subsampled every two days and the transition lag corresponds to three such subsampled steps.

The resulting eigenvalues and eigenvectors should therefore be interpreted as reduced RP resonances and reduced Kolmogorov modes associated with the observable pair $(\Delta T,T_{\rm ave})$, rather than as a full spectral resolution of the stochastic EBM. This distinction is important: the modes shown below describe the relaxation geometry seen through these two observables. They nevertheless provide a useful diagnostic of how the dominant reduced relaxation structures change as the system approaches the critical transition. Figures~\ref{Fig_Kolmo_current} and \ref{Fig_Kolmo_tipping} show representative reduced Kolmogorov modes in the current-climate and near-tipping regimes, respectively, while Fig.~\ref{Fig_RP_decay_rates} reports the corresponding reduced RP resonances and decay rates.

Three features are worth emphasizing. First, the dominant reduced RP eigenvalues are real, or numerically very close to real, in both regimes. This is consistent with the fact that the transition studied here is organized by relaxation toward noisy steady climate states, rather than by an oscillatory instability. In the RP language, the leading reduced blocks are therefore relaxation blocks: their eigenvalues primarily encode decay rates, while their eigenfunctions encode the reduced directions along which probability anomalies relax in the observable plane $(\Delta T,T_{\rm ave})$.

Second, the near-tipping reduced Kolmogorov modes display a marked geometric reorganization. In the current-climate regime, the dominant modes separate regions of enhanced and depleted probability inside the warm-state cloud, but their nodal structures remain relatively localized and mode-dependent. In physical terms, the leading relaxation patterns distinguish fluctuations in the global mean temperature and in the subtropical versus mid--high-latitude thermal contrast, without yet collapsing onto a single dominant transition direction.

Near tipping, by contrast, several leading modes become aligned along a common large-scale direction in the reduced state space. We refer to this effect as a harmonization of the reduced Kolmogorov modes. This harmonization is physically meaningful because the two reduced coordinates have direct climate interpretations: $T_{\rm ave}$ measures the bulk thermal state, while $\Delta T$ measures the area-weighted contrast between subtropical and mid--high latitudes. The organization of several modes along a common direction therefore indicates that fluctuations of the mean climate state and fluctuations of the meridional thermal contrast are increasingly coupled along the slow transition geometry.

Third, the approach to tipping is not visible only through the leading reduced spectral gap.  The gap itself does move, and substantially: expressed as a rate, the leading reduced resonance weakens from $|\Re\lambda_2|=0.436$~yr$^{-1}$ in the current climate, a relaxation time $\tau=2.3$~yr, to $0.066$~yr$^{-1}$ near tipping, $\tau=15.2$~yr, a slowing by a factor of about seven. But the gap is not the whole of the change. Figure~\ref{Fig_RP_decay_rates} shows that a group of dominant decay rates becomes compressed toward the origin. This compression is marked directly on the figure: the shaded band collects the resonances with $|\Re\lambda|<1$~yr$^{-1}$, and their number rises from three in the current-climate regime to eight near tipping, while the six modes whose spatial patterns are shown in Figs.~\ref{Fig_Kolmo_current} and~\ref{Fig_Kolmo_tipping} are color-matched between the two panels of each row. In other words, the reduced spectrum develops a small cluster of slow relaxation scales. This is important for the present EBM because the approach to the transition is not seen only by one scalar variable. The loss of resilience involves both the global thermal coordinate and the meridional redistribution coordinate. Several reduced modes may therefore contribute to the observed warning, depending on how the observable combines $T_{\rm ave}$ and $\Delta T$.

This last point is where the reduced-RP interpretation is useful but must be made carefully. A small reduced RP denominator is not, by itself, a universal early-warning signal. It becomes visible only if the measured variables have a non-negligible projection on the corresponding reduced Kolmogorov mode. Likewise, the response to a forcing depends on how the forcing couples to that same reduced block. In the present experiment, the chosen reduced coordinates are not arbitrary: $T_{\rm ave}$ tracks the bulk displacement of the warm climate state, while $\Delta T$ tracks the large-scale meridional thermal contrast affected by the ice-albedo feedback and by the imposed radiative perturbation. Their joint organization in Fig.~\ref{Fig_Kolmo_tipping} is therefore evidence that the reduced slow modes are seen by physically relevant observables.

This is the numerical counterpart of the residue-conditioned viewpoint developed in \citep{Chekroun_Lucarini26_theoretic}: early warnings are not properties of the spectrum alone, but of a spectral block seen through a chosen observable and perturbation direction. Here the observable plane $(\Delta T,T_{\rm ave})$ sees a coherent slow subspace associated with a bundle of modes that emerge near tipping. The corresponding eigenvalues move closer to the origin, producing longer relaxation, correlation, and response times, while the eigenfunctions become geometrically organized along the transition direction.

The numerical results in Figs.~\ref{Fig_Kolmo_current}--\ref{Fig_RP_decay_rates} therefore suggest the following interpretation. As the EBM approaches the critical transition, recovery in the reduced climate state becomes increasingly anisotropic: anomalies along the harmonized direction decay more slowly than anomalies transverse to it. This slow direction combines bulk cooling of the warm state with changes in the low-latitude versus mid--high-latitude thermal contrast. The near-tipping regime is thus not merely characterized by a smaller leading spectral gap; it is characterized by a reorganization of the dominant reduced Kolmogorov modes and by the compression of several relaxation scales. This provides a spectral explanation for why early-warning indicators may become stronger, more coherent, and more spatially interpretable as the system approaches the transition.


\subsection{Consequences for Green Functions and Response}
\label{Sec_EBM_Green_consequences}
The harmonization of the reduced Kolmogorov modes near tipping has a direct implication for Green functions. Recall from Section~\ref{Sec_Kolmo_tipping} that the Green function $\cG_{\Psi,G}(t)$ associated with an observable $\Psi$ and a forcing pattern ${\bm G}$ admits the RP expansion
\be
\cG_{\Psi,G}(t)
\approx
\Theta(t)
\sum_{j=1}^{N}
\sum_{\ell=0}^{m_j-1}
\frac{\alpha_{j\ell}(\Psi)}{\ell!}
e^{\lambda_j t}t^\ell .
\label{eq:EBM_green_RP_expansion_recall}
\ee
The coefficients $\alpha_{j\ell}$ were defined in Eq.~\eqref{Eq_alpha}. For the present discussion, it is useful to rename them as response residues:
\be
\mathcal R_{j\ell}^{\Psi,G}
\equiv
\alpha_{j\ell}(\Psi)=
\langle \Phi_j^\ast,\Psi\rangle_\mu
\int
L_{\bm G}\log(\mu)\big
(\mathcal L_K-\lambda_j\Id)^\ell
\Phi_j(\x)
\mu(\d\x).
\label{eq:EBM_response_residue_def}
\ee
Thus $\mathcal R_{j\ell}^{\Psi,G}$ measures the visibility of the $j$th Kolmogorov mode in the Green function associated with the observable-perturbation  pair $(\Psi,{\bm G})$. The factor $\langle \Phi_j^\ast,\Psi\rangle_\mu$ measures how the observable projects onto the dual mode, while the second factor measures how the forcing pattern excites the corresponding Kolmogorov mode through $L_{\bm G}\log(\mu)$.
With this notation,
\be
\cG_{\Psi,G}(t)
\approx
\Theta(t)
\sum_{j=1}^{N}
\sum_{\ell=0}^{m_j-1}
\frac{\mathcal R_{j\ell}^{\Psi,G}}{\ell!}
e^{\lambda_j t}t^\ell .
\label{eq:EBM_green_residue_expansion}
\ee
In the semisimple case, $m_j=1$, this reduces to
\be
\cG_{\Psi,G}(t)
\approx
\Theta(t)
\sum_{j=1}^{N}
\mathcal R_{j}^{\Psi,G}
e^{\lambda_j t},
\qquad
\mathcal R_{j}^{\Psi,G}=
\langle \Phi_j^\ast,\Psi\rangle_\mu
\int L_{\bm G}\log(\mu)\Phi_j(\x)\mu(\d\x).
\label{eq:EBM_green_semisimple_residues}
\ee
This formula makes clear that the Green function is controlled by two ingredients: the RP decay rates $\lambda_j$ and the response residues $\mathcal R_{j}^{\Psi,G}$. A slow mode affects the response only if it is seen by both the observable and the forcing.

The reduced spectra in Fig.~\ref{Fig_RP_decay_rates} indicate that, near tipping, several dominant reduced RP resonances move closer to the origin. This produces longer-lived contributions to $\cG_{\Psi,G}(t)$ whenever the corresponding residues do not vanish. The harmonization observed in Fig.~\ref{Fig_Kolmo_tipping} adds a second piece of information: the slow modes are not merely slow, they become geometrically organized along a common direction in the reduced state space. Hence, for observables $\Psi$ and forcing patterns ${\bm G}$ that see this direction, the Green function is expected to be dominated by a coherent slow subspace rather than by unrelated relaxation channels.
Let $\mathscr S$ denote the cluster of reduced modes that both slow down and harmonize near tipping. Then the leading Green function can be written schematically as
\be
\cG_{\Psi,G}(t)
\approx
\Theta(t)
\sum_{j\in\mathscr S}
\mathcal R_{j}^{\Psi,G}e^{\lambda_j t}
+
\cG_{\rm fast}(t),
\label{eq:EBM_green_slow_cluster}
\ee
in the semisimple case, in which $\cG_{\rm fast}$ collects faster decaying contributions. 

When the residues $\mathcal R_j^{\Psi,G}$ have compatible signs and comparable spatial structure, the slow terms add constructively. The impulse response then develops a long, coherent tail.
Integrating the Green function gives the static susceptibility:
\be
\chi_{\Psi,G}
\equiv
\int_0^\infty
\cG_{\Psi,G}(t)\d t .
\label{eq:EBM_static_susceptibility_def}
\ee
Using Eq.~\eqref{eq:EBM_green_residue_expansion}, one obtains
\be
\chi_{\Psi,G}
\approx
\sum_{j=1}^{N}
\sum_{\ell=0}^{m_j-1}
\frac{(-1)^\ell}{\lambda_j^{\ell+1}}
\mathcal R_{j\ell}^{\Psi,G},
\label{eq:EBM_static_susceptibility_residues}
\ee
up to the residual spectral contribution. In the semisimple case this becomes
\be
\chi_{\Psi,G}
\approx
\sum_{j=1}^{N}
\frac{\mathcal R_{j}^{\Psi,G}}{\lambda_j}.
\label{eq:EBM_static_susceptibility_semisimple}
\ee
Thus the near-tipping compression of several reduced decay rates can amplify the static response, but only through the modes for which $\mathcal R_j^{\Psi,G}\neq0$.
This gives a response-level interpretation of the harmonization in Fig.~\ref{Fig_Kolmo_tipping}. In the current-climate regime, different observables and forcings may sample different reduced modes, leading to Green functions with more mode-dependent shapes. Near tipping, the dominant reduced modes become organized around the same slow direction. Consequently, Green functions associated with observables and forcings aligned with that direction should share a common delayed-recovery tail after suitable amplitude rescaling. In physical terms, the EBM becomes increasingly responsive along a preferred large-scale direction of the warm-state climate anomaly.

The qualifier about residues is essential. The harmonization of Kolmogorov modes does not imply that every Green function is amplified. If $\langle \Phi_j^\ast,\Psi\rangle_\mu$ is small, or if the forcing pattern has little overlap with the mode through $L_{\bm G}\log(\mu)$, then the corresponding residue $\mathcal R_j^{\Psi,G}$ is small and the mode remains response-invisible. Thus the numerical observation should be interpreted in a residue-conditioned sense: near tipping, the EBM develops a coherent reduced slow subspace, and Green functions are strongly affected precisely for observable-perturbation  pairs that project onto this subspace. This is the finite-dimensional numerical counterpart of the RP early-warning viewpoint of \cite{Chekroun_Lucarini26_theoretic}.

In the frequency domain, the same mechanism implies a coherent low-frequency enhancement. Since the dominant near-tipping reduced RP resonances are real or nearly real, their Green-function contributions concentrate near zero frequency. The slow cluster in Eq.~\eqref{eq:EBM_green_slow_cluster} therefore produces a low-frequency susceptibility controlled by several nearby real poles. This goes beyond a single-mode critical-slowing-down picture: the warning is carried by a reduced slow subspace whose modes are both long-lived and geometrically aligned.


\section{Extreme tails: accessibility and persistence of the escape direction}
\label{Sec_EVT}
The RP analysis above describes relaxation and response through the bulk spectral properties of the reduced stochastic dynamics. We now complement this viewpoint by probing the tails of the invariant distribution. This is useful because, near a critical transition, the system may not only relax more slowly; it may also visit the edge of its metastable state more often and remain there for longer times. Extreme Value Theory (EVT) provides a natural diagnostic of this tail behavior.

The analysis is performed on annual data of the yearly global mean temperature anomaly,
\be
\delta T_{\rm AVE} = T_{\rm AVE} - \overline{T_{\rm AVE}},
\ee
using simulations of length $3\times 10^6$ years. For each parameter value, the record is divided into $1000$ non-overlapping blocks, each containing $3000$ annual values, and the largest value in each block is retained. This is the block-maxima method: instead of fitting the tail to every large fluctuation in a strongly dependent record, one extracts one extreme from each long block and fits the asymptotic extreme-value law to the resulting sequence of block extremes. Positive extremes correspond to warm excursions, $\delta T_{\rm AVE}>0$. For cold extremes, $\delta T_{\rm AVE}<0$, we apply exactly the same construction to $-\delta T_{\rm AVE}$, so that each retained maximum corresponds to the coldest anomaly in its block.
Classical EVT states that, under suitable mixing assumptions, the statistics of block maxima converges, when the limit is non-degenerate, to a Generalized Extreme Value (GEV) distribution \citep{Gnedenko1943,dehaan2006extreme},
\be
F_G(x;\gamma,\sigma,\xi) = 
\exp\left\{
-\left[
1+\xi
\left(
\frac{x-\gamma}{\sigma}
\right)
\right]^{-1/\xi}
\right\},
\label{eq:GEV_distribution}
\ee
with $1+\xi(x-\gamma)/\sigma>0$. Here $\gamma$ is the location parameter, $\sigma>0$ is the scale parameter, and $\xi$ is the shape parameter, or tail index. The sign of $\xi$ determines the tail class: $\xi<0$ corresponds to a bounded Weibull-type tail, $\xi=0$ to the Gumbel limit, and $\xi>0$ to a heavy Fr\'echet-type tail.

Alongside the GEV shape parameter, we estimate the extremal index $\theta$, which quantifies temporal clustering of extremes \citep{lucarini2016extremes}. For a weakly dependent sequence, $0<\theta\leq1$, and $1/\theta$ is approximately the mean cluster size expressed in the sampling time unit. Since the present series is annual, smaller $\theta$ means that extreme annual anomalies occur in longer-lived clusters rather than as isolated events.

\begin{figure}
\centerline{\includegraphics[width=.8\textwidth, height=.45\textwidth]{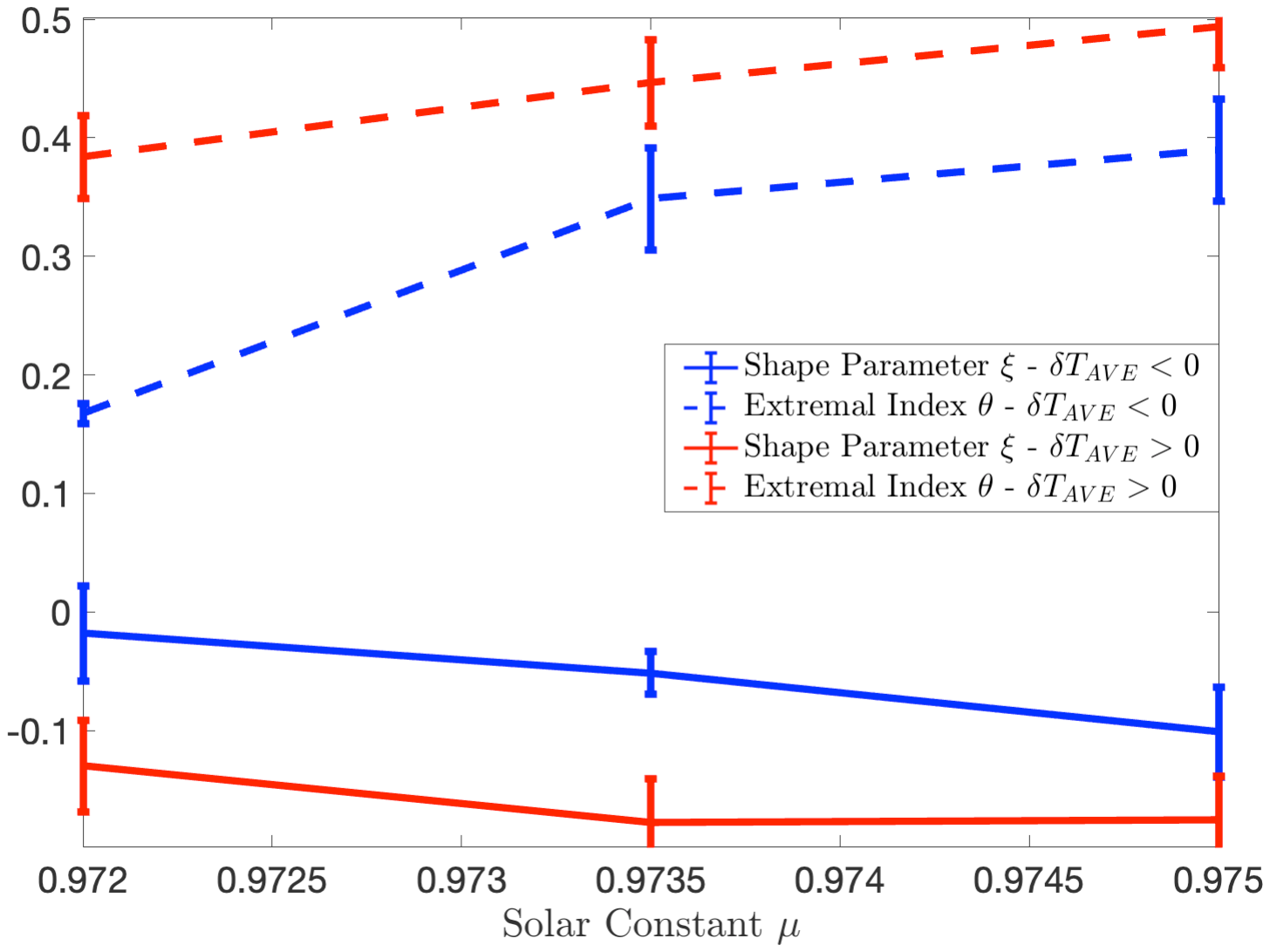}}
\caption{{\bf Tail shape and clustering of global-mean temperature extremes.}
Symbols report the estimated GEV shape parameter $\xi$ and extremal index $\theta$ at the sampled values of the solar-forcing factor $\mu$. Warm extremes are maxima of $\delta T_{\rm AVE}$ and cold extremes are analyzed as maxima of $-\delta T_{\rm AVE}$. The shape parameter measures the effective tail class: movement of the cold-side estimate toward the Gumbel limit $\xi=0$ means that the cold tail becomes less sharply bounded. The extremal index measures clustering: a decrease of the cold-side $\theta$ means that cold extreme episodes become more persistent, with $1/\theta$ approximately equal to the mean cluster size in years for the annual series. A vertical marker identifies the warm-to-cold transition threshold, allowing the directional changes in the tail diagnostics to be read relative to the loss of the warm regime.}
\label{Fig_EVT}
\end{figure}

Figure~\ref{Fig_EVT} reports the estimated shape parameter $\xi$ and the extremal index $\theta$ for positive and negative extremes of $\delta T_{\rm AVE}$ as the solar constant is varied. The shape parameter should not be read alone as the frequency of extremes. Rather, it characterizes the tail class and the effective sharpness of the tail. In the present simulations, the most significant change occurs for the cold extremes: as $\mu$ is decreased toward the warm-to-cold transition, the cold-side shape parameter moves toward the Gumbel limit $\xi=0$. This indicates that the cold tail becomes less sharply bounded, consistent with the warm-state distribution feeling the nearby edge state \citep{Bodai2015}. In physical terms, the system fluctuates more freely toward colder global temperatures than toward warmer ones. The extremal-index estimate changes in the same directional sense: $\theta$ decreases most clearly for the cold extremes, showing that these excursions become more temporally clustered as well as less sharply bounded. Thus the approach to the transition alters both how far the system explores the cold tail and how long it tends to remain there once such an excursion occurs.

As in \citep{Faranda2014}, where extremes of turbulent energy were used to anticipate transitions between different dynamical regimes, the asymmetry between positive and negative extremes is expected. The reference climate is the warm metastable state, and the relevant transition is toward a colder climate. As the critical transition is approached, cold excursions probe the direction of escape from the warm state. Warm excursions, by contrast, point deeper into the warm basin and remain more strongly constrained. Thus the EVT statistics are not symmetric: the cold tail is the dynamically relevant one for the impending transition.

This interpretation is consistent with the RP picture but addresses a different question. RP resonances and Kolmogorov modes identify the dominant relaxation directions and response channels in the bulk of the invariant dynamics. EVT, instead, probes the geometry and temporal organization of the tails. The shape parameter $\xi$ indicates how the accessible range of extremes changes, while the extremal index $\theta$ indicates whether extremes occur as isolated events or as persistent clusters. Hence EVT does not replace the reduced RP analysis; it complements it by showing that the near-tipping regime is also marked by an asymmetric and increasingly persistent cold tail.

A full synthesis of EVT and RP theory would require studying threshold-exceedance observables, such as $\mathds{1}_{\{\delta T_{\rm AVE}<-u\}}$, and decomposing their correlations or responses into RP modes. We do not pursue this here. Instead, we use EVT as an independent tail diagnostic. Together with the reduced RP results, it supports the same physical conclusion: as the warm EBM state approaches the transition, both the bulk relaxation geometry and the tail statistics become increasingly organized along the cold escape direction.

The complementary roles of the diagnostics are summarized in table~\ref{tab:diagnostic_tradeoffs}. The point of the comparison is not to rank the methods, but to make explicit that each gains access to a different physical aspect of the approach to tipping and therefore carries a different interpretive limitation.

\begin{table}
\caption{Complementary strengths and limitations of the three early-warning lenses used in this Perspective. The entries emphasize the physical question each method is best suited to answer rather than treating the methods as competing estimators of a single universal warning variable.}
\label{tab:diagnostic_tradeoffs}
\small
\begin{ruledtabular}
\begin{tabular}{p{0.17\textwidth}p{0.36\textwidth}p{0.36\textwidth}}
{\bf Diagnostic lens} & {\bf Main advantage} & {\bf Main limitation}\\ \hline
\raggedright Reduced RP modes and response &
\raggedright They identify the dominant relaxation and response channels and connect slow time scales to the geometry of physically chosen observables and forcings. &
\raggedright The result is projection-dependent: a reduced observable space can miss relevant full-state modes, and transfer-operator estimation requires adequate sampling of the selected coordinates. \tabularnewline[4pt]
\raggedright Extreme Value Theory &
\raggedright It targets the transition-relevant tails directly, separating how far excursions reach through $\xi$ from how persistently they cluster through $\theta$. &
\raggedright It does not by itself identify the relaxation mechanism producing the tail changes; reliable estimates also require long records and a defensible choice of blocks or thresholds. \tabularnewline[4pt]
\raggedright Data-Adaptive Harmonic Modes &
\raggedright They retain the multichannel field and resolve coherent spatial patterns and their relative phases frequency by frequency, thereby diagnosing field-level organization and compressibility. &
\raggedright They are less localized in time than a wavelet transform and become more demanding as the number of channels grows; by themselves they do not quantify escape probability or causal response. \tabularnewline
\end{tabular}
\end{ruledtabular}
\end{table}

\section{Full-field organization: spectral compression and phase decoherence}
\label{Sec_Multivariate}
\subsection{Data-adaptive harmonic modes (DAHMs) of variability}\label{Sec_DAHD}

The physical question behind the DAHM analysis is simple: at a given frequency, which latitude bands fluctuate together, and what are their relative phases? A standard EOF is very effective at finding spatial patterns that explain variance, but a single EOF can mix several time scales. A Fourier spectrum separates frequencies cleanly, but by itself it does not tell us which multichannel spatial pattern carries a particular frequency. Data-Adaptive Harmonic Modes (DAHMs) combine these two pieces of information by decomposing the time-lagged cross-correlations of the full field \citep{chekroun2017data}. Each DAHM is tied to a resolved frequency, while its amplitude and phase across the latitude channels are inferred from the data. In the present problem this gives exactly the information needed to ask whether the temperature field remains collectively organized as tipping is approached, rather than asking only whether one scalar record becomes slower or redder. The method has been used on both satellite observations and model simulations over a broad range of spatio-temporal scales \citep{kondrashov2017data,MASIE_paper,Kondrashov_al2018_QG,kondrashov2018data}.

The comparison with classical wavelet analysis is useful because the two methods share a spectral viewpoint but emphasize different information. Wavelets localize variability jointly in time and scale, which makes them especially well suited to determining \emph{when} a transient or intermittent oscillation becomes active. DAHMs are instead designed to determine \emph{which multivariate spatial pattern} carries a resolved frequency and how its channels are phased relative to one another. Thus a wavelet transform is naturally stronger for time localization, whereas a DAHM decomposition is naturally stronger for extracting frequency-specific cross-channel geometry and phase coherence. The two approaches are complementary rather than interchangeable: in a strongly non-stationary problem, wavelets can identify the time intervals in which a scale is active, while DAHMs can characterize the multichannel pattern and phase organization associated with that scale.

We now recall from \citep{chekroun2017data} the main steps needed to compute the DAHMs.
First, given a multivariate time series $\mathbf{X}(t)$ constituted of $L$ channels, $x_{\ell}$, one forms the following block-Hankel matrix 
\be
\mathbf{H}=\begin{bmatrix}
H_{11} & H_{12} & \cdots & H_{1L}\\
H_{12} & H_{22}  &  \cdots & H_{2L}\\
\vdots & \vdots &  \cdots & \vdots\\
H_{1L} & H_{2L} &  \cdots &H_{LL}\\
 \end{bmatrix}.
\ee 
Here, each $H_{\ell \ell'}$ is a $M\times M$ {\bf Hankel matrix} formed from  time-lagged (up to lag $\tau_{M-1}$) cross-correlations, $C_{\ell \ell'}$, between channel $\ell$ and $\ell'$ (here $x_{\ell}$ and $x_{\ell'}$ in Eq.~\eqref{Eq_EBM}, after discretization) according to:
\be\label{Hij}
H_{\ell \ell'}=\left( \begin{array}{ccccc}
C_{\ell \ell'}(0)& C_{\ell \ell'}(\tau_1) & C_{\ell \ell'}(\tau_2) &  \cdots &C_{\ell \ell'}(\tau_{M-1}) \\
C_{\ell \ell'}(\tau_1) & C_{\ell \ell'}(\tau_2) &  \dots  &   C_{\ell \ell'}(\tau_{M-1})  &  C_{\ell \ell'}(0) \\
C_{\ell \ell'}(\tau_2)  &    \vdots &  \adots &  C_{\ell \ell'}(0) & C_{\ell \ell'}(\tau_1)  \\
  \vdots &  \adots&  \adots &  \adots & \vdots \\
C_{\ell \ell'}(\tau_{M-1})  & C_{\ell \ell'}(0) &  C_{\ell \ell'}(\tau_1)  & \ldots & C_{\ell \ell'}(\tau_{M-2})
\end{array} \right).
\ee
The DAHMs are then extracted by solving the $(ML)\times(ML)$ spectral problem 
\bes
\mathbf{H}\mathbf{W}=\lambda \mathbf{W}.
\ees
As demonstrated in \citep[Theorem V.1]{chekroun2017data}, the eigenvectors of $\mathbf H$ possess a frequency-ranked structure that we recall here. Let $\Delta t$ denote the sampling interval used to form the lagged correlations, and let
\be
\tau_M=M\Delta t,
\ee
be the corresponding periodization window. The resolved Fourier frequencies are
\be
f_k=\frac{k}{\tau_M},
\qquad
\boldsymbol{\omega}_k=2\pi f_k=\frac{2\pi k}{\tau_M},
\qquad
0\leq k\leq M-1 .
\label{Eq_DAH_frequencies}
\ee

At each nonzero frequency $\boldsymbol{\omega}_k$, the DAHM spectrum is organized into phase-quadrature eigenpairs. More precisely, there are $L$ such pairs at each resolved nonzero frequency. We use the following indexing convention: $k$ labels the resolved frequency $\boldsymbol{\omega}_k$, $a=1,\ldots,L$ labels the DAHM eigenpair, or spectral branch, at that frequency, and $\ell=1,\ldots,L$ labels the physical channel. Thus, for each $\boldsymbol{\omega}_k$, the spectrum contains the $L$ positive--negative eigenvalue pairs
\be
\left(
+\lambda_a(\boldsymbol{\omega}_k),\mathbf W^{k,a}
\right),
\qquad
\left(
-\lambda_a(\boldsymbol{\omega}_k),\overline{\mathbf W}^{k,a}
\right),
\qquad
a=1,\ldots,L,
\label{Eq_DAH_eigenvalue_pairs}
\ee
in which $\lambda_a(\boldsymbol{\omega}_k)\geq0$ is the $a$th singular value of the symmetrized cross-spectral matrix at frequency $\boldsymbol{\omega}_k$; see \citep[Theorem V.1]{chekroun2017data}. The two eigenvectors in each pair are related by a phase quadrature relationship clarified below.

Notably,  by plotting the absolute values of the DAHM eigenvalues against frequency, equivalently the positive branches
$\lambda_a(\boldsymbol{\omega}_k)$, we obtain a multivariate power spectrum. Sharp or broadband peaks in this spectrum indicate significant recurrent spatio-temporal variability, in direct analogy with peaks in a univariate power spectral density, but now resolved across multichannel harmonic patterns.

These patterns are organized as follows. 
For the positive-branch, the mode $\mathbf W^{k,a}$ possesses a useful expression given,  
in the tensor product of lagged coordinates and channel space, by
\bea\label{Eq_DAHM}
\mathbf W^{k,a}
&=
(\mathbf E_1^{k,a},\cdots,\mathbf E_L^{k,a})^{\rm T},
\\
\mathbf{E}_\ell^{k,a}(\tau)
&=
B_\ell^{k,a}
\cos
\left(
\boldsymbol{\omega}_k\tau+\boldsymbol{\theta}_\ell^{k,a}
\right).
\eea
in which $\tau$ denotes the lags used to construct the DAHM correlation matrix, $0\leq\tau\leq\tau_{M-1}$, while $\ell$ runs across the $L$ channels. The notation
\be
B_\ell^{k,a}
\equiv
B_\ell^a(\boldsymbol{\omega}_k),
\qquad
\boldsymbol{\theta}_\ell^{k,a}
\equiv
\boldsymbol{\theta}_\ell^a(\boldsymbol{\omega}_k),
\label{Eq_DAH_amplitude_phase_notation}
\ee
makes explicit that both the amplitude and the phase depend on the resolved frequency and on the DAHM spectral branch. The amplitude coefficient $B_\ell^{k,a}$ measures the contribution of channel $\ell$ in the mode $\mathbf W^{k,a}$, stemming from the $a$th DAHM branch at frequency $\boldsymbol{\omega}_k$. The parameter $\boldsymbol{\theta}_\ell^{k,a}$ specifies where channel $\ell$ sits within the oscillatory cycle of that same multivariate harmonic pattern, providing this way a phase information. This way each mode $\mathbf W^{k,a}$ encodes a data-adaptive harmonic feature ($B_\ell^{k,a}$,$\boldsymbol{\theta}_\ell^{k,a}$) carried out at frequency $\boldsymbol{\omega}_k$ and the $a$th branch.

In particular, the phase differences
\be
\boldsymbol{\theta}_\ell^{k,a}-\boldsymbol{\theta}_{\ell'}^{k,a},
\label{Eq_DAH_phase_difference}
\ee
encode lead--lag relations between channels $\ell$ and $\ell'$ within the same DAHM branch at frequency $\boldsymbol{\omega}_k$. In this sense, the DAHM phases are data-adaptive cross-spectral phases: they describe the timing structure of the multivariate signal within each frequency-resolved coherent pattern.

\begin{wrapfigure}[22]{h!}{0.46\textwidth}
\centering
\includegraphics[width=.42\textwidth, height=.32\textwidth]{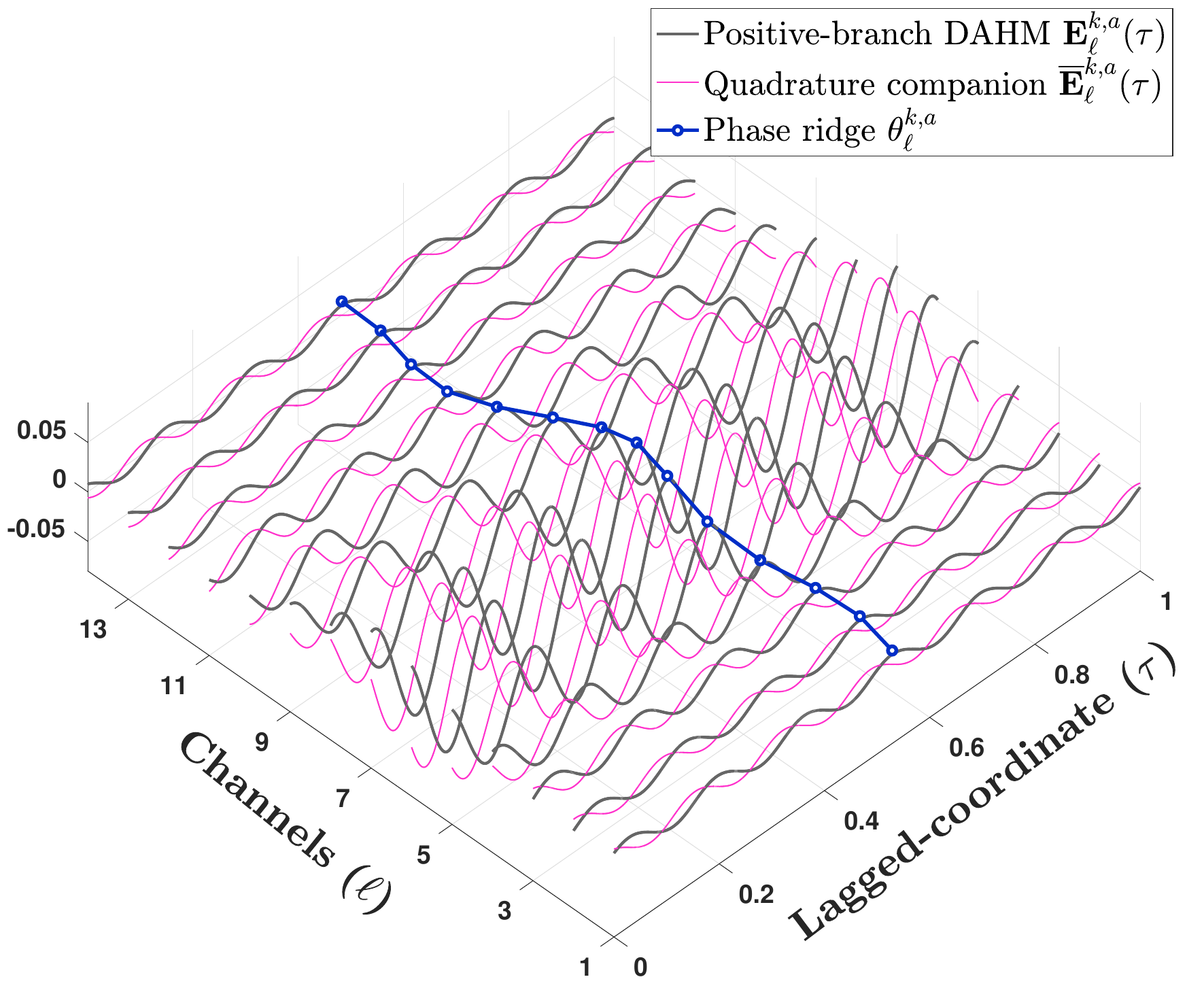} 
\caption{{\bf Anatomy of a Data-Adaptive Harmonic Mode pair.}
A representative positive-branch DAHM $\mathbf{E}_{\ell}^{k,a}(\tau)$ is shown across $L=14$ channels together with its quadrature companion $\overline{\mathbf{E}}_{\ell}^{k,a}(\tau)$, shifted by $\pi/2$. The locus of maxima across channels traces a phase ridge determined by the data-adaptive phases $\boldsymbol{\theta}_{\ell}^{k,a}$. Its nonuniform dependence on channel index illustrates the property used below: at a fixed frequency, a DAHM carries a multichannel pattern of amplitudes and relative phases inferred from the data rather than imposing a rigid spatial phase relation.}
\label{Fig_DAHM}
\end{wrapfigure}

The companion DAHM $\overline{\mathbf W}^{k,a}$ in Eq.~\eqref{Eq_DAH_eigenvalue_pairs}, associated with $-\lambda_a(\boldsymbol{\omega}_k)$, is shifted by one quarter of the period $1/f_k$, equivalently by a phase shift of $\pi/2$. Thus, up to normalization and sign convention,
\be
\overline{\mathbf E}_\ell^{k,a}(\tau)
=
B_\ell^{k,a}
\cos
\left(
\boldsymbol{\omega}_k\tau
+
\boldsymbol{\theta}_\ell^{k,a}
+
\frac{\pi}{2}
\right).
\label{Eq_DAHM_quadrature_pair}
\ee
This quadrature property along with the phase ridge tied to $\boldsymbol{\theta}_\ell^{k,a}$ across the channels, are illustrated schematically in Fig.~\ref{Fig_DAHM}. 

Because $\mathbf{H}$ is real and symmetric, the full collection of DAHMs forms an orthogonal basis, provided their quadrature pairing is correctly accounted for. This orthogonality naturally yields powerful reconstruction capabilities. Specifically, their distinct harmonic structure enables the targeted reconstruction of isolated frequency bands within the original spatio-temporal field. Furthermore, the emergence of a spectral gap between a few dominant DAHM pairs and the bulk spectrum within a given frequency band ensures optimal signal compression; in such cases, the macroscopic dynamics can be accurately recovered using only these leading pairs. The robustness of this data-adaptive reconstruction has been validated across diverse physical systems, including Arctic sea ice dynamics \citep{kondrashov2017data}, baroclinic turbulence \citep{Kondrashov_al2018_QG,kondrashov2020turbulence}, and solar wind--magnetosphere coupling \citep{kondrashov2018data}.

From this perspective, DAHMs represent statistically significant, persistent spatio-temporal patterns extracted directly from the singular-value structure of a multivariate cross-spectral density matrix. As established in Theorem V.1 of \citep{chekroun2017data}, the DAHM eigenvalues at any given frequency correspond to the singular values of the symmetrized cross-spectral matrix, while the eigenvectors capture the associated multichannel harmonic patterns and their frequency-resolved phase relationships. Unlike standard EOF analysis \citep{Lorenz1956}---which achieves strong data compression but frequently suffers from undesirable frequency-mixing across modes \citep{Groth_Ghil2011,dror_2021}---DAHMs overcome this limitation without sacrificing compression skill \citep{Kondrashov_al2018_QG,kondrashov2020turbulence}. Instead, they distribute this compression cleanly across the native frequency bands of the multivariate signal.
This dual capability---simultaneous spatial compression and spectral isolation---places DAHMs alongside modern spectral EOF techniques \citep{schmidt2019spectral} as a premier tool for dissecting multivariate signals into frequency-ranked empirical modes \citep{zerenner2021harmonic,Kondrashov_al2018_QG,kondrashov2020turbulence,kondrashov2026accurate}.

\begin{figure}
\centerline{\includegraphics[width=.95\textwidth, height=.45\textwidth]{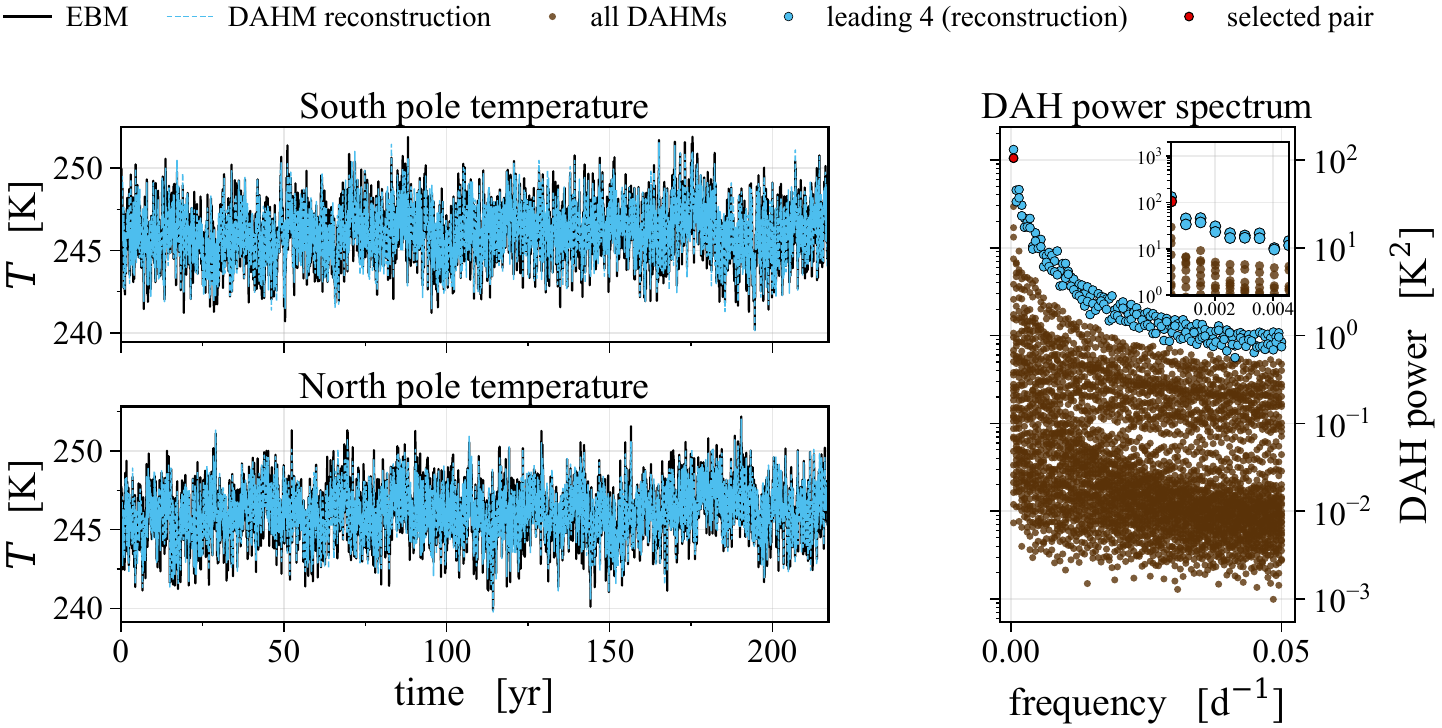}}
\caption{{\bf Full-field DAHM spectrum and reconstruction in the current climate, strong-noise case.}
DAHMs are computed from the complete latitude-discretized energy balance model (EBM) temperature field rather than from the two reduced observables used in the RP analysis. The reconstruction panels show representative South- and North-Pole temperatures obtained from the dominant DAHM pairs. The DAHM decomposition is performed on anomalies, but the reconstruction panels are shown in absolute temperature so that they can be read directly against Fig.~\ref{Fig_EBM_dynamics}; time is in years and the ordinate in kelvin. The spectral panel displays the positive DAHM branches as a function of frequency. A clear separation between a small set of dominant branches and the broadband background over much of the resolved spectrum indicates strong multivariate compressibility: a limited number of coherent frequency-resolved patterns accounts for a large fraction of the field variability. The inset magnifies the low-frequency range used for comparison with the near-tipping trajectory.}
\label{Fig_DAH_pwd_a}
\end{figure}

\begin{figure}
\centerline{\includegraphics[width=.95\textwidth, height=.45\textwidth]{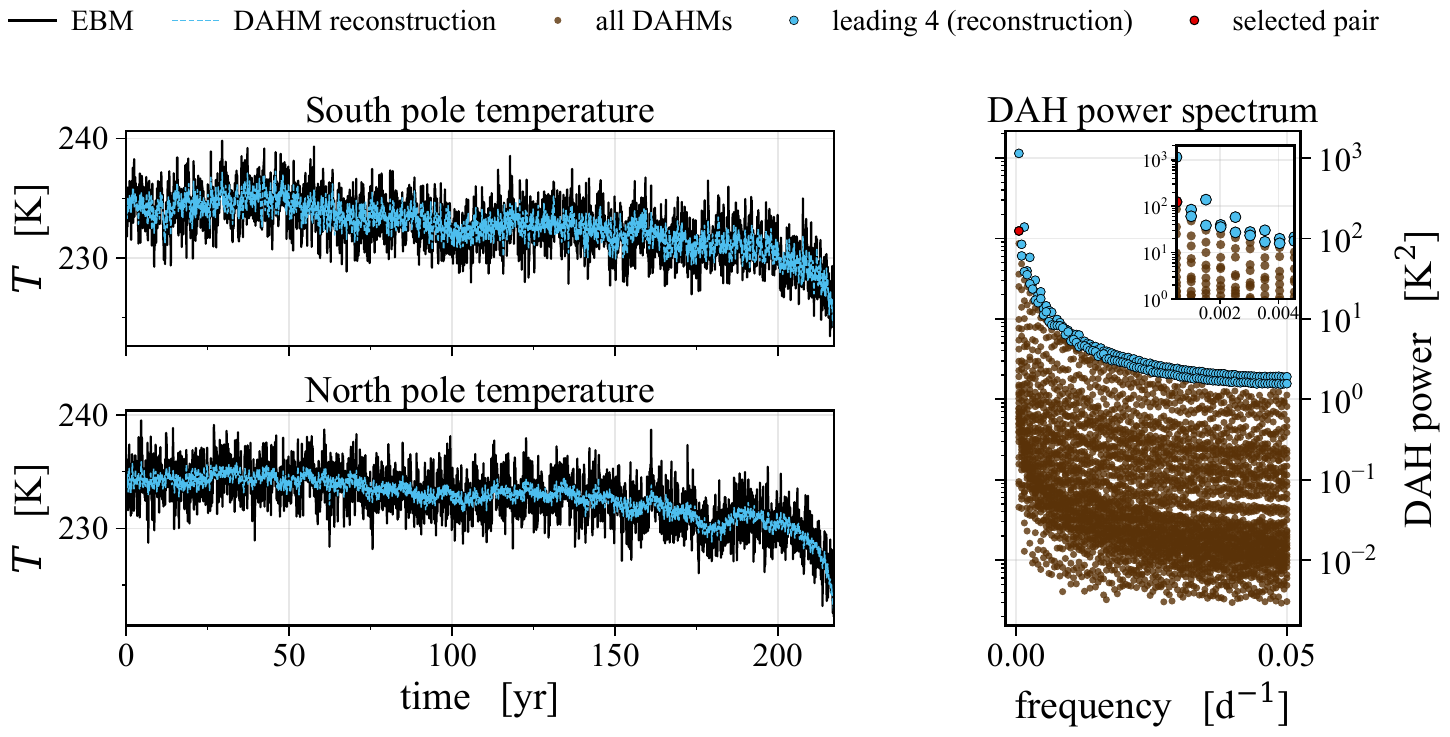}}
\caption{{\bf Full-field DAHM spectrum and reconstruction along the near-tipping trajectory, strong-noise case.}
The construction is the same as in Fig.~\ref{Fig_DAH_pwd_a}, but the input field follows the trajectory approaching the warm-to-cold transition. Two changes occur simultaneously. The broadband background gains relative spectral weight, so a fixed number of DAHM pairs provides a less complete reconstruction of the total multivariate variability; at the same time, the leading branch becomes much more clearly separated in the lowest-frequency band and the reconstruction retains the large-scale cooling component. This dual behavior is central to the full-field interpretation: the transition develops a dominant slow macroscopic component while the remaining spatial variability becomes less compressible.}
\label{Fig_DAH_pwd_b}
\end{figure}

The strong low-frequency component in Fig.~\ref{Fig_DAH_pwd_b} also invites a rare-event interpretation. In a noisy multistable system, transition trajectories often concentrate statistically around preferred escape pathways, commonly called instantons in the weak-noise large-deviation setting \citep{Ghil2020,Chekroun_Lucarini26_theoretic}. A field that is beginning to organize around such a preferred macroscopic transition direction can therefore acquire a conspicuous low-frequency component even while the remaining variability becomes less compressible. The DAHM result is consistent with this picture, but it should not be identified with an instanton itself: doing so would require an ensemble of rare transition paths and an explicit large-deviation or action-based analysis. Here the point is more modest and more directly observable: the full field develops a dominant transition-scale component at the same time as its residual multivariate organization becomes more diffuse.


\subsection{Multivariate Power Spectral Gap}
\label{Sec_shriking}

We now apply the DAHM decomposition to the full discretized EBM temperature field in the strong-noise regime. This differs from the reduced RP analysis of Section~\ref{Sec_EBM_reduced_Modes}: the latter was performed in the two-dimensional observable plane $(\Delta T,T_{\rm ave})$, whereas the DAHM analysis uses the multivariate space-time field itself. The goal is therefore not to estimate reduced RP resonances, but to ask whether the full field admits a low-rank, frequency-organized representation by coherent spatio-temporal modes.

Figure~\ref{Fig_DAH_pwd_a} shows the current-climate case. The multivariate DAHM power spectrum exhibits a distinct \textit{global spectral gap} separating a group of dominant branches (cyan) from the broadband background (brown) across the entire resolved frequency range. Retaining these dominant DAHM pairs yields an accurate reconstruction of representative polar channels. This indicates that the current-climate signal is strongly compressible: most of the full-field variability is organized by a relatively small number of coherent harmonic patterns. Furthermore, as detailed in the inset, the low-frequency spectrum shows that the dominant and subdominant branches lead the dynamics, and remain roughly an order of magnitude away from the background.

Figure~\ref{Fig_DAH_pwd_b} shows the same analysis for a trajectory approaching the critical transition. Two critical features stand out. First, the DAHM reconstruction successfully captures the large-scale trend, including the slow cooling visible in both polar channels. Because the trajectory is not strictly stationary---the system is actively drifting toward a transition---the ability of the DAHM reconstruction to track this cooling demonstrates the method's robustness for the data-adaptive decomposition of transient, non-stationary signals.

Second, the spectrum reveals a striking divergence in variance distribution, characterized by a dual-gap behavior. On one hand, the \textit{global spectral gap} shrinks: the energy of the broadband background rises, meaning the overall variance is distributed across a broader range of modes. Consequently, retaining the same number of modes yields a visibly less complete reconstruction than in the current-climate case, signifying a loss of fixed-rank compressibility. On the other hand, a profoundly different behavior emerges near the zero frequency; see inset of Fig.~\ref{Fig_DAH_pwd_b}. The \textit{low-frequency spectral gap} increases dramatically, with the leading branch separating distinctly from the subdominant one over the low-frequency band. This inflated low-frequency gap indicates that while the total field is less compressible overall, its slowest dynamics are becoming highly organized and dominant, due to the strong cooling trend. This observation seamlessly parallels the emergence of a coherent slow subspace associated with a bundle of (reduced) Kolmogorov modes near tipping, as discussed in Sec.~\ref{Sec_EBM_reduced_Modes}.

These dual-gap signature offer a powerful diagnostic interpretation. Approaching a critical transition does not merely inflate the total variance or low-frequency power; it fundamentally reorganizes the multivariate structure of the field. On one hand, the expanding low-frequency gap signals the emergence of a dominant, highly coherent component driving the large-scale transition. On the other hand, the shrinking global gap and the resulting degradation of fixed-rank reconstruction reveal that the residual variability is becoming less compressible, as additional degrees of freedom become dynamically excited.

This interplay sets the stage for the phase analysis in Section~\ref{Sec_decoherence}. While the power spectrum confirms the rise of a dominant slow component, the degraded reconstruction highlights that this leading mode cannot fully constrain the system's escalating variance. This naturally raises the question: do the spatial phases of these harmonic components remain tightly organized? As we demonstrate below, the approach to a transition triggers a distinct broadening of the phase distributions, a phenomenon we interpret as a fundamental loss of multivariate phase coherence---or \textit{decoherence}---across the full field.


\subsection{Multivariate Phase Spectrum and Decoherence}
\label{Sec_decoherence}
The phase information used below is already contained in the frequency-ranked DAHM representation introduced above. To make this point explicit, recall that the DAH construction starts from the block-Hankel matrix $\mathbf H$ of lagged cross-correlations. After periodization over the window $\tau_M=M\Delta t$, each resolved angular frequency $\boldsymbol{\omega}_k$ is associated with an $L\times L$ symmetrized cross-spectral matrix
\be
\mathfrak{S}_{\ell\ell'}(\boldsymbol{\omega}_k) =
\begin{cases}
\widehat{C_{\ell\ell'}}(\boldsymbol{\omega}_k),
&
\ell'\geq \ell,
\\
\widehat{C_{\ell'\ell}}(\boldsymbol{\omega}_k),
&
\ell'<\ell,
\end{cases}
\label{Eq_DAH_cross_spectral_matrix}
\ee
in which $C_{\ell\ell'}$ denotes the lagged cross-correlation between channels $\ell$ and $\ell'$ used to form $\mathbf H$. Theorem V.1 of \citep{chekroun2017data} shows that the singular values of $\mathfrak S(\boldsymbol{\omega}_k)$ give the $L$ positive--negative DAHM eigenvalue pairs in Eq.~\eqref{Eq_DAH_eigenvalue_pairs}, while the corresponding eigenvectors provide the multichannel harmonic patterns in Eq.~\eqref{Eq_DAHM}.

Thus the DAHM eigenvalues define a multivariate power spectrum, whereas the DAHM eigenvectors define a frequency-resolved phase structure. For the branch $a$ at frequency $\boldsymbol{\omega}_k$, the channel phases
\be
\boldsymbol{\theta}_\ell^{k,a}
=
\boldsymbol{\theta}_\ell^a(\boldsymbol{\omega}_k),
\qquad
\ell=1,\ldots,L,
\ee
specify where each channel sits within the oscillatory cycle of the same multivariate harmonic pattern. The physically meaningful quantities are therefore phase differences within a fixed frequency and branch:
\be
\Delta\boldsymbol{\theta}_{\ell\ell'}^{k,a}
=
\boldsymbol{\theta}_\ell^{k,a}
-
\boldsymbol{\theta}_{\ell'}^{k,a}.
\label{Eq_DAH_phase_difference_ka}
\ee
They measure whether channels $\ell$ and $\ell'$ lead, lag, or fluctuate nearly in phase within the same DAHM branch. Equivalently, modulo the period $2\pi/\boldsymbol{\omega}_k$, this phase difference corresponds to a relative time shift
\be
\Delta t_{\ell\ell'}^{k,a}
=
-
\frac{
\Delta\boldsymbol{\theta}_{\ell\ell'}^{k,a}
}{
\boldsymbol{\omega}_k
},
\label{Eq_DAH_phase_time_delay}
\ee
up to the sign convention used to orient the phase. The absolute phase of a DAHM may depend on sign and quadrature conventions, but phase differences, phase clustering, and phase broadening are robust diagnostics of multivariate organization.

For the EBM application, the channels are latitude bands. The phase $\boldsymbol{\theta}_{\ell}^{k,a}$ therefore measures the timing of latitude band $\ell$ within the $a$th coherent temperature pattern at frequency $\boldsymbol{\omega}_k$. A sharply clustered phase distribution means that latitude bands participate with stable phase relations, so the field fluctuates as a phase-organized spatio-temporal structure. A broad distribution means that channels, or branches at the same frequency, no longer share sharply defined phase relations. This loss of phase organization is what we refer to below as multivariate phase decoherence.

To summarize phase organization across the, $L$, DAHM branches at a fixed frequency, we introduce the channel-wise Kuramoto-type order parameter
\be
\mathbf{Z}_{\ell}(\boldsymbol{\omega}_k)
\equiv
\rho_{\ell}(\boldsymbol{\omega}_k)
e^{i\Phi_{\ell}(\boldsymbol{\omega}_k)}
=
\frac{1}{L}
\sum_{a=1}^{L}
e^{i\boldsymbol{\theta}_{\ell}^{k,a}} .
\label{Eq_DAH_phase_order_parameter}
\ee
The magnitude $0\leq\rho_{\ell}(\boldsymbol{\omega}_k)\leq1$ measures the concentration of phases for channel $\ell$ across DAHM branches at frequency $\boldsymbol{\omega}_k$. If $\rho_{\ell}\simeq1$, the phases are clustered and the channel participates coherently in that frequency band. If $\rho_{\ell}\simeq0$, the phases are broadly distributed and the channel has no preferred phase relation across branches. The argument $\Phi_{\ell}(\boldsymbol{\omega}_k)$ is the corresponding circular mean phase.

The same information can be represented through empirical phase distributions. For each frequency $\boldsymbol{\omega}_k$, we collect the phases $\boldsymbol{\theta}_{\ell}^{k,a}$ over channels $\ell$ and branches $a$, and estimate their density on the circle. We do this separately for the two quadrature families, referred to below as odd and even modes. Concentrated phase ridges indicate phase organization across the multivariate field; broad distributions indicate phase dispersion. We interpret the latter as multivariate phase decoherence.

\begin{figure}
\centerline{\includegraphics[width=.95\textwidth, height=.35\textwidth]{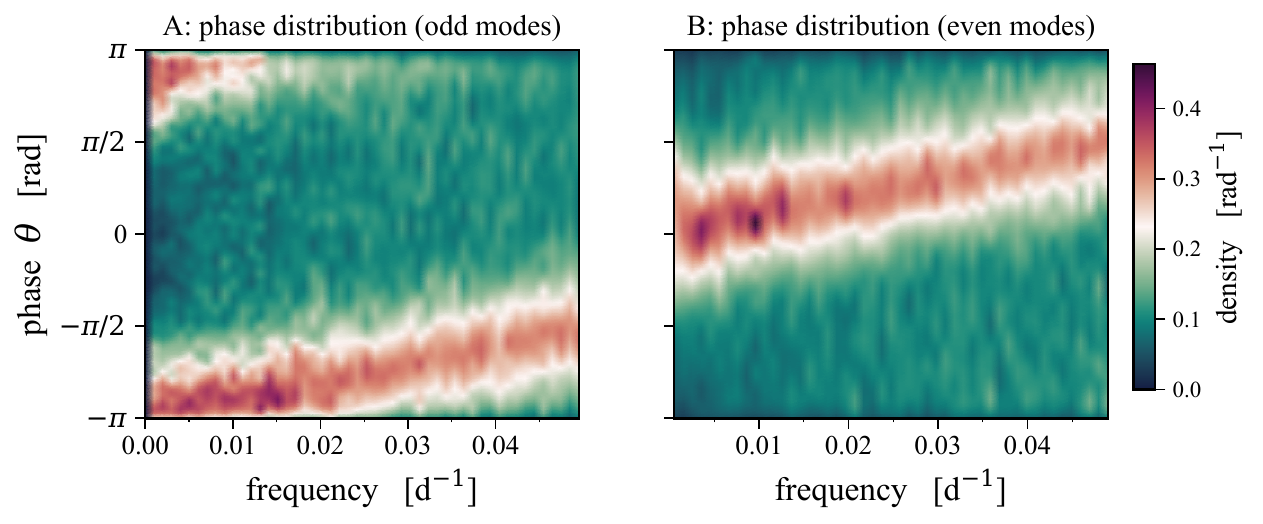}}
\caption{{\bf DAHM phase organization in the current climate, strong-noise case.}
Empirical density of the channel-dependent DAHM phases as a function of resolved frequency, displayed separately for the two quadrature families, panel~(A) for the odd modes and panel~(B) for the even ones. The horizontal axis is frequency in d$^{-1}$, the vertical axis the phase $\theta$ on the circle in radians, and the color scale the phase density in rad$^{-1}$. Relatively narrow phase ridges indicate stable lead--lag relationships across the latitude channels and therefore a strongly phase-organized multivariate temperature field.}
\label{Fig_Phase_current}
\end{figure}

\begin{figure}
\centerline{\includegraphics[width=.95\textwidth, height=.35\textwidth]{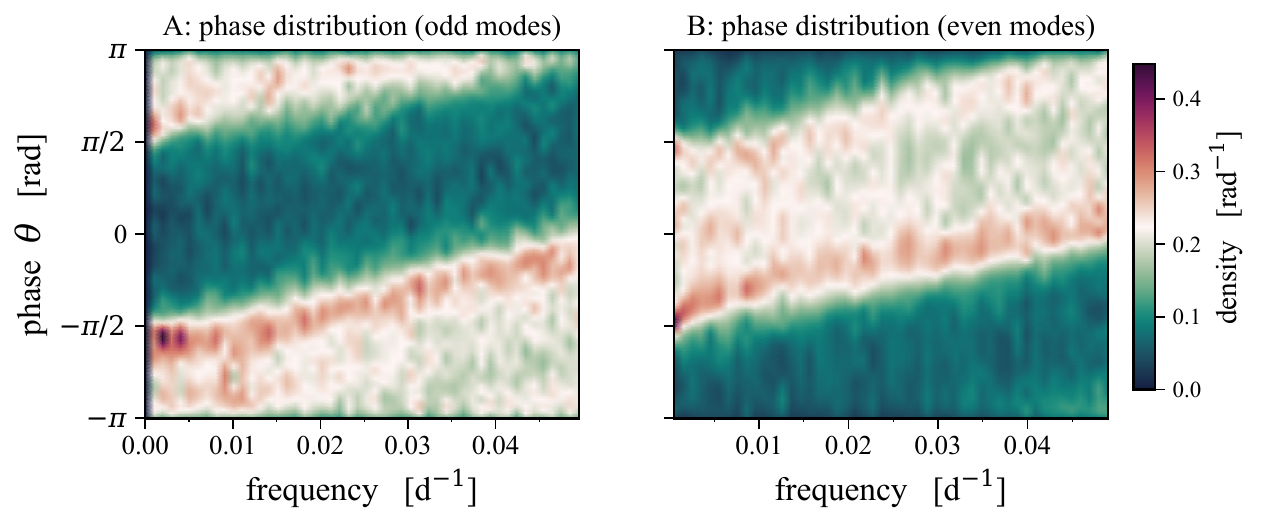}}
\caption{{\bf DAHM phase organization near tipping, strong-noise case.}
The diagnostic is identical to Fig.~\ref{Fig_Phase_current} but is evaluated along the trajectory approaching the warm-to-cold transition. The phase ridges broaden over low and intermediate frequencies, so the channel phases occupy a larger portion of the circle and are less tightly locked. We refer to this broadening as multivariate phase decoherence. It coexists with the dominant slow transition component in Fig.~\ref{Fig_DAH_pwd_b}: the large-scale transition direction becomes more pronounced while the residual full-field variability becomes less sharply phase-organized.}
\label{Fig_Phase_tipping}
\end{figure}

Figures~\ref{Fig_Phase_current} and \ref{Fig_Phase_tipping} show the resulting phase distributions for the current-climate and near-tipping trajectories in the strong-noise case. The use of strong noise is important here because the diagnostic is not based on a single deterministic trajectory or on a scalar observable. The noise excites a broad set of spatial channels and frequencies, allowing the DAHM analysis to probe how the full field organizes its phase relationships under stochastic variability. Since the comparison is made at the same noise level, the observed broadening of the phase distribution should be interpreted as a change in the multivariate organization of the field, not simply as a change in noise amplitude.
In the current-climate regime, the phase distributions are organized around relatively sharp ridges. This indicates that the dominant DAHM pairs carry coherent lead--lag relations across the latitudinal temperature field. In other words, the full-field variability can be represented by a small number of phase-organized multivariate harmonic patterns.

Near tipping, the phase distributions broaden. The ridges remain visible, so the field is not completely phase-randomized, but the phases occupy a wider portion of the circle and are less tightly clustered. This is the sense in which the approaching transition is accompanied by decoherence. The coherent large-scale component remains visible in the DAHM power spectrum, but the phase relations among channels and DAHM pairs become less sharply organized.

This observation complements the reconstruction results of Section~\ref{Sec_shriking}. Figure~\ref{Fig_DAH_pwd_b} showed that the DAHM reconstruction captures the large-scale trend along the transition path, while requiring more modes to reach the same reconstruction quality as in the current-climate case. The phase-spectrum analysis explains why: the near-tipping field contains a dominant coherent component, but the remaining variability is distributed across less phase-locked multivariate patterns. Thus the signal becomes simultaneously more dominated by a transition trend and less compressible at fixed DAHM rank.

This is also distinct from the reduced RP picture of Section~\ref{Sec_EBM_reduced_Modes}. There, in the observable plane $(\Delta T,T_{\rm ave})$, the reduced Kolmogorov modes harmonize near tipping: several slow modes align with a common transition direction. Here, by contrast, the full temperature field exhibits phase decoherence. These two statements are not contradictory. The reduced observables identify the emerging slow transition geometry, while the full-field DAHM analysis shows that, along the actual transition path, the spatial phase organization of the temperature field becomes less coherent. The transition is therefore visible both as a reduced slow spectral organization and as a full-field loss of multivariate phase coherence.

The DAHM phase spectrum therefore provides a diagnostic that scalar Early-Warning-Signal (EWS )indicators do not capture. Autocorrelation, variance, and reduced RP gaps describe temporal slowing or reduced-state relaxation. The phase spectrum measures whether the spatial channels of the full field still oscillate with coherent phase relations at each frequency. The broadening observed near tipping indicates that the onset of the transition is accompanied by a loss of organized multivariate phase structure. We refer to this loss as signal decoherence.

\section{Summary: what should an early warning tell us?}
\label{Sec_Synthesis}

The main physical lesson of the analysis is not simply that several indicators are preferable to one. The stronger conclusion is that different early-warning diagnostics answer different dynamical questions, and that these questions should not be collapsed into a single notion of ``proximity to tipping.'' Critical slowing down asks whether recovery within the observed regime is becoming slower. Extreme-value statistics ask whether fluctuations are reaching farther and persisting longer in the direction of a competing regime. Full-field harmonic analysis asks whether the spatially distributed degrees of freedom remain organized into a small number of coherent patterns. These are related manifestations of an approaching transition, but they are not interchangeable.

The first question concerns recovery. In the classical picture, one imagines a single soft eigenmode whose restoring rate tends to zero. The RP viewpoint generalizes this intuition to stochastic invariant dynamics: relaxation is organized by a collection of decay channels, and the relevant warning may be carried by a slow subspace rather than by one isolated rate. Just as importantly, a slow mode is not automatically visible in every measured variable. Its contribution depends on how the observable projects onto the corresponding Kolmogorov mode, and its contribution to a forced response also depends on how the perturbation couples to that mode. The companion RP theory makes this statement precise through response residues: a small spectral denominator can permit a large response, but only a nonzero residue produces it \citep{Chekroun_Lucarini26_theoretic}. Physically, this means that early warning is inseparable from the choice of what is measured and from the kind of perturbation whose consequences one wishes to anticipate.

The deterministic geometry of Figs.~\ref{Fig_GS_surface} and~\ref{Fig_GS_bifurcation} already contains both halves of this distinction in a form simple enough to check by inspection. Between the two experimental settings the linear relaxation time of the warm state roughly doubles, which is critical slowing down in its textbook form; over the same interval the separation between the warm state and the edge state halves, which is a statement about escape, not about recovery. The two effects have different magnitudes, are carried by different observables, and would be conflated by any single scalar indicator.

The second question concerns escape. A metastable state can become easier to leave even when local recovery inside it has not yet collapsed. This distinction is central for noisy systems, where a rare fluctuation may carry the trajectory toward an edge state or across a basin boundary before the deterministic attracting state disappears \citep{Bodai2015,Ghil2020,Chekroun_Lucarini26_theoretic}. This phenomenology is currently being explored for a variety of climatic processes \citep{Romanou2025}. The cold-tail EVT diagnostics should therefore be read as information about directional access to the transition channel and the persistence of excursions there, rather than as another estimator of the same relaxation rate measured by RP resonances. The theoretical companion sharpens this distinction by separating loss of local recovery from increasing transition risk \citep{Chekroun_Lucarini26_theoretic}. In practical terms, a tail becoming broader or more clustered is evidence that the system explores the escape direction differently; it is not, by itself, proof of local critical slowing down.

The third question concerns spatial organization. The reduced RP modes and the full-field DAHMs show why ``coherence'' itself is scale dependent. A transition can become increasingly organized in a low-dimensional macroscopic coordinate while the detailed spatial field becomes less phase-locked and less compressible. There is no contradiction in a clearer large-scale transition direction coexisting with a loss of fine-grained multivariate coherence. Rather, the two effects describe the emergence of a preferred macroscopic route through state space together with a redistribution of variability among the remaining spatial degrees of freedom.

These distinctions also clarify why observable design matters so much in high-dimensional systems. The same dynamics can appear quiescent in a coordinate that is nearly orthogonal to the transition channel and highly critical in one aligned with it. Recent work emphasizing edge-state-aligned observables reaches the same physical conclusion from a complementary direction \citep{Chekroun_Lucarini26_theoretic}. An early-warning analysis should therefore not begin by asking which scalar statistic is universally best. It should begin by asking which physical process one wants to diagnose---loss of recovery, increased escape risk, or reorganization of the field---and then select observables whose geometry is appropriate to that process.

A robust warning strategy is consequently one of triangulation rather than redundancy. Evidence of slower recovery in a transition-relevant coordinate, increasingly persistent excursions toward the competing regime, and a simultaneous reorganization of the spatial field carries more physical information than three versions of the same scalar autocorrelation statistic. Agreement among these views strengthens the interpretation of an approaching transition, while disagreement is itself informative because it can reveal that one aspect of the dynamics is changing before another. The take-home message is therefore a change of question: rather than asking only whether the system is slowing down, one should ask \emph{what is reorganizing, through which observable, toward which transition channel, and at which level of description}.

\section{Structural instability and parameterization uncertainties}
\label{Sec_Structural}

Everything above has treated the bifurcation diagram of Fig.~\ref{Fig_GS_bifurcation} as a fixed backdrop against which indicators are to be judged. That assumption deserves scrutiny, because the diagram is not an observed object but a property of a particular closure, and closures are uncertain. The question of how the multiplicity of steady states responds to a small change in the model itself is the classical problem of perturbed, or imperfect, bifurcation \citep{KeenerKeller1973,GolubitskySchaeffer1979}. For semilinear parabolic problems in one or two space dimensions it has a sharp and, in this EWS literature, largely unremarked but important answer: it is always possible to find an arbitrarily small perturbation of the nonlinear term, supported in a narrow window of the state variable, such that the perturbed problem acquires a multiple point or an additional fold as rigorously proved in \citep{chekroun2018topological}. The mechanism is a discontinuity of the branch of minimal solutions \citep{CazenaveEscobedoPozio2002}, and it is specific to one and two dimensions; in three or more the same construction requires a perturbation that is not small. The structural instability result \citep[Theorem 3.1]{chekroun2018topological} is proved for a semilinear Dirichlet problem with a monotone nonlinearity, and therefore does not literally cover the Ghil--Sellers operator, whose boundary conditions are of zero-flux type, whose diffusion is state-dependent and degenerate at the poles, and whose net radiative term is not monotone in temperature---that non-monotonicity being precisely what generates the bistability in the first place. Whether the mechanism is nonetheless operative here is a question that can be settled numerically, and it is worth settling, because a one-dimensional diffusive energy balance model is exactly the class of object to which the theorem speaks.

We therefore perturb the albedo parameterization by a narrow band in temperature,
\be\label{Eq_pert}
\alpha(x,T_S)\;\longrightarrow\;\alpha(x,T_S)-\delta\alpha\,
\exp\!\left[-\frac{(T_S-T_*)^2}{2w^2}\right],
\ee
which enters the energy budget of Eq.~\eqref{Eq_EBM} as an additional heating $\mu Q(x)\,\delta\alpha\exp[\cdot]$. The object is deliberately ordinary: a correction of this shape is what a cloud or surface scheme that responds over a few kelvin does to a model's radiation budget, and the band that turns out to matter is centred near $T_*\simeq297$~K, within the range of tropical sea-surface temperatures over which marine low-cloud cover, and with it the local albedo, varies most sharply. The amplitudes at issue, $\delta\alpha$ of order $10^{-2}$, amount to a few per cent of the planetary albedo and to a comparable change in low-cloud fraction, of a magnitude compatible with plausible parameterization differences, which are dominated precisely by the treatment of marine low cloud and its dependence on sea-surface temperature \citep{KleinHartmann1993,BonyDufresne2005,boucher_2013clouds,Zelinka2020}, and comparable to the cloud-cover changes observed over the tropical and subtropical oceans during the recent warming period \citep{liu2023opposing}.

Where the band is placed matters for a reason peculiar to a spatially extended model: at the equator $\partial_xT_S$ vanishes by symmetry, so a narrow window in temperature corresponds to the widest possible band in latitude, and the perturbation accordingly bites hardest when $T_*$ lies near the equatorial temperature of the warm branch. Centered instead near the freezing point, at $T_*=290$~K, the same perturbation merely displaces the existing fold without altering the number of folds. Structural fragility, in this model, is not distributed uniformly over the temperature range; it is concentrated where the meridional profile is flat.

\begin{figure}
  \centerline{\includegraphics[width=\textwidth]{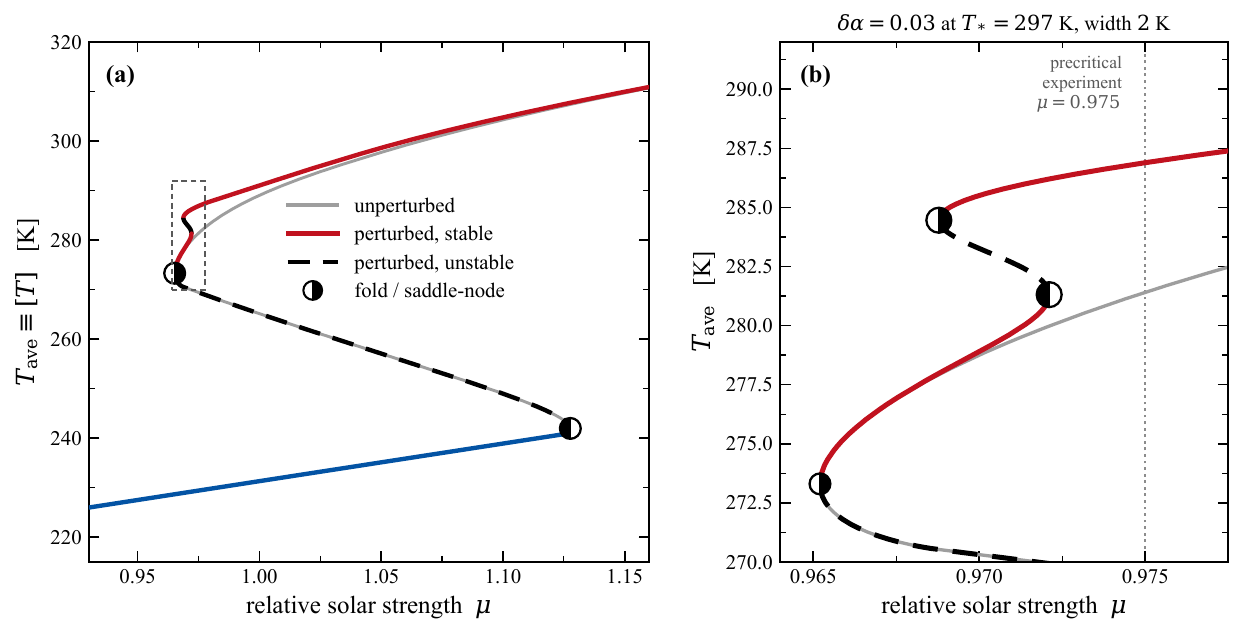}}
  \caption{{\bf A small, narrow-band change to the albedo parameterization adds a pair of folds.} (a) The $(\mu,T_{\rm ave})$ bifurcation diagram of the reference model (grey) and of the same model perturbed by a narrow-band albedo correction of amplitude $\delta\alpha=0.03$ centered at $T_*=297$~K with width $2$~K. Stable portions of the perturbed family are drawn in the branch colors of Fig.~\ref{Fig_GS_bifurcation}, unstable portions black dashed, and folds carry the half-filled circles of Arnold's convention. (b) Magnification of the boxed region. The original warm fold survives, unmoved, at the bottom; above it a new pair of folds encloses a new unstable limb and a new intermediate stable climate. The dotted line marks the solar forcing $\mu=0.975$ of the precritical stochastic experiment, which now lies markedly closer to the first fold it would encounter.}
\label{Fig_GS_imperfect}
\end{figure}

The outcome is shown in Fig.~\ref{Fig_GS_imperfect}. With $\delta\alpha=0.03$ over a window of width $2$~K---a peak radiative perturbation of $12.8$~W\,m$^{-2}$ and a peak temperature derivative of $3.9$~W\,m$^{-2}$K$^{-1}$ at the equator---the perturbed model has four folds where the reference model has two. Three aspects of this deserve emphasis. The two original folds are unchanged to eight significant figures, at $\mu_{w\to c}=0.96520$ and $\mu_{c\to w}=1.12767$: the perturbation does not displace the existing structure but adds to it. The new pair is a genuine saddle-node pair rather than a numerical artifact, the leading eigenvalue of the linearized operator being positive throughout the limb between the two new folds and negative on either side, so that a new unstable state and a new stable one are created together. And the new stable limb is a new climate in its own right, intermediate between the warm state and the edge state and existing over $\mu\in[0.9652,0.9721]$. The dynamical consequence is that the warm-to-cold transition ceases to be a single event: the warm state is now lost at $\mu=0.96878$, the system settles on the intermediate climate, and only at $\mu=0.96520$ does it tip to the snowball.  A peak albedo correction of $\delta\alpha=0.03$, that is three percentage points of albedo at the centre of the band and less away from it, confined to a two-kelvin window, has replaced one abrupt transition by a two-step cascade.

This bears directly on how far the precritical experiment of Section~\ref{Sec_EBM_num_setting} sits from the transition. Measured against the first fold it would actually encounter on a slow approach, $\mu=0.975$ lies $0.0098$ above threshold in the reference model but only $0.0062$ above it in the perturbed one, a reduction of thirty-seven per cent in the nominal distance to tipping, produced by a change to the model that no diagnostic applied to the reference model could detect. The amplitude required is moreover not a floor. Bisecting on $\delta\alpha$ for a sequence of window widths $w$ gives a threshold that falls as $\delta\alpha_{\min}\propto w^{0.53}$ (Table~\ref{Tab_threshold}), an exponent indistinguishable from one half over the range tested, so that narrowing the window drives the necessary amplitude towards zero, consistent with the structural-instability mechanism established by \citet[Theorem~3.1]{chekroun2018topological}. What does not fall is the local feedback strength: the peak temperature derivative needed rises from $2.4$ to $6.4$~W\,m$^{-2}$K$^{-1}$ as the window is narrowed from $2$ to $0.25$~K, because in a strongly diffusive medium it is the steepness of the narrow-band response, not its amplitude, that must overcome the background radiative damping. These thresholds are converged, changing by less than one per cent under refinement of both the continuation step and the spatial grid.

\begin{table}
\caption{Smallest albedo perturbation $\delta\alpha_{\min}$ in Eq.~\eqref{Eq_pert} that adds a pair of folds to the warm branch, as a function of the width $w$ of the temperature window, with $T_*=297$~K. The last two columns give the peak radiative perturbation $\delta F$ and its peak temperature derivative, both evaluated at the equator. The thresholds obey $\delta\alpha_{\min}\propto w^{0.53}$ over this range: the amplitude required falls as the window narrows, while the local feedback strength that must be supplied rises.}
\label{Tab_threshold}
\begin{ruledtabular}
\begin{tabular}{rrrr}
$w$ [K] & $\delta\alpha_{\min}$ & peak $\delta F$ [W\,m$^{-2}$] & peak $\mathrm{d}\,\delta F/\mathrm{d}T$ [W\,m$^{-2}$K$^{-1}$]\\ \hline
$2.00$ & $0.01882$ & $8.01$ & $2.43$\\
$1.00$ & $0.01332$ & $5.67$ & $3.44$\\
$0.50$ & $0.00934$ & $3.98$ & $4.82$\\
$0.25$ & $0.00624$ & $2.66$ & $6.44$\\
\end{tabular}
\end{ruledtabular}
\end{table}

None of this implies that Fig.~\ref{Fig_GS_bifurcation} is wrong. Both of its folds survive the perturbation untouched, and its qualitative content---two stable climates separated by an edge state, with hysteresis between them---is robust, resting on the ice--albedo feedback rather than on the details of any closure. The narrower and less comfortable conclusion is that the number of folds, and hence the multiplicity of accessible climates, is not itself a robust feature of a model of this class, and that the additional structure a slightly different closure may carry can sit squarely between the reference model's fold and the parameter values at which one is conducting experiments. The birth of new stable and unstable folds in EBMs following alterations of the ice--albedo feedback has been studied in the past to try to interpret some unexplained features of the deep paleoclimatology of our planet \citep{Abbott2011,Samuelberg2025}. Note also that the existence of a hierarchy of tipping processes, in which lower-rung criticalities are associated with smaller-scale features, has been documented when investigating the global stability properties of the ocean circulation \citep{Lohmannetal2024}. 

Early-warning strategies that require a threshold to be located---fitting a trend in an estimated relaxation rate, extrapolating it to zero, and reporting the implied critical forcing, as in the recent investigation of Atlantic tipping phenomena performed by \citet{Ditlevsen2023}---inherit this fragility directly, since the target they extrapolate towards is precisely the object that is not robust. Diagnostics of the system's present fluctuation statistics, which is what the resonance, tail and phase measures of the preceding sections provide, are not exposed in the same way: they characterize the state the system is in rather than the location of a boundary it has not yet reached. Two limitations of the argument should be stated plainly. It is deterministic, and whether a fold pair spanning $\Delta\mu\simeq0.003$ survives additive noise of the amplitude used here is a separate question that these calculations do not settle. And the underlying mechanism, as proved, is restricted to one and two spatial dimensions, so its status in a general circulation model remains open.

\section{Discussion and Outlook}
\label{Sec_Discussion_Outlook}

The present results suggest a general methodological lesson that is transferable well beyond the Ghil--Sellers model: early-warning analysis should not be organized only around the question ``is the system slowing down?'' A more informative question is \emph{which component of the dynamics is reorganizing, and at which level of description?} The synthesis above positions critical slowing down as one important part of that broader picture rather than as a principle to be discarded. Its great strength is that it turns loss of resilience into an observable change of time scale; its limitation is that, in a stochastic spatial system, the measured time scale is already a projection of a much richer evolution operator. The natural extension is therefore not simply a larger catalogue of scalar indicators, but a decomposition of the warning problem into recovery, escape, and spatial organization.

This changes how familiar early-warning statistics should be interpreted. Rising variance, increasing autocorrelation, spectral reddening, and enhanced susceptibility are not automatically independent pieces of evidence, nor do they necessarily diagnose the same mechanism. They are different observable manifestations of an evolving stochastic dynamics and can emphasize different relaxation or transition channels. A redder spectrum, for example, is first an observation that low-frequency variability has gained weight; it becomes a physical warning only after one asks what process generated that weight and whether the measured variable is actually aligned with the transition-relevant dynamics. In this sense, the three lenses used here are not an exercise in stacking indicators. They are a form of dynamical triangulation: RP analysis asks how recovery and response are reorganized, EVT asks how access to the escape direction changes, and DAHMs ask how the distributed field reorganizes while those changes take place \citep{Chekroun_Lucarini26_theoretic}.

This perspective also places an important limit on what can be inferred from the present experiments. The reduced RP spectrum is estimated from selected observables of the warm metastable regime, and the EVT calculation probes its global-mean tails. We do not explicitly construct committors, conditioned local dynamics, or separate intrawell and interwell spectral blocks. Consequently, the present paper should not be read as an operational classification of the simulated transition into pure bifurcation-induced or noise-induced tipping. What it does show is why that distinction matters empirically: the signatures of slower recovery and the signatures of increasingly accessible cold excursions are not the same object. The companion theory provides the formal spectral framework in which these two mechanisms can be separated \citep{Chekroun_Lucarini26_theoretic}.

The same caution applies to the emerging low-frequency DAHM component. Its growth is physically compatible with the concentration of transition trajectories around a preferred route toward the edge state, as expected from instanton ideas in noisy multistable systems \citep{Ghil2020,Chekroun_Lucarini26_theoretic}. But a data-adaptive harmonic mode is not a most-probable path, and the present decomposition does not evaluate a large-deviation action. What it establishes is an observable field-level precursor: as the large-scale transition component becomes more pronounced, the residual latitude-dependent variability becomes less compactly organized in amplitude and phase. Establishing whether that dominant component converges statistically toward an instanton is a natural next step requiring ensembles of rare transitions.

The approach is not tied to the specific EBM considered here. The model is useful because it provides a transparent laboratory for metastable tipping, but the three questions are generic. Reduced RP resonances can be estimated from sufficiently sampled stochastic or chaotic systems whenever meaningful observables can be chosen. EVT can probe scalar or spatially aggregated quantities whose extremes approach the boundary of a metastable regime. DAHMs apply to multichannel records and can therefore diagnose changes in the organization of spatial fields, networks, ecological patches, or other distributed systems. The common requirement is not that every system possess an equilibrium fold, but that one can identify physically meaningful aspects of recovery, escape, and collective organization.

A further reason to prefer diagnostics of the present state over estimates of a threshold's location is that the threshold may not be a robust property of the model at all. Section~\ref{Sec_Structural} showed that a narrow-band change to the albedo parameterization, small enough to sit well within the spread of defensible closures, adds a pair of folds to the diagram of the Ghil--Sellers model and converts a single warm-to-cold transition into a two-step cascade, without displacing the folds already present. This is the imperfect-bifurcation phenomenon of \citet{KeenerKeller1973} and \citet{GolubitskySchaeffer1979}, in the sharp form established for one- and two-dimensional semilinear parabolic problems by \citet{chekroun2018topological}. Its methodological consequence is that a computed distance-to-threshold carries an irreducible structural uncertainty in addition to its statistical one, and that this uncertainty is not reduced by observing the system for longer. How much of it survives stochastic forcing, which may render narrow folds dynamically irrelevant, and whether the mechanism operates in models of higher spatial dimension, are open questions worth pursuing.

Nor should the framework be restricted to steady-state tipping. The attracting object may be periodic, chaotic, spatially patterned, or only statistically defined \citep{Feudel2008_complex_multistability,alkhayuon2018rate,chekroun2024effective,kumar2025pace}. In such cases the warning need not correspond to a real eigenvalue approaching zero. It may instead appear as a change in a family of RP decay rates, a frequency-selective response, a reorganization of transition probabilities, or a loss of multivariate phase coherence. This is especially relevant in fluid mechanics, where transitions frequently involve turbulent or oscillatory regimes rather than the disappearance of a single equilibrium.

Several directions follow directly. One is to move from manually chosen reduced observables toward observables selected for their coupling to the transition-relevant spectral subspace. The residue viewpoint suggests that the quality of an observable should be judged not only by variance explained, but by whether it actually sees the slow response channel of interest. A second is to connect EVT and RP theory more directly by treating exceedance indicators as observables and asking which relaxation blocks control their persistence. A third is to convert DAHM phase broadening into quantitative field-level warning measures based on phase order, phase entropy, or frequency-localized coherence. These developments would retain the physical separation emphasized here while making the three diagnostic levels more directly comparable from data.

A practical implication concerns the scalability of the DAHM analysis. In the EBM application considered here, the number of spatial channels is modest enough that the lagged block-Hankel formulation is computationally accessible. In higher-dimensional geophysical fields, however, explicitly forming and diagonalizing the full grand Hankel matrix rapidly becomes prohibitive. A more direct route is to work frequency by frequency with the multivariate cross-spectral density matrix (Eq.~\eqref{Eq_DAH_cross_spectral_matrix}). Theorem~V.1 of \citep{chekroun2017data} provides the operator-theoretic justification for this implementation: at each resolved frequency, the DAHM computation reduces to the spectral decomposition of a much smaller cross-spectral matrix, whose eigenvectors encode the multichannel harmonic modes and phases. This frequency-domain formulation has already been implemented in high-dimensional oceanic and sea-ice applications \citep{kondrashov2020turbulence,kondrashov2026accurate}; see also \citep{zerenner2021harmonic}. These implementations make the full-field DAHM diagnostics used here---power-spectrum separation, reconstruction skill, and phase coherence---available well beyond low-dimensional EBM settings.


A second computational direction concerns the estimation of RP resonances and Kolmogorov modes themselves. In the present work, these quantities are obtained from a reduced Markov approximation in a physically selected observable plane. This is transparent and interpretable, but it is not the only possible route. Data-driven Koopman methods provide an alternative way to approximate spectral elements from a single long-term time series or many short-term simulated trajectories, bypassing explicit Markov-state discretizations. In particular, extended dynamic mode decomposition (EDMD) enriches classical DMD \citep{rowley2009spectral,Schmid2010} by projecting the dynamics onto a finite dictionary of observables \citep{tu2014dynamic,williams2015data}. In favorable cases, such methods can approximate Koopman eigenfunctions and eigenvalues, and hence provide a data-driven route toward the spectral objects underlying RP expansions.

The main practical difficulty is the choice of observables. The computed spectrum is not independent of the space in which the operator is represented: different dictionaries, embeddings, or measured variables can reveal or hide different spectral components \citep{brunton2022modern}. This issue is especially acute in high-dimensional and strongly nonlinear systems, where Koopman or Kolmogorov modes may have sharp gradients near separatrices, unstable manifolds, or transition regions \citep{budivsic2012applied,chekroun2025kolmogorov}. Knowledge of the global stability landscape of the system can guide the choice of the observable \citep{Lohmann2025}. Neural-network extensions of EDMD address part of this difficulty by learning the dictionary from data \citep{li2017extended,yeung2019learning}, while autoencoder-based approaches seek lower-dimensional latent coordinates in which the evolution is closer to linear and the Koopman representation more compact \citep{lusch2018deep}. Such approaches could therefore provide scalable alternatives to Markov-matrix estimation for computing reduced RP spectra and their associated modes.

There remains, however, a separate sampling problem. Even with a good observable dictionary, the data must sufficiently cover the dynamically relevant regions of phase space. For metastable systems under weak noise, this requirement is particularly demanding: the modes most relevant for tipping may be controlled by rare excursions toward transition channels that are poorly sampled by typical trajectories. In high dimension, obtaining enough data in these regions is an intrinsic challenge, closely related to the general difficulty of sampling rare events in nonequilibrium systems \citep{bouchet2019rare,li2019computing}. Thus machine-learning approximations of Koopman or Kolmogorov spectra are promising, but their use for early-warning diagnostics must be coupled to careful choices of observables and sampling strategies.


 A complementary route is to combine these ML-based spectral tools with theory-guided model reduction. Instead of estimating RP resonances directly in a high-dimensional state space, one may first derive an effective reduced dynamics on variables spanned by the unstable modes and a few selected stable modes of the linearized problem, and then compute the RP spectrum of the resulting reduced stochastic model. This perspective is naturally aligned with stochastic invariant- and center-manifold ideas \citep{DLS03,Caraballo_al09,DuanJ2015,CLW15_vol1,CLW15_vol2}, with the empirical mode reduction method \citep{kkg06,santos2021reduced}, and with their data-informed generalization through optimal parameterizing manifolds (OPMs) \citep{CLM19_closure}. The OPM approach derives reduced models in which the unresolved variables are replaced by parameterizations designed from the model equations and calibrated against simulation data through minimization of a parameterization defect. It has shown its relevance for deriving reduced models of noise-induced transitions in stochastic partial differential equations (SPDEs), in particular when the stochastic forcing acts in the unresolved, or orthogonal, component of the dynamics \citep{chekroun2023optimal,chekroun2023transitions,chekroun2025unravel}.

In such settings, stochastic OPM parameterization formulas show that fluctuations in the eliminated sector are transmitted to the resolved large-scale modes as non-Markovian stochastic coefficients factoring nonlinear terms in the reduced equations \citep{CLW15_vol2,chekroun2023transitions,chekroun2025unravel}. These coefficients depend explicitly on the history of the noise path. Their memory content is determined self-consistently by the noise intensity and by the way unresolved fluctuations interact with the resolved variables through the nonlinear terms of the SPDE, making them an effective tracer of noise-induced events \citep{chekroun2023optimal,chekroun2025unravel}. The important point for the present discussion is that these effective non-Markovian reduced systems can be recast, by augmenting the state with the relevant memory variables, as Markovian It\^o diffusions on an extended state space \citep{CLW15_vol2,chekroun2025unravel}. They are therefore compatible with the Kolmogorov operator-theoretic framework used here: once such a reduced stochastic model has been obtained, its RP resonances and Kolmogorov modes can be analyzed by the same spectral-response machinery as in the original formulation.

Such a program would turn the computation of reduced RP resonances into a two-stage problem: first construct an effective low-dimensional model that retains the transition-relevant geometry, then analyze the Kolmogorov or transfer spectrum of that model. The advantage is that the reduced dynamics may inherit more of the nonlinear transition structure than a purely empirical Markov discretization in a small observable plane. This is especially relevant when tipping is organized not only by equilibria, but by more complex invariant objects. Recent reductions of delay differential equations, for example, show how effective low-dimensional models can capture bifurcations of periodic orbits, including saddle-node bifurcations of periodic orbits, homoclinic structures, and tipping solution paths \citep{chekroun2020efficient,chekroun2024effective}. Related stochastic reductions further show how state-dependent noise interacting with delay-induced oscillations can generate shear-induced chaos \citep{Young2016,Chekroun_al22SciAdv} and other higher-dimensional stochastic attractors \citep{Chekroun_al22SciAdv}. In such settings, early-warning spectra should be computed on reduced dynamics that retain the relevant invariant objects (limit cycles, etc.), not only on coordinates designed for equilibrium-like slowing down.

This hybrid direction is promising but should not be viewed as automatic. The reduced model must still be validated at the level of the statistics that matter for tipping: invariant measure, correlations, tail excursions, transition paths, and response residues. Otherwise, its RP resonances may be accurate for the reduced model but not for the original dynamics. The opportunity is nevertheless substantial. The rare-event sampling issue raised above suggests one concrete role for stochastic OPM parameterizations: by reducing the effective dynamics while retaining the nonlinear channels through which unresolved fluctuations drive large-scale excursions, they may lower the computational cost of sampling transition pathways that are poorly visited by typical trajectories. Such reduced models could then be used  to generate or organize transition-relevant data for ML-based Koopman or Kolmogorov methods, helping identify observables and spectral coordinates that are sensitive to the rare excursions controlling tipping. In this sense, theory-guided stochastic reduction and ML-based spectral estimation should be viewed as complementary: the former constrains and accelerates the exploration of transition channels, while the latter can extract the associated spectral observables and response-relevant modes, provided that the reduced statistics remain faithful to those of the original system.

Many tipping problems are central in disciplines beyond climate, especially in ecology, where transitions often involve cyclic or multistable regimes rather than the collapse of a single equilibrium branch. Recent work on phase tipping and rate-induced phase tipping shows that ecological or biochemical oscillators may tip only for specific combinations of forcing rate and oscillation phase, including transitions between coexisting limit cycles without crossing an autonomous bifurcation point \citep{AlkhayuonTysonWieczorek2021_phase_tipping,kumar2025pace}. This opens a natural direction for the RP spectra and DAHM approaches developed here: RP resonances can identify which relaxation and response channels change as an oscillatory regime loses resilience, while DAHMs can diagnose whether spatial patches, trophic components, or oscillatory subpopulations remain phase-organized or begin to decohere. Bridging rate-induced phase tipping with RP spectra, linear response for nonautonomous systems \citep{Lucarini2026}, and DAHM phase diagnostics may therefore provide early-warning tools for ecological transitions governed by phase loss, desynchronization, and switching between coexisting dynamical regimes.  In all of these settings the structural sensitivity pointed out in Section~\ref{Sec_Structural} travels with the diagnostics: since the number and the location of the folds of a multistable or oscillatory model are themselves sensitive to the closures employed, measures of the present fluctuation statistics are the more portable objects, and any estimated distance to a threshold should be reported together with the structural uncertainty attached to it.

\bigskip
\bigskip
\paragraph*{\bf Acknowledgements.}
MDC and VL are grateful for inspiring discussions with M.~Ghil, D.~Kondrashov, J.C.~McWilliams, D.~Neelin, M.~Santos-Guti\'errez, and N. Zagli.

\medskip
\paragraph*{\bf Funding.}
This work has been supported by the European Research Council (ERC) under the European Union's Horizon 2020 research and innovation program (grant agreement No. 810370), by the Institute for Environmental Sustainability (IES) at the Weizmann Institute of Science, and by the Planning and Budgeting Committee (VATAT) of the Council for Higher Education in Israel, as part of a grant for the climate research center titled "Integrating Climate Dynamics, Clouds, and Extreme Events through Teleconnections in Climate Networks."
This work has been also partially supported by the Office of Naval Research (ONR) Multidisciplinary University Research Initiative (MURI) Grant N00014-20-1-2023, and by the National Science Foundation Grants DMS-2407484.
VL acknowledges the partial support provided by the Horizon Europe Projects Past2Future (Grant No. 101184070) and ClimTIP (Grant No. 100018693), by the ARIA SCOP-PR01-P003 - Advancing Tipping Point Early Warning AdvanTip project, by the European Space Agency Project PREDICT (Contract 4000146344/24/I-LR), and by the NNSFC  International Collaboration Fund for Creative Research Teams (Grant No. W2541005). VL acknowledges the support received as Simons Fellow of the I. Newton Institute (FCPW01 programme).

\medskip
\paragraph*{\bf Declaration of Interests.}
The authors report no conflict of interest.

\medskip
\paragraph*{\bf Data Availability Statement.}
The data and codes that support the findings of this study are available upon request.

\medskip
\paragraph*{\bf Author Contributions.}
MDC and VL conceptualized the approach and mathematical details. MDC performed most of the numerical simulations and data analysis. VL contributed to the numerical simulations and data analysis. MDC and VL interpreted the main results and wrote the manuscript.

\appendix


\section{Model parameters}\label{App_params}

Table~\ref{Tab_params} collects the constants entering the Ghil--Sellers model \eqref{Eq_EBM}. We follow the units of \citet{Ghil1976} throughout (cal, cm, s, K). The latitude-dependent functions $c(x)$, $Q(x)$, $b(x)$, $z(x)$, $k_1(x)$ and $k_2(x)$ are not given in closed form: they are empirical profiles tabulated at $10^\circ$ intervals by \citet{Ghil1976} and interpolated by cubic splines constrained to have vanishing slope at the poles. We use them exactly as in \citep{Bodai2015}, including two amendments made there. The first is the value $b(85^\circ)=2.912$, which corrects an apparent typographical error in the original table. The second concerns the two negative tabulated values of the latent-heat diffusivity $k_2$ nearest the equator, which would render the total diffusivity $k=k_1+k_2\,g(T)$ negative for $\mu\gtrsim1.03$ and hence violate the second law; these are replaced by small positive values. We emphasize that the transport closure is held fixed as $\mu$ is varied, so that the approach to tipping studied here is driven by the radiative balance and the ice--albedo feedback alone.

\begin{table}[!htb]
\caption{Constants of the Ghil--Sellers model, following \citet{Ghil1976} and \citet{Bodai2015}. The maximal albedo $\alpha_{\max}=0.6$ is the value adopted by \citet{Bodai2015} in place of Ghil's original $0.85$; it reduces the range of bistability and raises the temperature of the cold state, bringing both closer to what is found in general circulation models.}
\label{Tab_params}
\begin{ruledtabular}
\begin{tabular}{llll}
Symbol & Value & Units & Meaning\\ \hline
$c_1$            & $0.009$                          & K$^{-1}$                              & sensitivity of albedo to temperature\\
$c_2$            & $0.0065$                         & K\,m$^{-1}$                           & lapse rate, sea level to ground\\
$c_3$            & $1.9\times10^{-15}$              & K$^{-6}$                              & greenhouse modulation\\
$c_4$            & $6.105\times0.75\,e^{19.6}$      & dyn\,K\,cm$^{-2}$                     & latent-heat coefficient in $g(T)$\\
$c_5$            & $5350$                           & K                                     & latent-heat exponent in $g(T)$\\
$\sigma$         & $1.356\times10^{-12}$            & cal\,cm$^{-2}$s$^{-1}$K$^{-4}$        & Stefan--Boltzmann constant\\
$m$              & $0.5$                            & --                                    & greenhouse amplitude\\
$T_m$            & $283.16$                         & K                                     & snow-free reference temperature\\
$\alpha_{\max}$  & $0.6$                            & --                                    & upper albedo cut-off\\
$\alpha_{\min}$  & $0.25$                           & --                                    & lower albedo cut-off\\
$\mu$            & $1$                              & --                                    & present-day solar strength\\
\end{tabular}
\end{ruledtabular}
\end{table}

\section{Numerical construction of Figs.~\ref{Fig_GS_surface} and~\ref{Fig_GS_bifurcation}}\label{App_numerics}

The two deterministic figures require the \emph{unstable} branch and the two folds, neither of which is reachable by simply integrating the model forward in time. We summarize the principles used; no claim of novelty is made, and the underlying techniques are standard \citep{Dijkstra05}.

\emph{Steady states.} We discretize the model in its flux form, so that the meridional transport term appears as the divergence of a flux that carries a factor $\cos(\pi x/2)$. This factor vanishes identically at the poles, so the no-flux boundary condition is satisfied by construction and the coordinate singularity that afflicts the second-order form never appears. Equatorial symmetry lets us work on one hemisphere. Steady states are then the roots of a system of algebraic equations, obtained by Newton iteration.

\emph{Continuation.} Because the diagram is S-shaped, $\mu$ cannot be used to follow the family of steady states through the folds. We use instead the fact, noted by \citet{Bodai2015}, that the mean temperature parametrizes the family monotonically, so that $\mu$ is a single-valued function of $T_{\rm ave}$. Prescribing $T_{\rm ave}$ and treating $\mu$ as an unknown yields a system that remains well posed \emph{at} the folds, and a single monotone sweep in $T_{\rm ave}$ delivers the warm, edge and cold branches together. This is the pinning device familiar from numerical continuation \citep{Keller1977,Dijkstra05}, in the form best adapted to the present problem.

\emph{Stability and folds.} The stability of each state is determined from the spectrum of the linearization about it, rather than inferred from the geometry of the diagram. The folds are located independently, as the points where the leading eigenvalue changes sign; that these coincide with the extrema of $\mu$ along the branch is a useful check on the computation.

\emph{The transient sheet.} The surface of Fig.~\ref{Fig_GS_surface} is generated by taking the edge state at each $\mu$, displacing it infinitesimally along its unstable eigendirection in both senses, and integrating forward. Each such trajectory relaxes almost immediately onto the one-dimensional heteroclinic orbit connecting the edge state to one of the attractors, and the two branches join into a single curve running from the cold state, through the edge state, to the warm state. The union of these curves over $\mu$ is the sheet. This is the deterministic counterpart of the edge-tracking algorithm introduced for shear flows by \citet{Skufca2006}.

\emph{Resolution.} The steady states shown in Figs.~\ref{Fig_GS_surface} and~\ref{Fig_GS_bifurcation} were computed with $210$ finite-volume cells on the hemisphere, a latitudinal resolution of about $0.43^\circ$. The continuation sweep prescribes $T_{\rm ave}$ in steps of $0.25$~K from $318$~K down to $205$~K, so that each of the curves in Fig.~\ref{Fig_GS_bifurcation}(a) is assembled from $452$ separate Newton solves. The transient sheet of Fig.~\ref{Fig_GS_surface} is built from $81$ values of $\mu$ spanning the bistability window; at each of them the edge state is displaced by $0.02$~K along its unstable eigendirection, and the resulting orbit is refined until successive samples are close in the $(T_{\rm ave},\Delta T)$ plane, which requires between $3300$ and $5200$ points per orbit. The fold positions are insensitive to the spatial resolution: halving and quartering the number of cells moves $\mu_{w\to c}$ and $\mu_{c\to w}$ by less than $10^{-4}$, and even at the $5^\circ$ resolution used for the stochastic integrations of \S\ref{Sec_EBM_num_setting} they shift by less than $5\times10^{-5}$ in $\mu$ and $0.2$~K in $T_{\rm ave}$. The deterministic geometry displayed here is therefore the one on which the stochastic experiments actually live, and not an artefact of the finer grid used to draw it.

\emph{Validation.} The computation reproduces the bifurcation diagrams in \citep{Bodai2015} quantitatively, including the positions of both folds, the mean temperatures and contrasts of the three states at $\mu=1$, and the reversal of the vertical ordering of the warm and edge branches between $T_{\rm ave}$ and $\Delta T$. It was additionally cross-checked against the numerical code distributed with that paper.

\bibliographystyle{unsrtnat}
\bibliography{Super_Bib_v2MC_VL_JFM_clean}

\begin{thebibliography}{148}
\providecommand{\natexlab}[1]{#1}
\providecommand{\url}[1]{\texttt{#1}}
\expandafter\ifx\csname urlstyle\endcsname\relax
  \providecommand{\doi}[1]{doi: #1}\else
  \providecommand{\doi}{doi: \begingroup \urlstyle{rm}\Url}\fi

\bibitem[Boers et~al.(2025)Boers, Liu, Bathiany, Ben-Yami, Blaschke, Bochow,
  Boulton, Lenton, Morr, Nian, Rypdal, and Smith]{Boers2025}
Niklas Boers, Teng Liu, Sebastian Bathiany, Maya Ben-Yami, Lana~L. Blaschke,
  Nils Bochow, Chris~A. Boulton, Timothy~M. Lenton, Andreas Morr, Da~Nian,
  Martin Rypdal, and Taylor Smith.
\newblock Destabilization of {E}arth system tipping elements.
\newblock \emph{Nature Geoscience}, 18\penalty0 (10):\penalty0 949--960, 2025.
\newblock \doi{10.1038/s41561-025-01787-0}.

\bibitem[Lenton et~al.(2025)Lenton, Milkoreit, Willcock, Abrams, {Armstrong
  McKay}, Buxton, Donges, Loriani, Wunderling, Alkemade, Barrett, Constantino,
  Powell, Smith, Boulton, Pinho, Dijkstra, Pearce-Kelly, Roman-Cuesta, and
  Dennis]{Lenton2025}
T.~M. Lenton, M.~Milkoreit, S.~Willcock, J.~F. Abrams, D.~I. {Armstrong McKay},
  J.~E. Buxton, J.~F. Donges, S.~Loriani, N.~Wunderling, F.~Alkemade,
  M.~Barrett, S.~Constantino, T.~Powell, S.~R. Smith, C.~A. Boulton, P.~Pinho,
  H.~Dijkstra, P.~Pearce-Kelly, R.~M. Roman-Cuesta, and D.~Dennis, editors.
\newblock \emph{The {G}lobal {T}ipping {P}oints {R}eport 2025}.
\newblock University of Exeter, October 2025.

\bibitem[Ashwin et~al.(2012)Ashwin, Wieczorek, Vitolo, and
  Cox]{ashwin2012tipping}
Peter Ashwin, Sebastian Wieczorek, Renato Vitolo, and Peter Cox.
\newblock Tipping points in open systems: bifurcation, noise-induced and
  rate-dependent examples in the climate system.
\newblock \emph{Philosophical Transactions of the Royal Society A},
  370\penalty0 (1962):\penalty0 1166--1184, 2012.
\newblock \doi{10.1098/rsta.2011.0306}.

\bibitem[Alkhayuon and Ashwin(2018)]{alkhayuon2018rate}
H.M. Alkhayuon and P.~Ashwin.
\newblock {Rate-induced tipping from periodic attractors: Partial tipping and
  connecting orbits}.
\newblock \emph{Chaos}, 28\penalty0 (3):\penalty0 033608, 2018.

\bibitem[Feudel et~al.(2018)Feudel, Pisarchik, and Showalter]{Feudel2018}
Ulrike Feudel, Alexander~N. Pisarchik, and Kenneth Showalter.
\newblock Multistability and tipping: {F}rom mathematics and physics to climate
  and brain.
\newblock \emph{Chaos}, 28:\penalty0 033501, 2018.
\newblock \doi{10.1063/1.5027718}.

\bibitem[Halekotte and Feudel(2020)]{Halekotte2020}
Lukas Halekotte and Ulrike Feudel.
\newblock Minimal fatal shocks in multistable complex networks.
\newblock \emph{Scientific Reports}, 10:\penalty0 11783, 2020.
\newblock \doi{10.1038/s41598-020-68805-6}.

\bibitem[Alkhayuon et~al.(2021)Alkhayuon, Tyson, and
  Wieczorek]{AlkhayuonTysonWieczorek2021_phase_tipping}
H.~Alkhayuon, R.C. Tyson, and S.~Wieczorek.
\newblock Phase tipping: how cyclic ecosystems respond to contemporary climate.
\newblock \emph{Proceedings of the Royal Society A: Mathematical, Physical and
  Engineering Sciences}, 477\penalty0 (2254):\penalty0 20210059, 2021.
\newblock \doi{10.1098/rspa.2021.0059}.

\bibitem[Hastings et~al.(2026)Hastings, Petrovskii, Lucarini, and
  Morozov]{Hastings2026}
Alan Hastings, Sergei Petrovskii, Valerio Lucarini, and Andrew Morozov.
\newblock Tipping points in complex ecological systems.
\newblock \emph{Nature Reviews Physics}, 2026.
\newblock (in press);
  \href{https://arxiv.org/abs/2602.20702}{arXiv:2602.20702}.

\bibitem[Scheffer et~al.(2012)Scheffer, Carpenter, Lenton, Bascompte, Brock,
  Dakos, Van~de Koppel, Van~de Leemput, Levin, Van~Nes,
  et~al.]{scheffer2012anticipating}
M.~Scheffer, S.R. Carpenter, T.M. Lenton, J.~Bascompte, W.~Brock, V.~Dakos,
  J.~Van~de Koppel, I.A. Van~de Leemput, S.A. Levin, E.H. Van~Nes, et~al.
\newblock Anticipating critical transitions.
\newblock \emph{Science}, 338\penalty0 (6105):\penalty0 344--348, 2012.

\bibitem[Boers and Rypdal(2021)]{boers2021critical}
N.~Boers and M.~Rypdal.
\newblock Critical slowing down suggests that the western greenland ice sheet
  is close to a tipping point.
\newblock \emph{Proceedings of the National Academy of Sciences}, 118\penalty0
  (21):\penalty0 e2024192118, 2021.

\bibitem[Held and Kleinen(2004)]{Held2004}
H.~Held and T.~Kleinen.
\newblock Detection of climate system bifurcations by degenerate
  fingerprinting.
\newblock \emph{Geophysical Research Letters}, 31\penalty0 (23), 2004.
\newblock \doi{https://doi.org/10.1029/2004GL020972}.

\bibitem[Hasselmann(1976)]{hasselmann1976}
K.~Hasselmann.
\newblock {Stochastic climate models Part I. Theory}.
\newblock \emph{Tellus}, 28\penalty0 (6):\penalty0 473--485, 1976.
\newblock ISSN 0040-2826.
\newblock \doi{10.3402/tellusa.v28i6.11316}.

\bibitem[Majda et~al.(2005)Majda, Abramov, and Grote]{majda2005information}
A.~Majda, R.V. Abramov, and M.J. Grote.
\newblock \emph{{Information Theory and Stochastics for Multiscale Nonlinear
  Systems}}, volume~25.
\newblock American Mathematical Soc., 2005.

\bibitem[Lucarini and Chekroun(2023)]{LucariniChekroun2023}
V.~Lucarini and M.D. Chekroun.
\newblock {Theoretical tools for understanding the climate crisis from
  Hasselmann's programme and beyond}.
\newblock \emph{{Nature Reviews Physics}}, 2023.
\newblock \doi{10.1038/s42254-023-00650-8}.
\newblock URL \url{https://doi.org/10.1038/s42254-023-00650-8}.

\bibitem[Lucarini and Chekroun(2024)]{Lucarini_Chekroun_PRL24}
V.~Lucarini and M.D. Chekroun.
\newblock {Detecting and attributing change in climate and complex Systems:
  Foundations, Green's functions, and nonlinear fingerprints}.
\newblock \emph{Physical Review Letters}, 133:\penalty0 224201, 2024.
\newblock
  \href{https://doi.org/10.1103/PhysRevLett.133.244201}{doi:10.1103/PhysRevLett.133.244201}.

\bibitem[Chekroun et~al.(2025{\natexlab{a}})Chekroun, Zagli, and
  Lucarini]{chekroun2025kolmogorov}
M.D. Chekroun, N.~Zagli, and V.~Lucarini.
\newblock Kolmogorov modes and linear response of jump-diffusion models.
\newblock \emph{Reports on Progress in Physics}, 88\penalty0 (12):\penalty0
  127601, 2025{\natexlab{a}}.
\newblock
  \href{https://iopscience.iop.org/article/10.1088/1361-6633/ae2206}{doi:10.1088/1361-6633/ae2206}.

\bibitem[Kubo(1966)]{kubo1966}
R.~Kubo.
\newblock {The fluctuation-dissipation theorem}.
\newblock \emph{Reports on Progress in Physics}, 29\penalty0 (1):\penalty0
  255--284, 1966.

\bibitem[Santos~Guti{\'e}rrez and Lucarini(2022)]{Santos2022}
M.~Santos~Guti{\'e}rrez and V.~Lucarini.
\newblock On some aspects of the response to stochastic and deterministic
  forcings.
\newblock \emph{Journal of Physics A: Mathematical and Theoretical},
  55\penalty0 (42):\penalty0 425002, oct 2022.
\newblock \doi{10.1088/1751-8121/ac90fd}.
\newblock URL \url{https://dx.doi.org/10.1088/1751-8121/ac90fd}.

\bibitem[Chekroun et~al.(2020{\natexlab{a}})Chekroun, Tantet, Dijkstra, and
  Neelin]{Chekroun_al_RP2}
M.D. Chekroun, A.~Tantet, H.~Dijkstra, and J.D. Neelin.
\newblock {Ruelle-Pollicott Resonances of Stochastic Systems in Reduced State
  Space. Part I: Theory}.
\newblock \emph{J.~Stat.~Phys.}, 179:\penalty0 1366--1402, 2020{\natexlab{a}}.
\newblock \href{https://doi.org/10.1007/s10955-020-02535-x}{doi:
  10.1007/s10955-020-02535-x}.

\bibitem[Eckmann and Hairer(2003)]{eckmann2003spectral}
J.-P. Eckmann and M.~Hairer.
\newblock Spectral properties of hypoelliptic operators.
\newblock \emph{{Communications in Mathematical Physics}}, 235\penalty0
  (2):\penalty0 233--253, 2003.

\bibitem[Dyatlov and Zworski(2015)]{dyatlov2015stochastic}
S.~Dyatlov and M.~Zworski.
\newblock {Stochastic stability of Pollicott--Ruelle resonances}.
\newblock \emph{Nonlinearity}, 28\penalty0 (10):\penalty0 3511, 2015.

\bibitem[Chekroun and Lucarini(2026)]{Chekroun_Lucarini26_theoretic}
M.~D. Chekroun and V.~Lucarini.
\newblock {Ruelle--Pollicott Theory for Metastable Systems: A Unified Framework
  for Tipping Transitions}.
\newblock \emph{arXiv preprint}, 2026.
\newblock \href{https://arxiv.org/abs/2609.02614}{arXiv:2609.02614}.

\bibitem[Ghil(1976)]{Ghil1976}
M.~Ghil.
\newblock {Climate stability for a Sellers-type model}.
\newblock \emph{J. Atmos. Sci.}, 33:\penalty0 3--20, 1976.

\bibitem[B{\'o}dai et~al.(2015)B{\'o}dai, Lucarini, Lunkeit, and
  Boschi]{Bodai2015}
T.~B{\'o}dai, V.~Lucarini, F.~Lunkeit, and R.~Boschi.
\newblock {Global instability in the Ghil--Sellers model}.
\newblock \emph{Climate Dynamics}, 44\penalty0 (11-12):\penalty0 3361--3381,
  2015.

\bibitem[Lucarini et~al.(2022)Lucarini, Serdukova, and
  Margazoglou]{Lucarini2022NPG}
V.~Lucarini, L.~Serdukova, and G.~Margazoglou.
\newblock {L\'evy noise versus Gaussian-noise-induced transitions in the
  Ghil--Sellers energy balance model}.
\newblock \emph{Nonlinear Processes in Geophysics}, 29\penalty0 (2):\penalty0
  183--205, 2022.
\newblock \doi{10.5194/npg-29-183-2022}.
\newblock URL \url{https://npg.copernicus.org/articles/29/183/2022/}.

\bibitem[Faranda et~al.(2014)Faranda, Lucarini, Manneville, and
  Wouters]{Faranda2014}
D.~Faranda, V.~Lucarini, P.~Manneville, and J.~Wouters.
\newblock On using extreme values to detect global stability thresholds in
  multi-stable systems: The case of transitional plane couette flow.
\newblock \emph{Chaos, Solitons \& Fractals}, 64:\penalty0 26--35, 2014.
\newblock ISSN 0960-0779.
\newblock \doi{https://doi.org/10.1016/j.chaos.2014.01.008}.
\newblock URL
  \url{https://www.sciencedirect.com/science/article/pii/S0960077914000162}.
\newblock Nonequilibrium Statistical Mechanics: Fluctuations and Response.

\bibitem[Gnedenko(1943)]{Gnedenko1943}
B.~Gnedenko.
\newblock Sur la distribution limite du terme maximum d'une s{\'e}rie
  al{\'e}atoire.
\newblock \emph{Annals of Mathematics}, 44\penalty0 (3):\penalty0 423--453,
  1943.
\newblock ISSN 0003486X, 19398980.
\newblock URL \url{http://www.jstor.org/stable/1968974}.

\bibitem[de~Haan and Ferreira(2006)]{dehaan2006extreme}
L.~de~Haan and A.~Ferreira.
\newblock \emph{Extreme Value Theory: An Introduction}.
\newblock Springer, New York, NY, 2006.
\newblock ISBN 978-0387239460.
\newblock \doi{10.1007/0-387-34471-3}.

\bibitem[Lucarini et~al.(2016)Lucarini, Faranda, de~Freitas, de~Freitas,
  Holland, Kuna, Nicol, Todd, and Vaienti]{lucarini2016extremes}
V.~Lucarini, D.~Faranda, A.C.G.M.M. de~Freitas, J.M.M. de~Freitas, M.~Holland,
  T.~Kuna, M.~Nicol, M.~Todd, and S.~Vaienti.
\newblock \emph{Extremes and Recurrence in Dynamical Systems}.
\newblock John Wiley \& Sons, mar 2016.
\newblock ISBN 9781118632192.
\newblock \doi{10.1002/9781118632321}.

\bibitem[Chekroun and Kondrashov(2017)]{chekroun2017data}
M.D. Chekroun and D.~Kondrashov.
\newblock Data-adaptive harmonic spectra and multilayer {S}tuart-{L}andau
  models.
\newblock \emph{Chaos}, 27\penalty0 (9):\penalty0 093110, 2017.

\bibitem[Schmidt et~al.(2019)Schmidt, Mengaldo, Balsamo, and
  Wedi]{schmidt2019spectral}
O.T. Schmidt, G.~Mengaldo, G.~Balsamo, and N.P. Wedi.
\newblock Spectral empirical orthogonal function analysis of weather and
  climate data.
\newblock \emph{Monthly Weather Review}, 147\penalty0 (8):\penalty0 2979--2995,
  2019.

\bibitem[Peixoto and Oort(1992)]{Peixoto1992}
J.P. Peixoto and A.H. Oort.
\newblock \emph{{Physics of Climate}}.
\newblock AIP Press, New York, 1992.

\bibitem[Ghil and Lucarini(2020)]{Ghil2020}
M.~Ghil and V.~Lucarini.
\newblock The physics of climate variability and climate change.
\newblock \emph{Rev. Mod. Phys.}, 92:\penalty0 035002, Jul 2020.
\newblock \doi{10.1103/RevModPhys.92.035002}.
\newblock URL \url{https://link.aps.org/doi/10.1103/RevModPhys.92.035002}.

\bibitem[Hairer and Majda(2010)]{Hairer_Majda}
M.~Hairer and A.J. Majda.
\newblock {A simple framework to justify linear response theory}.
\newblock \emph{Nonlinearity}, 23\penalty0 (4):\penalty0 909--922, 2010.
\newblock \doi{10.1088/0951-7715/23/4/008}.

\bibitem[Majda et~al.(2001)Majda, Timofeyev, and
  Vanden-Eijnden]{majda2001mathematical}
A.J. Majda, I.~Timofeyev, and E.~Vanden-Eijnden.
\newblock A mathematical framework for stochastic climate models.
\newblock \emph{Comm.~Pure Appl. Math}, 54\penalty0 (8):\penalty0 891--974,
  2001.

\bibitem[Ruelle(2009)]{ruelle2009}
D.~Ruelle.
\newblock {A review of linear response theory for general differentiable
  dynamical systems}.
\newblock \emph{Nonlinearity}, 22\penalty0 (4):\penalty0 855--870, 2009.

\bibitem[Lucarini et~al.(2017)Lucarini, Ragone, and Lunkeit]{Lucarini2017}
V.~Lucarini, F.~Ragone, and F.~Lunkeit.
\newblock {Predicting Climate Change Using Response Theory: Global Averages and
  Spatial Patterns}.
\newblock \emph{Journal of Statistical Physics}, 166\penalty0 (3):\penalty0
  1036--1064, feb 2017.
\newblock \doi{10.1007/s10955-016-1506-z}.

\bibitem[Lucarini(2018)]{Lucarini2018JSP}
V.~Lucarini.
\newblock {Revising and Extending the Linear Response Theory for Statistical
  Mechanical Systems: Evaluating Observables as Predictors and Predictands}.
\newblock \emph{Journal of Statistical Physics}, 173\penalty0 (6):\penalty0
  1698--1721, dec 2018.
\newblock ISSN 1572-9613.
\newblock \doi{10.1007/s10955-018-2151-5}.

\bibitem[Ruelle(1998)]{ruelle_nonequilibrium_1998}
D.~Ruelle.
\newblock {Nonequilibrium statistical mechanics near equilibrium: computing
  higher-order terms}.
\newblock \emph{Nonlinearity}, 11\penalty0 (1):\penalty0 5--18, jan 1998.

\bibitem[Lucarini(2008)]{lucarini2008}
V.~Lucarini.
\newblock {Response theory for equilibrium and non-equilibrium statistical
  mechanics: Causality and generalized kramers-kronig relations}.
\newblock \emph{Journal of Statistical Physics}, 131\penalty0 (3):\penalty0
  543--558, 2008.
\newblock ISSN 00224715.
\newblock \doi{10.1007/s10955-008-9498-y}.

\bibitem[Lucarini and Colangeli(2012)]{LucariniColangeli2012}
V.~Lucarini and M.~Colangeli.
\newblock Beyond the linear fluctuation-dissipation theorem: the role of
  causality.
\newblock \emph{Journal of Statistical Mechanics: Theory and Experiment},
  2012:\penalty0 05013, 2012.

\bibitem[Baiesi and Maes(2013)]{Baiesi2013}
M.~Baiesi and C.~Maes.
\newblock An update on the nonequilibrium linear response.
\newblock \emph{New Journal of Physics}, 15\penalty0 (1):\penalty0 013004, jan
  2013.
\newblock \doi{10.1088/1367-2630/15/1/013004}.
\newblock URL \url{https://doi.org/10.1088/1367-2630/15/1/013004}.

\bibitem[Ruelle(1986)]{ruelle1986locating}
D.~Ruelle.
\newblock Locating resonances for axiom a dynamical systems.
\newblock \emph{{Journal of Statistical Physics}}, 44\penalty0 (3-4):\penalty0
  281--292, 1986.

\bibitem[Pollicott(1986)]{pollicott1986meromorphic}
M.~Pollicott.
\newblock Meromorphic extensions of generalised zeta functions.
\newblock \emph{{Inventiones Mathematicae}}, 85\penalty0 (1):\penalty0
  147--164, 1986.

\bibitem[Keller and Liverani(1999)]{keller1999stability}
G.~Keller and C.~Liverani.
\newblock Stability of the spectrum for transfer operators.
\newblock \emph{{Annali della Scuola Normale Superiore di Pisa-Classe di
  Scienze}}, 28\penalty0 (1):\penalty0 141--152, 1999.

\bibitem[Baladi(2000)]{baladi2000positive}
V.~Baladi.
\newblock \emph{{Positive Transfer Operators and Decay of Correlations}},
  volume~16.
\newblock World scientific, 2000.

\bibitem[Gaspard(2005)]{gaspard2005chaos}
P.~Gaspard.
\newblock \emph{Chaos, Scattering and Statistical Mechanics}, volume~9 of
  \emph{Cambridge Nonlinear Science Series}.
\newblock Cambridge University Press, 2005.

\bibitem[Sch{\"u}tte et~al.(1999)Sch{\"u}tte, Fischer, Huisinga, and
  Deuflhard]{schutte1999direct}
C.~Sch{\"u}tte, A.~Fischer, W.~Huisinga, and P.~Deuflhard.
\newblock A direct approach to conformational dynamics based on hybrid monte
  carlo.
\newblock \emph{J. Stat. Phys.}, 151\penalty0 (1):\penalty0 146--168, 1999.

\bibitem[Sch{\"u}tte et~al.(2001)Sch{\"u}tte, Huisinga, and
  Deuflhard]{schutte2001transfer}
C.~Sch{\"u}tte, W.~Huisinga, and P.~Deuflhard.
\newblock Transfer operator approach to conformational dynamics in biomolecular
  systems.
\newblock In \emph{Ergodic theory, analysis, and efficient simulation of
  dynamical systems}, pages 191--223. Springer, 2001.

\bibitem[Chekroun et~al.(2014)Chekroun, Neelin, Kondrashov, McWilliams, and
  Ghil]{Chek_al14_RP}
M.D. Chekroun, J.D. Neelin, D.~Kondrashov, J.C. McWilliams, and M.~Ghil.
\newblock {Rough parameter dependence in climate models: The role of
  Ruelle-Pollicott resonances}.
\newblock \emph{{Proc. Natl. Acad. Sci USA}}, 111\penalty0 (5):\penalty0
  1684--1690, 2014.
\newblock \doi{10.1073/pnas.1321816111}.

\bibitem[Mezi{\'c}(2005)]{mezic2005spectral}
I.~Mezi{\'c}.
\newblock Spectral properties of dynamical systems, model reduction and
  decompositions.
\newblock \emph{Nonlinear Dynamics}, 41\penalty0 (1--3):\penalty0 309--325,
  2005.
\newblock \doi{10.1007/s11071-005-2824-x}.

\bibitem[Budi{\v{s}}i{\'c} et~al.(2012)Budi{\v{s}}i{\'c}, Mohr, and
  Mezi{\'c}]{budivsic2012applied}
M.~Budi{\v{s}}i{\'c}, R.~Mohr, and I.~Mezi{\'c}.
\newblock {Applied Koopmanism}.
\newblock \emph{Chaos}, 22\penalty0 (4):\penalty0 047510, 2012.

\bibitem[Brunton et~al.(2022)Brunton, Budi{\v{s}}i{\'c}, Kaiser, and
  Kutz]{brunton2022modern}
S.L. Brunton, M.~Budi{\v{s}}i{\'c}, E.~Kaiser, and J.N. Kutz.
\newblock Modern koopman theory for dynamical systems.
\newblock \emph{SIAM Review}, 64\penalty0 (2):\penalty0 229--340, 2022.

\bibitem[Faure and Sj{\"o}strand(2011)]{faure2011upper}
F.~Faure and J.~Sj{\"o}strand.
\newblock {Upper bound on the density of Ruelle resonances for Anosov flows}.
\newblock \emph{Communications in Mathematical Physics}, 308\penalty0
  (2):\penalty0 325--364, 2011.

\bibitem[Giulietti et~al.(2013)Giulietti, Liverani, and
  Pollicott]{giulietti2013anosov}
P.~Giulietti, C.~Liverani, and M.~Pollicott.
\newblock Anosov flows and dynamical zeta functions.
\newblock \emph{Annals of Mathematics}, pages 687--773, 2013.

\bibitem[Dyatlov and Zworski(2019)]{dyatlov2019mathematical}
S.~Dyatlov and M.~Zworski.
\newblock \emph{{Mathematical Theory of Scattering Resonances}}, volume 200.
\newblock American Mathematical Soc., 2019.

\bibitem[Gaspard(2002)]{gaspard2002dynamical}
P.~Gaspard.
\newblock Dynamical theory of relaxation in classical and quantum systems.
\newblock In \emph{{Dynamics of Dissipation}}, pages 111--163. Springer,
  Berlin, 2002.

\bibitem[Butterley and Liverani(2007)]{butterley2007smooth}
O.~Butterley and C.~Liverani.
\newblock {Smooth Anosov flows: correlation spectra and stability}.
\newblock \emph{Journal of Modern Dynamics}, 1\penalty0 (2):\penalty0 301--322,
  2007.

\bibitem[Melbourne and Gottwald(2007)]{melbourne2007power}
I.~Melbourne and G.A. Gottwald.
\newblock Power spectra for deterministic chaotic dynamical systems.
\newblock \emph{Nonlinearity}, 21\penalty0 (1):\penalty0 179, 2007.

\bibitem[Lasota and Mackey(2013)]{lasota2013chaos}
A.~Lasota and M.C. Mackey.
\newblock \emph{Chaos, Fractals, and Noise: Stochastic Aspects of Dynamics},
  volume~97.
\newblock Springer Science \& Business Media, 2013.

\bibitem[Eisner et~al.(2015)Eisner, Farkas, Haase, and
  Nagel]{eisner2015operator}
T.~Eisner, B.~Farkas, M.~Haase, and R.~Nagel.
\newblock \emph{{Operator Theoretic Aspects of Ergodic Theory}}, volume 272.
\newblock Springer, 2015.

\bibitem[Froyland et~al.(2007)Froyland, Padberg, England, and
  Treguier]{froyland2007detection}
G.~Froyland, K.~Padberg, M.H. England, and A.M. Treguier.
\newblock Detection of coherent oceanic structures via transfer operators.
\newblock \emph{Physical review letters}, 98\penalty0 (22):\penalty0 224503,
  2007.

\bibitem[Froyland(2013)]{froyland2013analytic}
G.~Froyland.
\newblock An analytic framework for identifying finite-time coherent sets in
  time-dependent dynamical systems.
\newblock \emph{Physica D}, 250:\penalty0 1--19, 2013.

\bibitem[Froyland and Padberg-Gehle(2014)]{froyland2014almost}
G.~Froyland and K.~Padberg-Gehle.
\newblock Almost-invariant and finite-time coherent sets: directionality,
  duration, and diffusion.
\newblock In \emph{Ergodic Theory, Open Dynamics, and Coherent Structures},
  pages 171--216. Springer, 2014.

\bibitem[Froyland et~al.(2021)Froyland, Giannakis, Lintner, Pike, and
  Slawinska]{Froyland2021}
G.~Froyland, D.~Giannakis, B.R. Lintner, M.~Pike, and J.~Slawinska.
\newblock Spectral analysis of climate dynamics with operator-theoretic
  approaches.
\newblock \emph{Nature Communications}, 12\penalty0 (1):\penalty0 6570, 2021.
\newblock \doi{10.1038/s41467-021-26357-x}.
\newblock URL \url{https://doi.org/10.1038/s41467-021-26357-x}.

\bibitem[Gou{\"e}zel and Liverani(2006)]{gouezel2006banach}
S.~Gou{\"e}zel and C.~Liverani.
\newblock {Banach spaces adapted to Anosov systems}.
\newblock \emph{Ergodic Theory and dynamical systems}, 26\penalty0
  (1):\penalty0 189--217, 2006.

\bibitem[Baladi(2017)]{baladi2017quest}
V.~Baladi.
\newblock The quest for the ultimate anisotropic banach space.
\newblock \emph{Journal of Statistical Physics}, 166\penalty0 (3):\penalty0
  525--557, 2017.

\bibitem[Giulietti and Liverani(2019)]{giulietti2019parabolic}
P.~Giulietti and C.~Liverani.
\newblock {Parabolic dynamics and anisotropic Banach spaces}.
\newblock \emph{Journal of the European Mathematical Society}, 21\penalty0
  (9):\penalty0 2793--2858, 2019.

\bibitem[Froyland et~al.(2013)Froyland, Junge, and
  Koltai]{froyland2013estimating}
G.~Froyland, O.~Junge, and P.~Koltai.
\newblock Estimating long-term behavior of flows without trajectory
  integration: the infinitesimal generator approach.
\newblock \emph{SIAM Journal on Numerical Analysis}, 51\penalty0 (1):\penalty0
  223--247, 2013.

\bibitem[Engel and Nagel(2000)]{Engel_Nagel}
K.-J. Engel and R.~Nagel.
\newblock \emph{{One-parameter Semigroups for Linear Evolution Equations}},
  volume 194.
\newblock Springer Science \& Business Media, 2000.

\bibitem[Lucarini(2016)]{lucarini2016response}
V.~Lucarini.
\newblock Response operators for {M}arkov processes in a finite state space:
  radius of convergence and link to the response theory for axiom {A} systems.
\newblock \emph{J. Stat. Phys.}, 162\penalty0 (2):\penalty0 312--333, 2016.

\bibitem[Tantet et~al.(2015)Tantet, van~der Burgt, and
  Dijkstra]{tantet2015early}
A.~Tantet, F.R. van~der Burgt, and H.A. Dijkstra.
\newblock An early warning indicator for atmospheric blocking events using
  transfer operators.
\newblock \emph{Chaos}, 25\penalty0 (3):\penalty0 036406, 2015.

\bibitem[Tantet et~al.(2018{\natexlab{a}})Tantet, Lucarini, Lunkeit, and
  Dijkstra]{tantet2018crisis}
A.~Tantet, V.~Lucarini, F.~Lunkeit, and H.A. Dijkstra.
\newblock Crisis of the chaotic attractor of a climate model: a transfer
  operator approach.
\newblock \emph{Nonlinearity}, 31\penalty0 (5):\penalty0 2221,
  2018{\natexlab{a}}.

\bibitem[Bagheri(2014)]{bagheri2014effects}
S.~Bagheri.
\newblock {Effects of weak noise on oscillating flows: Linking quality factor,
  Floquet modes, and Koopman spectrum}.
\newblock \emph{Physics of Fluids}, 26\penalty0 (9):\penalty0 094104, 2014.

\bibitem[Tantet et~al.(2020{\natexlab{a}})Tantet, Chekroun, Dijkstra, and
  Neelin]{RP_Hopf}
A.~Tantet, M.D. Chekroun, H.~Dijkstra, and J.D. Neelin.
\newblock {Ruelle-Pollicott Resonances of Stochastic Systems in Reduced State
  Space. Part II: Stochastic Hopf Bifurcation}.
\newblock \emph{J.~Stat.~Phys.}, 179:\penalty0 1403--1448, 2020{\natexlab{a}}.
\newblock
  \href{https://doi.org/10.1007/s10955-020-02526-y}{doi:10.1007/s10955-020-02526-y}.

\bibitem[Crommelin and Vanden-Eijnden(2011)]{crommelin2011diffusion}
D.~Crommelin and E.~Vanden-Eijnden.
\newblock Diffusion estimation from multiscale data by operator eigenpairs.
\newblock \emph{Multiscale Modeling \& Simulation}, 9\penalty0 (4):\penalty0
  1588--1623, 2011.

\bibitem[Chekroun et~al.(2021)Chekroun, Liu, and
  McWilliams]{chekroun2021stochastic}
M.D. Chekroun, H.~Liu, and J.C. McWilliams.
\newblock Stochastic rectification of fast oscillations on slow manifold
  closures.
\newblock \emph{Proc. Natl. Acad. Sci. USA}, 118\penalty0 (48):\penalty0
  e2113650118, 2021.
\newblock \doi{10.1073/pnas.2113650118}.

\bibitem[Kondrashov et~al.(2018{\natexlab{a}})Kondrashov, Chekroun, and
  Berloff]{Kondrashov_al2018_QG}
D.~Kondrashov, M.D. Chekroun, and P.~Berloff.
\newblock Multiscale {S}tuart-{L}andau emulators: {A}pplication to wind-driven
  ocean gyres.
\newblock \emph{Fluids}, 3\penalty0 (1):\penalty0 21, 2018{\natexlab{a}}.
\newblock \doi{10.3390/fluids3010021}.

\bibitem[Matkowsky and Schuss(1981)]{matkowsky1981eigenvalues}
B.J. Matkowsky and Z.~Schuss.
\newblock {Eigenvalues of the Fokker--Planck operator and the approach to
  equilibrium for diffusions in potential fields}.
\newblock \emph{SIAM Journal on Applied Mathematics}, 40\penalty0 (2):\penalty0
  242--254, 1981.

\bibitem[Sch{\"u}tte and Sarich(2013)]{schutte2013metastability}
C.~Sch{\"u}tte and M.~Sarich.
\newblock \emph{Metastability and Markov State Models in Molecular Dynamics},
  volume~24.
\newblock American Mathematical Society, 2013.

\bibitem[Pavliotis(2014)]{pavliotis2014stochastic}
G.A. Pavliotis.
\newblock \emph{{Stochastic Processes and Applications: Diffusion processes,
  the Fokker-Planck and Langevin Equations}}, volume~60.
\newblock Springer, 2014.

\bibitem[Tantet et~al.(2018{\natexlab{b}})Tantet, Lucarini, and
  Dijkstra]{Tantet2018}
A.~Tantet, V.~Lucarini, and H.A. Dijkstra.
\newblock {Resonances in a Chaotic Attractor Crisis of the Lorenz Flow}.
\newblock \emph{Journal of Statistical Physics}, 170\penalty0 (3):\penalty0
  584--616, 2018{\natexlab{b}}.
\newblock ISSN 00224715.
\newblock \doi{10.1007/s10955-017-1938-0}.

\bibitem[Ashwin and von~der Heydt(2019)]{AshwinJSP}
P.~Ashwin and A.S. von~der Heydt.
\newblock {Extreme Sensitivity and Climate Tipping Points}.
\newblock \emph{Journal of Statistical Physics}, 2019.
\newblock \doi{10.1007/s10955-019-02425-x}.
\newblock URL \url{https://doi.org/10.1007/s10955-019-02425-x}.

\bibitem[Stanley(1971)]{stanley1971}
H.E. Stanley.
\newblock \emph{Introduction to Phase Transitions and Critical Phenomena}.
\newblock Clarendon Press, 1971.
\newblock ISBN 9780198512578.
\newblock URL \url{https://books.google.co.uk/books?id=CN5XAAAAYAAJ}.

\bibitem[Ulam(1960)]{ulam1960collection}
S.M. Ulam.
\newblock \emph{{A Collection of Mathematical Problems}}, volume~8.
\newblock Interscience Publishers, 1960.

\bibitem[Junge and Koltai(2009)]{junge2009discretization}
O.~Junge and P.~Koltai.
\newblock {Discretization of the Frobenius--Perron operator using a sparse Haar
  tensor basis: the sparse Ulam method}.
\newblock \emph{SIAM Journal on Numerical Analysis}, 47\penalty0 (5):\penalty0
  3464--3485, 2009.

\bibitem[Klus et~al.(2018)Klus, N{\"u}ske, Koltai, Wu, Kevrekidis, Sch{\"u}tte,
  and No{\'e}]{klus2018data}
S.~Klus, F.~N{\"u}ske, P.~Koltai, H.~Wu, I.~Kevrekidis, C.~Sch{\"u}tte, and
  F.~No{\'e}.
\newblock Data-driven model reduction and transfer operator approximation.
\newblock \emph{Journal of Nonlinear Science}, 28\penalty0 (3):\penalty0
  985--1010, 2018.

\bibitem[Froyland and Dellnitz(2003)]{froyland2003detecting}
G.~Froyland and M.~Dellnitz.
\newblock Detecting and locating near-optimal almost-invariant sets and cycles.
\newblock \emph{SIAM Journal on Scientific Computing}, 24\penalty0
  (6):\penalty0 1839--1863, 2003.

\bibitem[Froyland(1998)]{froylandapproximating1998}
G.~Froyland.
\newblock {Approximating physical invariant measures of mixing dynamical
  systems in higher dimensions}.
\newblock \emph{Nonlinear Analysis}, 32\penalty0 (7):\penalty0 831--860, 1998.

\bibitem[Fishman and Rahav(2002)]{fishman2002}
S.~Fishman and S.~Rahav.
\newblock {Relaxation and Noise in Chaotic Systems}.
\newblock In \emph{Dynamics of Dissipation}, pages 165--192. {Dynamics of
  Dissipation}, Berlin, 2002.

\bibitem[Crommelin and Vanden-Eijnden(2006)]{crommelin2006b}
D.T. Crommelin and E.~Vanden-Eijnden.
\newblock {Fitting timeseries by continuous-time Markov chains: A quadratic
  programming approach}.
\newblock \emph{Journal of Computational Physics}, 217\penalty0 (2):\penalty0
  782--805, 2006.
\newblock ISSN 10902716.
\newblock \doi{10.1016/j.jcp.2006.01.045}.

\bibitem[Dellnitz and Junge(1999)]{dellnitz1999approximation}
M.~Dellnitz and O.~Junge.
\newblock On the approximation of complicated dynamical behavior.
\newblock \emph{SIAM Journal on Numerical Analysis}, 36\penalty0 (2):\penalty0
  491--515, 1999.

\bibitem[Froyland et~al.(2010)Froyland, Lloyd, and Quas]{froyland2010coherent}
G.~Froyland, S.~Lloyd, and A.~Quas.
\newblock Coherent structures and isolated spectrum for perron--frobenius
  cocycles.
\newblock \emph{Ergodic Theory and Dynamical Systems}, 30\penalty0
  (3):\penalty0 729--756, 2010.

\bibitem[Tantet et~al.(2020{\natexlab{b}})Tantet, Chekroun, Dijkstra, and
  Neelin]{RP_ENSO}
A.~Tantet, M.D. Chekroun, H.A. Dijkstra, and J.D. Neelin.
\newblock {Ruelle-Pollicott resonances of stochastic systems in reduced state
  space. Part III: Application to the Cane--Zebiak model of the El
  Ni{\~n}o--Southern Oscillation}.
\newblock \emph{J. Stat. Phys.}, 179\penalty0 (5):\penalty0 1449--1474,
  2020{\natexlab{b}}.
\newblock \doi{10.1007/s10955-019-02444-8}.
\newblock URL \url{https://doi.org/10.1007/s10955-019-02444-8}.

\bibitem[Kondrashov et~al.(2017)Kondrashov, Chekroun, Yuan, and
  Ghil]{kondrashov2017data}
D.~Kondrashov, M.D. Chekroun, X.~Yuan, and M.~Ghil.
\newblock {Data-adaptive harmonic decomposition and stochastic modeling of
  Arctic sea ice}.
\newblock In \emph{Advances in nonlinear geosciences}, pages 179--205.
  Springer, 2017.

\bibitem[Kondrashov et~al.(2018{\natexlab{b}})Kondrashov, Chekroun, and
  Ghil]{MASIE_paper}
D.~Kondrashov, M.D. Chekroun, and M.~Ghil.
\newblock Data-adaptive harmonic decomposition and prediction of {A}rctic sea
  ice extent.
\newblock \emph{Dynamics and Statistics of the Climate System}, 3\penalty0
  (1):\penalty0 1--23, 2018{\natexlab{b}}.
\newblock \doi{10.1093/climsys/dzy001}.

\bibitem[Kondrashov and Chekroun(2018)]{kondrashov2018data}
D.~Kondrashov and M.D. Chekroun.
\newblock Data-adaptive harmonic analysis and modeling of solar
  wind-magnetosphere coupling.
\newblock \emph{Journal of Atmospheric and Solar-Terrestrial Physics},
  177:\penalty0 179--189, 2018.

\bibitem[Kondrashov et~al.(2020)Kondrashov, Ryzhov, and
  Berloff]{kondrashov2020turbulence}
D.~Kondrashov, E.A. Ryzhov, and P.~Berloff.
\newblock Data-adaptive harmonic analysis of oceanic waves and turbulent flows.
\newblock \emph{Chaos}, 30\penalty0 (6), 2020.

\bibitem[Lorenz(1956)]{Lorenz1956}
E.N. Lorenz.
\newblock Empirical orthogonal functions and statistical weather prediction.
\newblock {\it Sci. Rep. No. 1}, {\it Statistical Forecasting Project}, M.I.T,
  Cambridge, MA, 1956.
\newblock 48 pp.

\bibitem[Groth and Ghil(2011)]{Groth_Ghil2011}
A.~Groth and M.~Ghil.
\newblock Multivariate singular spectrum analysis and the road to phase
  synchronization.
\newblock \emph{Phys. Rev. E}, 84:\penalty0 036206 (10 pp.), 2011.
\newblock \doi{10.1103/PhysRevE.84.036206}.

\bibitem[Dror et~al.(2021)Dror, Chekroun, Altaratz, and Koren]{dror_2021}
T.~Dror, M.D. Chekroun, O.~Altaratz, and I.~Koren.
\newblock {Deciphering organization of GOES-16 green cumulus through the
  empirical orthogonal function (EOF) lens}.
\newblock \emph{Atmospheric Chemistry and Physics}, 21\penalty0 (16):\penalty0
  12261--12272, 2021.
\newblock \doi{10.5194/acp-21-12261-2021}.
\newblock URL \url{https://acp.copernicus.org/articles/21/12261/2021/}.

\bibitem[Zerenner et~al.(2021)Zerenner, Goodfellow, and
  Ashwin]{zerenner2021harmonic}
T.~Zerenner, M.~Goodfellow, and P.~Ashwin.
\newblock Harmonic cross-correlation decomposition for multivariate time
  series.
\newblock \emph{Physical Review E}, 103\penalty0 (6):\penalty0 062213, 2021.

\bibitem[Kondrashov et~al.(2026)Kondrashov, Sudakow, Livina, and
  Yang]{kondrashov2026accurate}
D.~Kondrashov, I.~Sudakow, V.~Livina, and Q.~Yang.
\newblock {Accurate and robust real-time prediction of September Arctic sea
  ice}.
\newblock \emph{Chaos}, 36\penalty0 (2), 2026.

\bibitem[Romanou et~al.(2025)Romanou, Hegerl, Seneviratne, Abis, Bastos,
  Conversi, Landolfi, Kim, Lerner, Mekus, Otto-Bliesner, Pausata, Pinto, and
  Suarez-Guiterrez]{Romanou2025}
A.~Romanou, G.~C. Hegerl, S.~I. Seneviratne, B.~Abis, A.~Bastos, A.~Conversi,
  A.~Landolfi, H.~Kim, P.~E. Lerner, J.~Mekus, B.~L. Otto-Bliesner, F.~S.~R.
  Pausata, I.~Pinto, and L.~Suarez-Guiterrez.
\newblock Extreme events contributing to tipping elements and tipping points.
\newblock \emph{Surveys in Geophysics}, 46\penalty0 (2):\penalty0 375--420,
  2025.
\newblock ISSN 1573-0956.
\newblock \doi{10.1007/s10712-024-09863-7}.
\newblock URL \url{https://doi.org/10.1007/s10712-024-09863-7}.

\bibitem[Keener and Keller(1973)]{KeenerKeller1973}
J.~P. Keener and H.~B. Keller.
\newblock {Perturbed Bifurcation Theory}.
\newblock \emph{Arch. Rational Mech. Anal.}, 50:\penalty0 159--175, 1973.

\bibitem[Golubitsky and Schaeffer(1979)]{GolubitskySchaeffer1979}
M.~Golubitsky and D.~Schaeffer.
\newblock {A Theory for Imperfect Bifurcation via Singularity Theory}.
\newblock \emph{Comm. Pure Appl. Math.}, 32\penalty0 (1):\penalty0 21--98,
  1979.

\bibitem[Chekroun(2018)]{chekroun2018topological}
M.D. Chekroun.
\newblock Topological instabilities in families of semilinear parabolic
  problems subject to nonlinear perturbations.
\newblock \emph{Discrete \& Continuous Dynamical Systems-B}, 23\penalty0
  (9):\penalty0 3723, 2018.
\newblock \doi{10.3934/dcdsb.2018075}.

\bibitem[Cazenave et~al.(2002)Cazenave, Escobedo, and
  Pozio]{CazenaveEscobedoPozio2002}
T.~Cazenave, M.~Escobedo, and A.~Pozio.
\newblock {Some Stability Properties for Minimal Solutions of $-\Delta
  u=\lambda g(u)$}.
\newblock \emph{Portugaliae Math.}, 59:\penalty0 373--391, 2002.

\bibitem[Klein and Hartmann(1993)]{KleinHartmann1993}
S.~A. Klein and D.~L. Hartmann.
\newblock The seasonal cycle of low stratiform clouds.
\newblock \emph{J. Climate}, 6\penalty0 (8):\penalty0 1587--1606, 1993.
\newblock
  \href{https://doi.org/10.1175/1520-0442(1993)006<1587:TSCOLS>2.0.CO;2}{doi:10.1175/1520-0442(1993)006<1587:TSCOLS>2.0.CO;2}.

\bibitem[Bony and Dufresne(2005)]{BonyDufresne2005}
S.~Bony and J.-L. Dufresne.
\newblock Marine boundary layer clouds at the heart of tropical cloud feedback
  uncertainties in climate models.
\newblock \emph{Geophys. Res. Lett.}, 32\penalty0 (20):\penalty0 L20806, 2005.
\newblock
  \href{https://doi.org/10.1029/2005GL023851}{doi:10.1029/2005GL023851}.

\bibitem[Boucher et~al.(2013)Boucher, Randall, Artaxo, Bretherton, Feingold,
  Forster, Kerminen, Kondo, Liao, Lohmann, et~al.]{boucher_2013clouds}
O.~Boucher, D.~Randall, P.~Artaxo, C.~Bretherton, G.~Feingold, P.~Forster,
  V.-M. Kerminen, Y.~Kondo, H.~Liao, U.~Lohmann, et~al.
\newblock Clouds and aerosols.
\newblock In \emph{Climate change 2013: the physical science basis.
  Contribution of Working Group I to the Fifth Assessment Report of the
  Intergovernmental Panel on Climate Change}, pages 571--657. Cambridge
  University Press, 2013.

\bibitem[Zelinka et~al.(2020)Zelinka, Myers, McCoy, Po-Chedley, Caldwell,
  Ceppi, Klein, and Taylor]{Zelinka2020}
M.~D. Zelinka, T.~A. Myers, D.~T. McCoy, S.~Po-Chedley, P.~M. Caldwell,
  P.~Ceppi, S.~A. Klein, and K.~E. Taylor.
\newblock Causes of higher climate sensitivity in {CMIP6} models.
\newblock \emph{Geophys. Res. Lett.}, 47\penalty0 (1):\penalty0 e2019GL085782,
  2020.
\newblock
  \href{https://doi.org/10.1029/2019GL085782}{doi:10.1029/2019GL085782}.

\bibitem[Liu et~al.(2023)Liu, Koren, Altaratz, and Chekroun]{liu2023opposing}
H.~Liu, I.~Koren, O.~Altaratz, and M.D. Chekroun.
\newblock Opposing trends of cloud coverage over land and ocean under global
  warming.
\newblock \emph{Atmospheric Chemistry and Physics}, 23\penalty0 (11):\penalty0
  6559--6569, 2023.
\newblock
  \href{https://doi.org/10.5194/acp-23-6559-2023}{doi:10.5194/acp-23-6559-2023}.

\bibitem[Abbot et~al.(2011)Abbot, Voigt, and Koll]{Abbott2011}
Dorian~S. Abbot, Aiko Voigt, and Daniel Koll.
\newblock The {J}ormungand global climate state and implications for
  {N}eoproterozoic glaciations.
\newblock \emph{Journal of Geophysical Research: Atmospheres}, 116\penalty0
  (D18), 2011.
\newblock \doi{https://doi.org/10.1029/2011JD015927}.
\newblock URL
  \url{https://agupubs.onlinelibrary.wiley.com/doi/abs/10.1029/2011JD015927}.

\bibitem[Samuelsberg and Jakobsen(2025)]{Samuelberg2025}
Aksel Samuelsberg and Per~Kristen Jakobsen.
\newblock {S}lushball {E}arth equilibria in a one-dimensional energy balance
  model.
\newblock \emph{Physica D: Nonlinear Phenomena}, 482:\penalty0 134866, 2025.
\newblock ISSN 0167-2789.
\newblock \doi{https://doi.org/10.1016/j.physd.2025.134866}.
\newblock URL
  \url{https://www.sciencedirect.com/science/article/pii/S0167278925003434}.

\bibitem[Lohmann et~al.(2024)Lohmann, Dijkstra, Jochum, Lucarini, and
  Ditlevsen]{Lohmannetal2024}
Johannes Lohmann, Henk~A. Dijkstra, Markus Jochum, Valerio Lucarini, and
  Peter~D. Ditlevsen.
\newblock Multistability and intermediate tipping of the {A}tlantic {O}cean
  circulation.
\newblock \emph{Science Advances}, 10\penalty0 (12):\penalty0 eadi4253, 2024.
\newblock \doi{10.1126/sciadv.adi4253}.
\newblock URL \url{https://www.science.org/doi/abs/10.1126/sciadv.adi4253}.

\bibitem[Ditlevsen and Ditlevsen(2023)]{Ditlevsen2023}
Peter Ditlevsen and Susanne Ditlevsen.
\newblock Warning of a forthcoming collapse of the {A}tlantic meridional
  overturning circulation.
\newblock \emph{Nature Communications}, 14\penalty0 (1):\penalty0 4254, 2023.
\newblock ISSN 2041-1723.
\newblock \doi{10.1038/s41467-023-39810-w}.
\newblock URL \url{https://doi.org/10.1038/s41467-023-39810-w}.

\bibitem[Feudel(2008)]{Feudel2008_complex_multistability}
U.~Feudel.
\newblock Complex dynamics in multistable systems.
\newblock \emph{International Journal of Bifurcation and Chaos}, 18\penalty0
  (6):\penalty0 1607--1626, 2008.

\bibitem[Chekroun and Liu(2024)]{chekroun2024effective}
M.D. Chekroun and H.~Liu.
\newblock {Effective reduced models from delay differential equations:
  Bifurcations, tipping solution paths, and ENSO variability}.
\newblock \emph{Physica D: Nonlinear Phenomena}, 460:\penalty0 134058, 2024.

\bibitem[Kumar~K et~al.(2025)Kumar~K, Alkhayuon, Wieczorek, and
  Sharathi~Dutta]{kumar2025pace}
R.~Kumar~K, H.~Alkhayuon, S.~Wieczorek, and P.~Sharathi~Dutta.
\newblock Pace in concert with phase: rate-induced phase-tipping in birhythmic
  oscillators.
\newblock \emph{Proceedings of the Royal Society A}, 481\penalty0
  (2311):\penalty0 20240629, 2025.

\bibitem[Rowley et~al.(2009)Rowley, Mezi{\'c}, Bagheri, Schlatter, and
  Henningson]{rowley2009spectral}
C.W. Rowley, I.~Mezi{\'c}, S.~Bagheri, P.~Schlatter, and D.S. Henningson.
\newblock Spectral analysis of nonlinear flows.
\newblock \emph{Journal of Fluid Mechanics}, 641:\penalty0 115--127, 2009.
\newblock \doi{10.1017/S0022112009992059}.

\bibitem[Schmid(2010)]{Schmid2010}
P.J. Schmid.
\newblock {Dynamic mode decomposition of numerical and experimental data}.
\newblock \emph{Journal of Fluid Mechanics}, 656:\penalty0 5--28, jul 2010.
\newblock ISSN 0022-1120.
\newblock \doi{10.1017/S0022112010001217}.
\newblock URL
  \url{http://www.journals.cambridge.org/abstract{\_}S0022112010001217}.

\bibitem[Tu et~al.(2014)Tu, Rowley, Luchtenburg, Brunton, and
  Kutz]{tu2014dynamic}
J.H. Tu, C.W. Rowley, D.M. Luchtenburg, S.L. Brunton, and J.N. Kutz.
\newblock {On dynamic mode decomposition: Theory and applications}.
\newblock \emph{Journal of Computational Dynamics}, 1\penalty0 (2):\penalty0
  391--421, 2014.

\bibitem[Williams et~al.(2015)Williams, Kevrekidis, and
  Rowley]{williams2015data}
M.O. Williams, I.G. Kevrekidis, and C.W. Rowley.
\newblock {{A data-driven approximation of the Koopman operator: Extending
  dynamic mode decomposition}}.
\newblock \emph{Journal of Nonlinear Science}, 25\penalty0 (6):\penalty0
  1307--1346, 2015.

\bibitem[Lohmann et~al.(2025)Lohmann, Hansen, Lovo, Chapman, Bouchet, and
  Lucarini]{Lohmann2025}
J.~Lohmann, A.B. Hansen, A.~Lovo, R.~Chapman, F.~Bouchet, and V.~Lucarini.
\newblock The role of edge states for early warning of tipping points.
\newblock \emph{Proceedings of the Royal Society A: Mathematical, Physical and
  Engineering Sciences}, 481\penalty0 (2312):\penalty0 20240753, 04 2025.
\newblock ISSN 1364-5021.
\newblock \doi{10.1098/rspa.2024.0753}.
\newblock URL \url{https://doi.org/10.1098/rspa.2024.0753}.

\bibitem[Li et~al.(2017)Li, Dietrich, Bollt, and Kevrekidis]{li2017extended}
Q.~Li, F.~Dietrich, E.M. Bollt, and I.G. Kevrekidis.
\newblock {Extended dynamic mode decomposition with dictionary learning: A
  data-driven adaptive spectral decomposition of the Koopman operator}.
\newblock \emph{Chaos}, 27\penalty0 (10):\penalty0 103111, 2017.

\bibitem[Yeung et~al.(2019)Yeung, Kundu, and Hodas]{yeung2019learning}
E.~Yeung, S.~Kundu, and N.~Hodas.
\newblock {Learning deep neural network representations for Koopman operators
  of nonlinear dynamical systems}.
\newblock In \emph{2019 American Control Conference (ACC)}, pages 4832--4839.
  IEEE, 2019.

\bibitem[Lusch et~al.(2018)Lusch, Kutz, and Brunton]{lusch2018deep}
B.~Lusch, J.N. Kutz, and S.L. Brunton.
\newblock Deep learning for universal linear embeddings of nonlinear dynamics.
\newblock \emph{Nature communications}, 9\penalty0 (1):\penalty0 1--10, 2018.

\bibitem[Bouchet et~al.(2019)Bouchet, Rolland, and Simonnet]{bouchet2019rare}
F.~Bouchet, J.~Rolland, and E.~Simonnet.
\newblock Rare event algorithm links transitions in turbulent flows with
  activated nucleations.
\newblock \emph{Physical Review Letters}, 122\penalty0 (7):\penalty0 074502,
  2019.

\bibitem[Li et~al.(2019)Li, Lin, and Ren]{li2019computing}
Q.~Li, B.~Lin, and W.~Ren.
\newblock Computing committor functions for the study of rare events using deep
  learning.
\newblock \emph{The Journal of Chemical Physics}, 151\penalty0 (5):\penalty0
  054112, 2019.

\bibitem[Duan et~al.(2003)Duan, Lu, and Schmalfuss]{DLS03}
J.~Duan, K.~Lu, and B.~Schmalfuss.
\newblock Invariant manifolds for stochastic partial differential equations.
\newblock \emph{Ann. Probab.}, 31:\penalty0 2109--2135, 2003.

\bibitem[Caraballo et~al.(2010)Caraballo, Duan, Lu, and
  Schmalfu\ss]{Caraballo_al09}
T.~Caraballo, J.~Duan, K.~Lu, and B.~Schmalfu\ss.
\newblock Invariant manifolds for random and stochastic partial differential
  equations.
\newblock \emph{Advanced Nonlinear Studies}, 10\penalty0 (1):\penalty0 23--52,
  2010.

\bibitem[Duan(2015)]{DuanJ2015}
J.~Duan.
\newblock \emph{{An Introduction to Stochastic Dynamics}}.
\newblock Cambridge University Press, 2015.
\newblock ISBN 9781107075399.

\bibitem[Chekroun et~al.(2015{\natexlab{a}})Chekroun, Liu, and
  Wang]{CLW15_vol1}
M.D. Chekroun, H.~Liu, and S.~Wang.
\newblock \emph{{Approximation of Stochastic Invariant Manifolds: Stochastic
  Manifolds for Nonlinear SPDEs I}}.
\newblock Springer Briefs in Mathematics, Springer, New York,
  2015{\natexlab{a}}.

\bibitem[Chekroun et~al.(2015{\natexlab{b}})Chekroun, Liu, and
  Wang]{CLW15_vol2}
M.D. Chekroun, H.~Liu, and S.~Wang.
\newblock \emph{{Stochastic Parameterizing Manifolds and Non-Markovian Reduced
  Equations: Stochastic Manifolds for Nonlinear SPDEs II}}.
\newblock Springer Briefs in Mathematics, Springer, 2015{\natexlab{b}}.

\bibitem[Kondrashov et~al.(2006)Kondrashov, Kravtsov, and Ghil]{kkg06}
D.~Kondrashov, S.~Kravtsov, and M.~Ghil.
\newblock Empirical mode reduction in a model of extratropical low-frequency
  variability.
\newblock \emph{J. Atmos. Sci.}, 63:\penalty0 1859--1877, 2006.

\bibitem[Santos~Guti{\'e}rrez et~al.(2021)Santos~Guti{\'e}rrez, Lucarini,
  Chekroun, and Ghil]{santos2021reduced}
M.~Santos~Guti{\'e}rrez, V.~Lucarini, M.D. Chekroun, and M.~Ghil.
\newblock {Reduced-order models for coupled dynamical systems: Data-driven
  methods and the Koopman operator}.
\newblock \emph{Chaos}, 31\penalty0 (5):\penalty0 053116, 2021.

\bibitem[Chekroun et~al.(2020{\natexlab{b}})Chekroun, Liu, and
  McWilliams]{CLM19_closure}
M.D. Chekroun, H.~Liu, and J.C. McWilliams.
\newblock {Variational approach to closure of nonlinear dynamical systems:
  Autonomous case}.
\newblock \emph{J. Stat. Phys.}, 179:\penalty0 1073--1160, 2020{\natexlab{b}}.
\newblock \doi{10.1007/s10955-019-02458-2}.

\bibitem[Chekroun et~al.(2023{\natexlab{a}})Chekroun, Liu, and
  McWilliams]{chekroun2023optimal}
M.D. Chekroun, H.~Liu, and J.C. McWilliams.
\newblock Optimal parameterizing manifolds for anticipating tipping points and
  higher-order critical transitions.
\newblock \emph{Chaos}, 33\penalty0 (9), 2023{\natexlab{a}}.

\bibitem[Chekroun et~al.(2023{\natexlab{b}})Chekroun, Liu, McWilliams, and
  Wang]{chekroun2023transitions}
M.D. Chekroun, H.~Liu, J.C. McWilliams, and S.~Wang.
\newblock {{Transitions in stochastic non-equilibrium systems: Efficient
  reduction and analysis}}.
\newblock \emph{Journal of Differential Equations}, 346:\penalty0 145--204,
  2023{\natexlab{b}}.

\bibitem[Chekroun et~al.(2025{\natexlab{b}})Chekroun, Liu, and
  McWilliams]{chekroun2025unravel}
M.D. Chekroun, H.~Liu, and J.C. McWilliams.
\newblock {Non-Markovian reduced models to unravel transitions in
  non-equilibrium systems}.
\newblock \emph{J. Phys. A: Math. Theor.}, 58\penalty0 (4):\penalty0 045204,
  2025{\natexlab{b}}.
\newblock
  \href{https://doi.org/10.1088/1751-8121/ada7ad}{doi:10.1088/1751-8121/ada7ad}.

\bibitem[Chekroun et~al.(2020{\natexlab{c}})Chekroun, Koren, and
  Liu]{chekroun2020efficient}
M.D. Chekroun, I.~Koren, and H.~Liu.
\newblock {Efficient reduction for diagnosing Hopf bifurcation in delay
  differential systems: Applications to cloud-rain models}.
\newblock \emph{Chaos}, 30\penalty0 (5):\penalty0 053130, 2020{\natexlab{c}}.
\newblock \doi{10.1063/5.0004697}.

\bibitem[Young(2016)]{Young2016}
L.-S. Young.
\newblock Generalizations of {SRB} measures to nonautonomous, random, and
  infinite dimensional systems.
\newblock \emph{J. Stat. Phys.}, 166:\penalty0 494--515, 2016.
\newblock \doi{10.1007/s10955-016-1639-0}.

\bibitem[Chekroun et~al.(2022)Chekroun, Koren, Liu, and
  Liu]{Chekroun_al22SciAdv}
M.D. Chekroun, I.~Koren, H.~Liu, and H.~Liu.
\newblock Generic generation of noise-driven chaos in stochastic time delay
  systems: Bridging the gap with high-end simulations.
\newblock \emph{Science Advances}, 8\penalty0 (46):\penalty0 eabq7137, 2022.
\newblock
  \href{https://doi.org/10.1126/sciadv.abq7137}{doi.org/10.1126/sciadv.abq7137}.

\bibitem[Lucarini(2026)]{Lucarini2026}
Valerio Lucarini.
\newblock Linear response and optimal fingerprinting for nonautonomous systems.
\newblock \emph{arXiv preprint}, 2026.
\newblock \href{https://arxiv.org/abs/2602.08022}{arXiv:2602.08022}.

\bibitem[Dijkstra(2005)]{Dijkstra05}
H.A. Dijkstra.
\newblock \emph{Nonlinear Physical Oceanography : A Dynamical Systems Approach
  to the Large-Scale Ocean Circulation and {E}l {N}i\~no}.
\newblock Springer, Dordrecht, The Netherlands, 2005.

\bibitem[Keller(1977)]{Keller1977}
H.~B. Keller.
\newblock {Numerical Solution of Bifurcation and Nonlinear Eigenvalue
  Problems}.
\newblock In P.~H. Rabinowitz, editor, \emph{{Applications of Bifurcation
  Theory}}, pages 359--384. Academic Press, New York, 1977.

\bibitem[Skufca et~al.(2006)Skufca, Yorke, and Eckhardt]{Skufca2006}
J.~D. Skufca, J.~A. Yorke, and B.~Eckhardt.
\newblock {Edge of Chaos in a Parallel Shear Flow}.
\newblock \emph{Phys. Rev. Lett.}, 96\penalty0 (17):\penalty0 174101, 2006.

\end{thebibliography}

\end{document}